\documentclass[acmtog,nonacm]{acmart}
\setcopyright{none}
\renewcommand\footnotetextcopyrightpermission[1]{}
\usepackage[T1]{fontenc}
\usepackage[utf8]{inputenc}
\usepackage{microtype}
\usepackage[normalem]{ulem}
\usepackage{siunitx}

\usepackage{amsfonts}
\usepackage{mathtools} %

\usepackage{ifthen}
\usepackage{float} %
\usepackage{hyperref}
\usepackage{cleveref}
\usepackage{csquotes}
\usepackage{calc}
\usepackage{placeins}
\usepackage{colortbl}

\usepackage{algorithmicx}
\usepackage{algpseudocode}
\usepackage{listingsutf8}

\usepackage{array}
\usepackage{booktabs}
\usepackage{multirow}
\usepackage{caption}
\usepackage{subcaption}
\usepackage{tabularx}

\usepackage{float}
\usepackage{wrapfig} %
\usepackage{graphicx}
\usepackage[percent]{overpic}
\usepackage{subcaption}

\definecolor{DaviColor}{rgb}{0,0.70,0.20}
\definecolor{ZizhouColor}{rgb}{0.34,0.1,0.25}
\definecolor{DanieleColor}{rgb}{0.56,0.34,0.62}
\definecolor{DenisColor}{rgb}{0.55,0.35,0.05}

\definecolor{Gray}{rgb}{0.9,0.9,0.9}

\newcommand{\DP}[1]{{\leavevmode\color{DanieleColor} Daniele: #1 $\square$}}
\newcommand{\DZ}[1]{{\leavevmode\color{DenisColor} Denis: #1 $\square$}}
\newcommand{\ZH}[1]{{\leavevmode\color{ZizhouColor} Zizhou: #1 $\square$}}
\newcommand{\DT}[1]{{\leavevmode\color{DaviColor} Davi: #1 $\square$}}

\newcommand{\nothing}[1]{}

\providecommand{\finalversion}{0} %
\ifthenelse{\equal{\finalversion}{1}}
{
	\renewcommand{\DP}[1]{}
	\renewcommand{\ZH}[1]{}
	\renewcommand{\DZ}[1]{}
	\renewcommand{\DT}[1]{}

}
{}

\definecolor{forestgreen}{rgb}{0.13,0.54,0.13}
\definecolor{darkblue}{rgb}{0,0,.5}
\hypersetup{
	unicode=true,
	colorlinks=true,
	citecolor=forestgreen, %
	linkcolor=darkblue, %
	urlcolor=darkblue, %
}

\algrenewcommand{\algorithmiccomment}[1]{{\footnotesize\color{forestgreen}\(\triangleright\) #1}}

\crefname{algocf}{alg.}{algs.}
\Crefname{algocf}{Algorithm}{Algorithms}

\crefname{appsec}{Appendix}{Appendices}

\DeclareOldFontCommand{\bf}{\normalfont\bfseries}{\mathbf}

\let\originalleft\left
\let\originalright\right
\renewcommand{\left}{\mathopen{}\mathclose\bgroup\originalleft}
\renewcommand{\right}{\aftergroup\egroup\originalright}

\renewcommand{\geq}{\geqslant}

\renewcommand{\vec}[1]{{\bm{{#1}}}}

\newcommand{\cE}{\mathcal{E}}

\renewcommand\vec[1]{{\mathbf #1}}

\DeclareFontFamily{U}{mathx}{\hyphenchar\font45}
\DeclareFontShape{U}{mathx}{m}{n}{<-> mathx10}{}
\DeclareSymbolFont{mathx}{U}{mathx}{m}{n}

\def\u{\vec{u}}

\def\p{\vec{p}}

\def\vf{\vec{f}}

\def\strain{\varepsilon}

\def\mtg{\mathcal{P}}
\def\mtgb{\mathcal{P}_b}

\graphicspath{{images/}}

\begin{document}
\title{Cut-Cell Microstructures for Two-scale Structural Optimization}

\author{Davi Colli Tozoni}
\affiliation{%
\institution{nTop / New York University}
\country{USA}
}
\email{davi.tozoni@nyu.edu}

% Author list
\author{Zizhou Huang}
\affiliation{%
\institution{New York University}
\country{USA}
}
\email{zizhou@nyu.edu}

\author{Daniele Panozzo}
\affiliation{%
\institution{New York University}
\country{USA}
}
\email{panozzo@nyu.edu}

\author{Denis Zorin}
\affiliation{%
\institution{New York University}
\country{USA}
}
\email{dzorin@cs.nyu.edu}

\begin{abstract}
Two-scale topology optimization, combined with the design of microstructure families with a broad range of effective material parameters, is increasingly widely used in many fabrication applications to achieve a target deformation behavior for a variety of objects.   
The main idea of this approach is to optimize the distribution of material properties in the object partitioned into relatively coarse cells, and then replace each cell  with microstructure geometry that mimics these material properties.  In this paper, we focus on adapting this approach to complex shapes in situations when  preserving the shape's surface is important.   

Our approach extends any regular (i.e. defined on a regular lattice grid) microstructure family to complex shapes, by enriching it with individually optimized cut-cell tiles adapted to the geometry of the cut-cell. 
We propose an automated and robust pipeline based on this approach, and we show that the performance of the regular microstructure family is only minimally affected by our extension while allowing its use on 2D and 3D shapes of high complexity.
\end{abstract}

\begin{teaserfigure}
	\centering
    \includegraphics[width=0.9\textwidth]{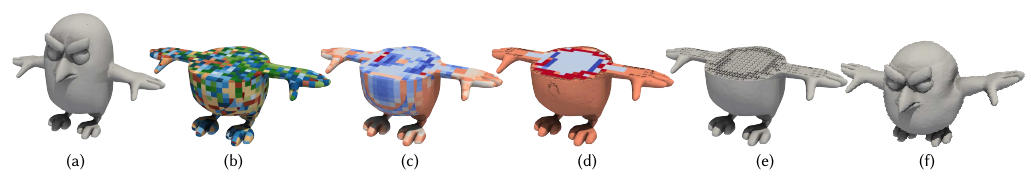}
    \caption{(a) Our method starts with an arbitrary initial surface mesh. (b) We generate a regular 3d grid enclosing the object, further subdivided into tetrahedra with the surface of the tetrahedral mesh conforming to the object's boundary. (c) We perform optimization for a particular deformation behavior (wings going up by a prescribed distance with pressure on the top) and obtain a material distribution (the color shows Young's modulus). (d) The cut-cell microstructure is inserted into boundary cells to match the  Young's modulus computed by the optimization, and then a second optimization is performed only on interior cells to compensate for the change in the deformation due to deviation of effective elastic properties of cut-cell microstructure from the optimized Young modulus (e) The full-cell microstructures  with effective Young modulus and Poisson ratio matching the optimized values are inserted into the interior cells (f) The deformation result on the final model. }
	\label{fig:teaser}
\end{teaserfigure}

\maketitle

\section{Introduction}
\label{sec:intro}

Additive manufacturing enables  easy fabrication of complex geometric structures designed with specific effective material properties in mind, with a broad range of applications from lightweight but strong structures in aerospace to shoe soles, prosthetic devices, and flexible robot parts~\cite{ntopology,carbon}. While the more complex geometry can be easily handled by modern 3D printing software and hardware, optimized design of such objects is a fast-growing research area. Most traditional engineering analysis and optimization software is not suitable for this task, due to limitations on the complexity and scale of geometry that can be handled. 

A popular approach is to directly optimize the shape of the entire structure by using topology optimization and/or shape optimization. However, the computation cost of these direct methods is high, especially for large objects, often requiring HPC clusters \cite{Aage:2017:GVC}. An alternative is to use a two-scale approach: (1) a material optimization is performed on a voxelized grid to compute material properties varying between voxels and resulting in desired target deformation behavior, and (2) each voxel is replaced by a geometrical structure (microstructure tile) with effective properties matching the computed solid material properties in the voxel. The advantage of this strategy, compared to whole object high-resolution topology optimization, is that step (1) can be very efficient and step (2) can be reduced to a lookup table: a family of microstructure tiles, mapping material properties to geometry, which can be precomputed using a cluster, and then used for multiple designs. 

\begin{figure}
    \centering
    \includegraphics[width=0.8\columnwidth]{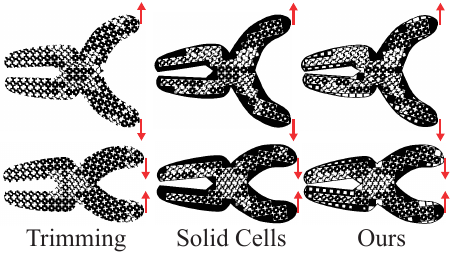}
    \caption{Two examples comparing simpler baselines, trimming (left) and solid cells (middle), with our approach (right). The scissor is optimized to close when the handles are pulled apart (top row) or pulled together (bottom row). Our approach succeeds in both cases, while trimming performs very poorly, and solid cells are somewhere in-between. We provide a detailed comparison against the solid cells baseline in Section \ref{sec:results}.}
    \label{fig:comparison-2D}
\end{figure}

\begin{figure}
    \centering
    \includegraphics[width=0.8\columnwidth]{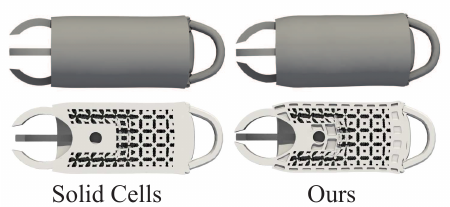}
    \caption{Comparison between the solid baseline (left) and our method (right) in 3D. The top row shows both versions seen from the outside (as someone using the gripper would see it), while the bottom row shows section cuts of both shapes. The structure is optimized so that its jaws close together under a compression force on the handle (gripper's right side). The baseline example fails to reach the prescribed deformation.}
    \label{fig:comparison-3D}
\end{figure}

Unavoidably, the voxelization of a shape leads to artifacts in the boundary, becoming problematic in applications where the shape's boundary matters, which is common in many settings, from moving engine parts to prosthetic devices.  Trivial approaches, such as trimming the microstructure to follow the boundary or filling the voxels on the shape's boundary, lead to a major change in effective microstructure properties in this cell, and  may result in significant deviations from the intended behavior under surface loads (see Figure~\ref{fig:comparison-2D} and ~\ref{fig:comparison-3D} for a comparison and Section \ref{sec:results} for a detailed analysis). 
To address this problem, these approaches have been extended, in 2D only, from regular grids to boundary-aligned structured meshes \cite{tozoni2020low}. Still, the strong requirement for mesh uniformity and quality makes them less robust than the grid-aligned methods, which are still the standard de facto in industry \cite{ntopology}. In 3D, such an extension is much more difficult: while planar quad mesh generation is possible (although quads may be far from parallelograms required by \cite{tozoni2020low}), generating boundary-aligned hex meshes is still an open problem, with a recent state-of-the-art method yielding a 50\% success rate on a recent benchmark \cite{liu2023locally}.

We propose a novel family of cut-cell microstructures to extend existing 2D or 3D regular tiling microstructure families to shapes with a complex boundary. Our construction ensures that the shape boundary exactly matches the input (with the exception of tunnels needed to remove the internal material after printing), and, in  combination with an interior structure, leads to accurate matching of the prescribed deformation behavior. Our algorithm is the first method enabling two-scale shape optimization on very complex shapes in 3D, in a fully automated and robust way, with the resulting complex geometric structures including the input boundary precisely. 

We demonstrate its effectiveness by combining our cut-cell family with a variant of the microstructure family proposed in \cite{tozoni2020low} to realize complex 3D shapes with controllable material properties, and evaluate their properties by fabricating them using 3D printing and testing them in a set of controlled load experiments.

To summarize, the contributions of this paper are:
\begin{itemize}
\item An extended microstructure construction that includes a structure for boundary cells which ensures their connection to the interior cells and supports arbitrary surface geometry. 
\item An algorithm for two-scale structural optimization, optimizing the material properties for interior and boundary cells, and a mapping from material to shape parameters of the microstructure, in two and three dimensions, with a refinement optimization pass. 
\item A robust geometric pipeline for constructing microstructure in-fill for arbitrary input shapes.  
\item Quantitative and qualitative physical evaluation of the optimized structures. 
\end{itemize}

\section{Related Work}
\label{sec:related}

\paragraph{Microstructure design and optimization}

There is extensive literature on the design of elastic and other types of metamaterials/microstructures. We refer to the books, e.g.,  \cite{cioranescu-book,milton-book,torquato-book} for the foundational work and the recent survey \cite{kadic20193d} for more up-to-date references. 
     
The approaches to the design of periodic microstructures with prescribed homogenized properties rely on topology optimization, going back to \cite{bendsoe1989optimal,bendsoe2003topology}, and shape optimization, as well as combinations of the two. Examples in computer graphics literature include \cite{schumacher2015microstructures} (topology optimization) and \cite{panetta2015elastic} for shape optimization, with many more in studies in material science and engineering. These works, as well as, e.g.,  \cite{Zhu:2017:TST}, consider constructing families spanning a broad range of elastic properties, with the potential maximally achievable ranges characterized in  \cite{Milton2017}. 

\paragraph{Global topology optimization}
In \cite{Wu:2016:ASF,Aage:2017:GVC,Liu:2018:NBT}, topology optimization  was scaled up to high-resolution uniform and adaptive 3D grids. \cite{Wu:2018:IOF} performs high-resolution topology optimization with additional constraints to create an evenly distributed porous small-scale structure minimizing compliance for specific loading scenarios. 
Even with these improvements, direct topology optimization at the microstructure level remains computationally expensive: e.g., a modest cell resolution of $32^3$ combined with an equally modest 
$32^3$ resolution at the coarse level requires over 1 billion cells to work with. To avoid these high computational costs, we favor a two-scale optimization approach.

\paragraph{Two-scale optimization using microstructures.}
Two-scale optimization uses microstructures to optimize the deformation behavior of objects efficiently by separating the problem into two scales through the partitioning of an object into cells. Fine-scale structures for individual cells are often precomputed to yield a particular range of effective material properties within the limit of infinitely fine cells. The coarse-scale optimization is performed by treating the whole cells as made of homogeneous material. Some works, e.g., \cite{liu2018parameterized}, do not precompute microstructures and optimize at both macro- and microscale simultaneously.  This approach, while most flexible, is computationally very expensive and does not support well typical workflows. Recent surveys \cite{wu2021topology} and \cite{plocher2019review} provide an overview of the approaches and recent work in this domain.  

All two-scale optimization approaches rely on tiling the input shapes with microstructures. The most common approach, and the current standard de facto in industry, is voxelization. Papers using voxelization focus on rectangular structures, as examples \cite{panetta2015elastic,zhu2017two}. Other works
\cite{rastegarzadeh2022two,xia2015multiscale,djourachkovitch2021multiscale} approximate arbitrary shapes  with collections of cubes.  This approach is unsuitable for applications that require preserving the shape of the surface (e.g., a wheel would not roll) or its appearance. This problem is exacerbated for relatively coarse grids of cells, which are often unavoidable due to limitations in the minimal beam thickness in many 3D printing technologies: an exception is \cite{sanders2021optimal}, where an expensive and slow printing process is used to reduce the size of voxels. Yet, despite the high resolution, the resulting surfaces are still not smooth.

The conforming tiling construction is introduced in \cite{tozoni2020low} for 2D and \cite{wang2021hierarchical} for 3D, where a voxelization is replaced by a boundary adapted, isotropic, low distortion quad or hexahedral tiling, and the domain of the microstructure families is extended accordingly. In addition to the increased complexity in the family design, the main drawback of these methods is that structured meshes with these properties are challenging to generate for arbitrary geometries. While some solutions exist for 2D \cite{tozoni2020low}, the solution to conforming hexahedral mesh generation  in 3D is a challenging open problem \cite{HexMe}, where the failure rate of state-of-the-art methods is shown to be around 50\%. %
Finally, the approach \cite{Groen:2019:HBS,Groen:2017:HBT,GilUreta:2019:SOS} for 2D and surface structures can be viewed as a partial conforming tiling construction: it uses a simple rectangular microstructure with two parameters, and a frame field to define the directions and scale for cells from a simple family.  The frame field orientations and other fields are chosen by optimization. Similarly to the previous category, this approach is also difficult to generalize reliably to 3D volumes due to a more complex structure of 3D fields \cite{Pietroni:2022}. 

Our approach shares the benefits of conforming tiling constructions, as it allows reproduction of boundaries for complex geometries, but is doing so by extending the voxelization methods using a cut-cell approach. Our algorithm combines the robustness and simplicity of the voxelization methods with the advantages of conforming tiling constructions without requiring a difficult-to-build quad or hexahedral mesh.

\paragraph{Special Shape-Adapted Microstructures}
\cite{konakovic2018rapid} uses a special type of 2D triangular auxetic structure to effect conformal surface deformations. This method requires domain meshing with triangles close to regular. Similarly, a recent paper \cite{Malomo:2018:FCD} uses 2D spiral microstructures.
As an alternative to periodic microstructures, \cite{martinez2016procedural,Martinez:2017:OKN} construct randomized printable structures with control over Young's moduli both for isotropic and anisotropic target properties. However, it cannot independently control the Poisson's ratio, and the behavior of randomized structures is less controllable overall. 

\paragraph{Variable Base Material.}
Our focus is fabrication methods supporting single base material due to their wide availability, lower cost, and support for stiff materials (metal). Having access to controllable base material further widens the range covered by microstructures: \cite{Bickel2010} designs and fabricates objects satisfying an input deformation using actual material properties variation, with fabrication done using a multi-material printer. 
\cite{skouras2013computational} applies discrete material optimization
to achieve desired deformations of complex characters, also using multi-material printing for fabrication. %

\section{Method}
\label{sec:method}

\subsection{Material Optimization Primer}
\label{sec:overview}

We provide a short introduction to material optimization and to two-scale optimization using microstructures to keep the paper self-contained. We refer to \cite{panetta2015elastic,schumacher2015microstructures} for more details.

Let $\Omega$ be the domain of a solid. Loads (Neumann boundary conditions) and/or Dirichlet boundary conditions are applied on some parts of $\Omega$, typically on the boundary. The deformed state is obtained by solving the linearized elasticity equation for the displacement $u$, 
\begin{equation}
    \label{eqn:elasticity}
  -\nabla \cdot [C: \strain(\u)] = 0 \quad
  \text{ in }\Omega,
\end{equation}
where $C_{ijkl}$ represents the 4th order elasticity tensor and $\strain(\u) \coloneqq \frac{1}{2}(\nabla \u + (\nabla \u)^T)$ is the linearized Cauchy strain tensor of the averaged deformation \footnote{The notation $A:B$ is used to denote  the contraction $\sum_{k,l} A_{ijkl}B_{kl}$.}.
The boundary conditions  have the form $\u |_{\partial\Omega_D} = \u_0$, where $\u_0$ are prescribed boundary displacements on the part of the boundary $\partial\Omega_D$ and  $C:\strain(\u) |_{\partial\Omega_N} = \vf_0$, where $\vf_0$ are external forces applied to the part of the surface $\Omega_N$. 

Our goal is to minimize a function $\cE$ of the displacement $u$. For example, to obtain a desired deformation of a part of the object's boundary under a certain load condition (Figure \ref{fig:teaser}), $\cE$ is the $L_2$ norm of the difference between $u$ and the target deformation.   We do this by changing the spatially varying material properties (Young's modulus and Poisson ratio) encoded in the spatially varying tensor C. The problem is constrained by the linearized elasticity PDE in \eqref{eqn:elasticity}:
\begin{equation}
\min_C \cE(\u, C),\; \mbox{subject to \eqref{eqn:elasticity}}.
\label{eq:general-min}
\end{equation}

This problem is closely related to widely used topology optimization formulations \cite{bendsoe2003topology}. 
Additional constraints are usually added to this problem to force, in every point in space, either  homogeneous material properties or zero density, thus obtaining an object with homogeneous material properties and possibly holes. These constraints make this large optimization problem even harder to solve.

\paragraph{The two-scale approach.}

\begin{figure}
    \centering
    \includegraphics[width=0.96\columnwidth]{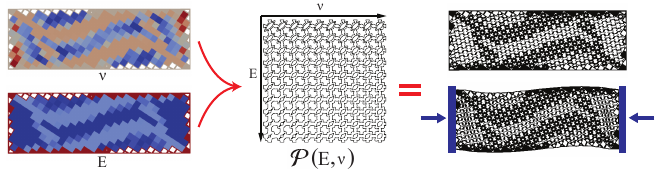}    
    \caption{Two-scale optimization pipeline. (left) we run material optimization on a bar with square cells; (middle) a map $\mtg$ is used to map material properties to geometry; and (right) we obtain the final shape, which deforms as expected.}
    \label{fig:two-scale-optimization}
\end{figure}

A two-scale optimization approach \cite{panetta2015elastic,tozoni2020low} uses a microstructure family $\mtg$ to make solving this problem orders of magnitude cheaper. $\mtg$ is a map from an elasticity tensor $C$ (or equivalently material properties such as Young's modulus and Poisson's ratio) to a periodic geometric pattern with the same homogenized stiffness tensor (see Figure \ref{fig:two-scale-optimization}). The periodic patterns are typically characterized by a set of shape parameters. 
Equipped with $\mtg$, the two-scale optimization  solves the problem in two steps.  First, a material optimization problem, in which the per-voxel value of the stiffness tensor is optimized, is solved on a voxel grid at a coarse scale:
\begin{equation}
\min_{C \in D}  \cE(\u, C),\; \mbox{subject to \protect\eqref{eqn:elasticity}},
\label{eq:material-min}
\end{equation}
where $D$ is the domain of the map $\mtg$, which is a subset of all stiffness tensors that can be realized by using a family of cell microstructures. After solving this small problem (e.g., this problem has around 500 dofs for the 2D examples in Figure~\ref{fig:2D_examples} and can be solved within seconds), the map $\mtg$ is used to replace the cell with the corresponding microstructure geometry; the resulting object with complex geometric structure can be fabricated using a single material. 

The two-scale approach described above has two important challenges: (1) designing the microstructure family $\mtg$  and (2) reducing the error introduced in the object's shape due to a coarse voxel grid, as fabrication constraints typically do not allow one to make the voxels too small. Our focus in this paper is on the latter problem, which has received little attention in the literature. For (1), we borrow and slightly modify the family proposed in  \cite{tozoni2020low}, as detailed in Section \ref{sec:cutcells}.

\subsection{Overview}

\begin{figure}[t!]
    \centering
    \begin{subfigure}[t]{0.3\columnwidth}
        \centering
        \includegraphics[width=\linewidth]{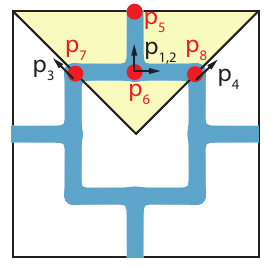}
        \caption{2D}
    \end{subfigure}
    ~ 
    \begin{subfigure}[t]{0.5\columnwidth}
        \centering
        \includegraphics[width=\linewidth]{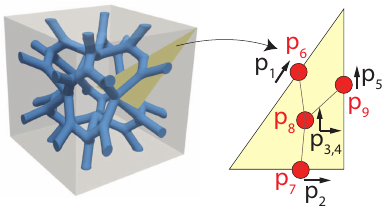}
        \caption{3D}
    \end{subfigure}
    \caption{Cell parameters for 2D and 3D, interior cells.}
\label{fig:parameters}
\end{figure}

We describe our approach to computing an infill structure producing desired deformation behavior for an object for 3D, but the same construction works for 2D: we highlight the required modifications in Section \ref{sec:2d}. In Section~\ref{sec:results}, we demonstrate how these steps contribute to the quality of the results.
The approach is composed of five steps:

\begin{enumerate}
\item  {\bf Cell Partition.} We produce a cell partition of the domain $\Omega$ by intersecting it with a regular grid; the partition consists of cubic (square, in 2D) cells in the interior and 
cut cells, obtained by intersecting a cubic cell with the interior of the object. Each cell is  partitioned into tetrahedra, with the same material property variable $C_i$ assigned to all tetrahedra corresponding to a cell.
\item  {\bf Initial Material Optimization.} We solve the problem defined in Equation ~\ref{eq:general-min}, with constant properties per cell,  computing  material parameters $C_i$ by minimizing the objective $\cE$ (see Equation~\ref{eq:material-objective}), with material parameter bounds for the interior cells determined by the domain of $\mtg$, and for boundary cells inferred from cell geometry.  For interior cells, $(E,\nu)$ are used as material parameters. For boundary cells, $E$ is the only optimized material parameter  and we keep $\nu$ fixed at the base material value. 
\item {\bf Generation of Cut-Cell Geometry.} For the boundary cells, we use a one-parametric family of microstructures, with a single parameter for each cell, the interior void size. We determine this parameter for each boundary cell so that the effective material properties approximately match the target material and compute the cell geometry for this void size (Section~\ref{sec:cutcells}).
\item  {\bf Interior Material Refinement.} We build a hybrid geometry using solid interior cells with variable material $C_i$ assigned to each interior cell but the actual microstructure geometry from Step 2 for the boundary cells. For this new volumetric mesh, with the complex structure on the boundary only, we repeat the material optimization for interior cells, but, this time, with material properties fixed to the base material for already inserted boundary cell geometry and variable material properties for the internal cells (Section~\ref{sec:refinement}).
\item {\bf Surface Extraction.} The geometry of the boundary cells is merged with the microstructure geometry produced from optimized material properties from the previous step via the map $\mtg$. Optionally, we add tunnels to the exterior to allow fabrication using  3D printing methods requiring residual material removal (Section~\ref{sec:surface_extraction}).
\end{enumerate}

\subsection{Step 1. Cell Partition}

The input for this step is a surface mesh and a background regular grid, with each grid cell corresponding to a microstructure tile in the final structure. The output is a tetrahedral mesh of the interior of the surface mesh, with the regular grid cells split into tetrahedra. This mesh is used for the first material optimization. %

This problem could be solved by computing a 3D arrangement \cite{cgal:hk-bonp3-22b} or by cut-cell meshing tools \cite{Tao:Mandoline:2019,Aftosmis:1998}, followed by postprocessing to tesselate the interior of each cell with tetrahedra while ensuring a conforming mesh at the cell boundaries. 

\begin{figure}
    \centering
    \includegraphics[width=0.9\linewidth]{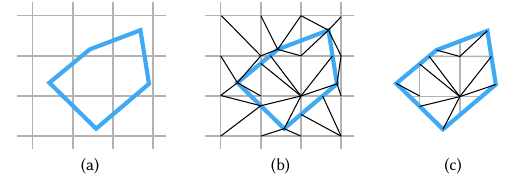}
    \caption{Illustration of cell partition. (a) Initial surface mesh with a background regular grid. (b) Triangular/Tetrahedral mesh of the bounding box. (c) Final mesh after removing triangles/tetrahedra outside of the surface mesh.}
    \label{fig:cell-partition}
\end{figure}

As shown in Figure \ref{fig:cell-partition}, we opted for a more direct and robust solution, creating a tetrahedral mesh with fTetWild \cite{hu2020fast}, while providing both the surface mesh and grid faces as input. The fTetWild tool outputs a tetrahedral mesh of the bounding box of the input mesh. We then filter the tetrahedra outside the object surface using the winding number \cite{Jacobson:2013}, and assign the resulting tetrahedra to the cell of the grid containing its barycenter. Eventually, we obtain a tetrahedral mesh of the surface, while each tetrahedron is inside a grid cell. We use a geometrical tolerance of $10^{-3}$ for fTetWild. This approach is insensitive to surface mesh imperfections and simultaneously allows us to maintain the high mesh quality needed for simulation. 

\begin{figure}
    \centering
    \includegraphics[trim={0.1 0 0 0.1},clip,width=0.7\columnwidth]{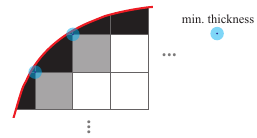}
    \caption{Interior Cells that are too close to the boundary may cause thin features in the shape. To avoid issues during printing and non-smoothness in displacement behavior, we consider those as boundary cells. The drawing shows boundary cells (in black) touching the (red) boundary and two quad cells (in dark gray) that are too close to the boundary, according to a min. thickness parameter.}
    \label{fig:close-to-boundary}
\end{figure}

To satisfy the minimum thickness requirement in 3D printing, the maximum size of the internal voids of boundary cells is restricted. However, some boundary cells might be too thin even without internal voids. If a thin boundary cell is adjacent to an interior cell, after the interior cell is replaced by the microstructure, this boundary cell may violate the minimum thickness. To avoid this, we  change the type of some interior cells into boundary cells, although they are not cut by the boundary surface. Specifically,  we compute the distance between the surface  cutting a boundary cell and its internal boundary faces, edges, and vertices, shared with other internal cells; if the distance is too small, we transform all internal cells sharing this element into a boundary cell. See Figure~\ref{fig:close-to-boundary} for an example.

\subsection{Step 2. Initial Material Optimization}
\label{sec:matopt}

\begin{figure}[t!]
    \centering
    \begin{subfigure}[t]{0.45\columnwidth}
        \centering
        \includegraphics[width=\linewidth]{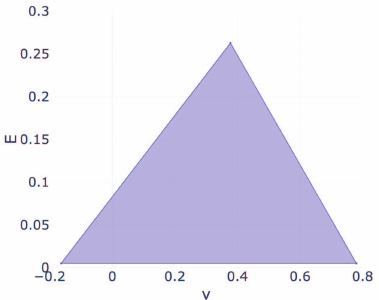}
        \caption{2D}
    \end{subfigure}
    ~ 
    \begin{subfigure}[t]{0.45\columnwidth}
        \centering
        \includegraphics[width=\linewidth]{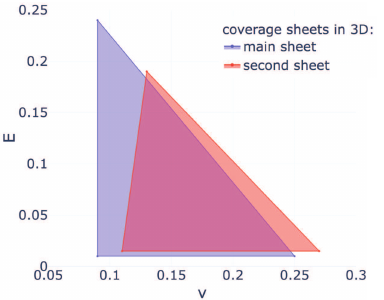}
        \caption{3D}
    \end{subfigure}
    \caption{Coverage of material parameter properties in the families we use. Our coverage regions are approximated to triangles, so they can be easily added as constraints in our optimization.}
\label{fig:coverage}
\end{figure}

We solve problem \eqref{eq:material-min} with the following functional:

\begin{equation}
    \cE(\u, C) = \int_{\partial \Omega^t} (\u(C) - \u^t)^2 dA  + w \int_{\Omega} 
    \|\nabla C\|^2 dA,
    \label{eq:material-objective}
\end{equation}
where $\partial \Omega^t$ is typically a part of the boundary of our domain for which we prescribe target displacements $\u^t$,
$w$ is a weight to adjust the smoothness of the material distribution. The constraints on the problem consist of a set of
linear bounds for the domain of  material properties. For interior cells, we constraint the material properties $((E,\nu)$ to be inside the domain of $\mtg$, shown in Figure~\ref{fig:coverage}. For boundary cells, we add constraints on Young's modulus, preventing the creation of features smaller than a predefined thickness. 

As seen in Equation~\ref{eq:material-objective}, our objective has two terms. The first corresponds to the deformation error, which measures the difference between the simulated displacement $\u(C)$ and the target $\u^t$. The second term is a regularizationterm, aiming to achieve a smoother spatial material variation, which  helps to reduce large differences in microstructure geometry when transitioning between neighboring cells, which may lead to local deformation artifacts.

The problem is discretized using the finite element method~\cite{polyfem}, and the adjoint method is used to compute the gradient of the target functional with respect to elasticity parameters \cite{huang2022differentiable}. The optimization itself is performed using L-BFGS with box constraints.

\subsection{Step 3. Generation of Cut-Cell Microstructure}
\label{sec:cutcells}

While we can use $\mtg$ to find microgeometry for the internal cells, the construction of the map from material properties to geometry is based on periodic-material homogenization and requires cells to be cubes: in this step, we provide a construction of a microstructure family for boundary cut cells with arbitrary geometry.

\begin{figure}
    \centering
    \includegraphics{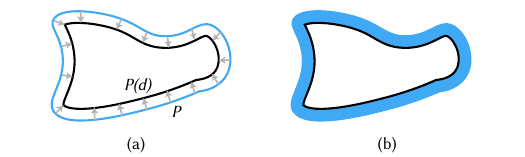}
    \caption{Illustration of creating cavity to reach a certain volume ratio based on the surface offset. (a) The surface $P$ is offset by $d$ to get $P(d)$. (b) Volume inside $P(d)$ is dropped to match target volume ratio $r$.}
    \label{fig:erosion}
\end{figure}

\begin{figure}
    \centering
    \includegraphics[width=\linewidth]{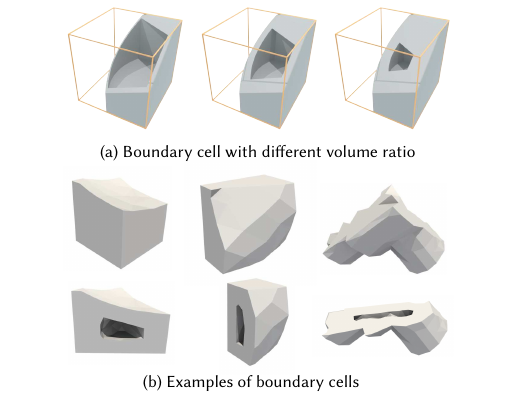}
    \caption{Boundary cell geometry in 3D. (a) The cross-section of a cut-cell enclosed by the grid partition in yellow. (b) The bottom row shows the cross-section of the top row.}
    \label{fig:boundary-cell}
\end{figure}

Our approach is to create a 1D bijective smooth map $\mtgb$ between volume fraction $r$ of the filled part of a cell and Young's modulus $E$, using the fact that there is usually a strong connection between stiffness and solid material volume fraction (see Subsection \ref{sec:volume-material}) in simple cell geometries with a single void, while the shape of the void has less impact. 
We can obtain such single-void one-parametric families for arbitrary cell shapes by defining our microstructure geometry to be solid between the polygonal boundary $P$ of the cell and an offset $P(d)$ of $P$ (Figure~\ref{fig:erosion}), where $d\geq d_{\min}$ is a parameter interpolating between the minimal wall thickness ($d_{\min}$) and a fully filled cell. We construct $P(d)$ for an arbitrary cell shape using a signed distance function (SDF). The cell boundary mesh $P$ is converted to an SDF (we use 
OpenVDB \cite{Museth:2013}), and then we call the \textbf{offset} function in OpenVDB, and finally convert the offset SDF back to a mesh. Figure~\ref{fig:boundary-cell} shows three examples of boundary cell geometry in real models.
Notice that, in 3D, we may need to add additional openings to avoid enclosed voids, as discussed in Section \ref{sec:surface_extraction}.

Next, we determine the correspondence between volume fraction $r$ and the offset $d$ for a specific cut cell geometry. Denote the volume enclosed by surface $P$ as $V$, and the volume enclosed by surface $P(d)$ as $V(d)$. Given a fixed $r$, our goal is to find the offset value $d$ so that $$V(d)=(1-r)V.$$ This is a 1D root-finding problem for a monotonically increasing function, so we use a binary search to find the root $d$ with a relative tolerance of $0.01$.  

\begin{figure}
    \centering
    \begin{subfigure}[t]{0.48\columnwidth}
        \centering
        \includegraphics[width=1.0\columnwidth]{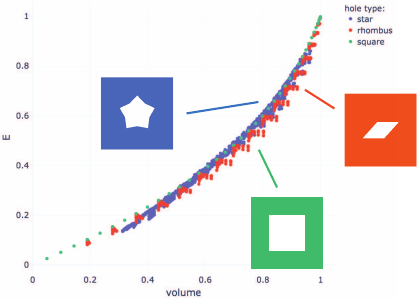}
        \caption{2D}
    \end{subfigure}
    \begin{subfigure}[t]{0.48\columnwidth}
        \centering
        \includegraphics[width=\linewidth]{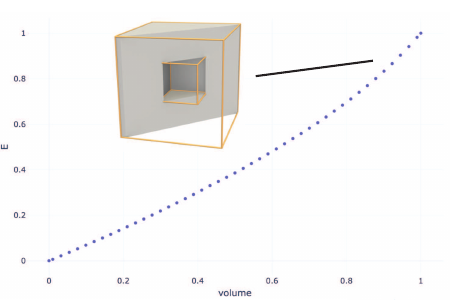}
        \caption{3D}
    \end{subfigure}
    \caption{Volume ratio to average Young's modulus map. The left figure represents the map for 2D microstructures, where each color represents a different shape of the opening, while the figure on the right represents the map in 3D.}
    \label{fig:volume-Y}
\end{figure}

\noindent{\bf From material properties to volume fraction.}  
\label{sec:volume-material}
The relationship between the effective Young's modulus and volume ratio, in general, depends on the shape of the cell and, in the case of non-tileable cells, may even be difficult to define precisely, as it is unclear how to perform correct homogenization in this case (homogenization requires periodic tiling). However, a rough approximation of this dependence can be obtained from a one-parametric family of complete cubic cells, which can be tiled periodically.  In this case, we can compute the effective Young's modulus by averaging Young's modulus on orthogonal directions ($x$, $y$, $z$) (obtained from the homogenized elasticity tensor), for a variety of interior void shapes, to verify that  the dependence on the shape of the void is not strong (Figure~\ref{fig:volume-Y}). 
As a result, we have a function $E(r)$ which we invert to map the optimized Young modulus values $C_i$ (obtained at Step 2 for boundary cells) to the volume ratios $r$.   For a specific choice of cell boundary, we determine the offset $d$ from $r$ as described above, building a complete map from $C_i$ to a boundary cell microstructure. 

The need for precise knowledge of the effective material properties of a cell is made less critical for this stage by the next step,  which compensates for the errors introduced in the boundary cells.

\subsection{Step 4. Interior Material Refinement}
\label{sec:refinement}
To increase the accuracy of the solution, we run a second round of material optimization fixing the geometry and material properties of the boundary cells, and optimizing only the regular, interior cells.

Given a target volume ratio for each boundary cell determined by their target Young modulus computed in Step 1, we
determine the offset for the cell microstructure in Step 3.  We perform the offsets for all boundary cells in parallel, append the offset surface mesh to the original mesh, and then the approach of Step 2 is used again to generate a new volumetric mesh.

On this mesh (which has explicitly meshed cut-cell microstructures on the boundary and regular grid cells in the interior), we solve a material optimization problem with fixed base material in boundary cells (as their microstructure is already integrated into mesh geometry). While the overall complexity of the optimization increases, as the simulation needs to be done on a finer mesh in boundary cells, it is still much lower than the complexity of direct microstructure shape or topology optimization, as the number of optimization variables is lower.  %
In principle, the process can be iterated, by switching boundary and interior cell optimization, but using explicit geometry for interior cells is expensive, and little additional benefit can be expected.

Note that converting the Young's modulus to volume fraction in boundary cells is not precise resulting in the change of deformations relative to the optimized ones. As a consequence, 
the objective value obtained by the first optimization, can be significantly increased by this transformation. While the second optimization brings it back closer to the originally obtained 
valiues,  even after the second optimization the objective may be higher than after the first one (Table~\ref{tab:examples-numerical}) -- but it is alaways lower than the objetive after the cut-cell microstructure insertion.

\subsection{Step 5. Surface Extraction}
\label{sec:surface_extraction}

To extract the surface from implicit volumetric functions
(see Section~\ref{sec:hom}) 
of internal and boundary cells, we first compute the union of all interior cell SDFs and intersect it with the SDF obtained in Step 5 which included boundary cell geometry only.  The surface mesh is extracted from the resulting SDF using the marching cubes method.  We use a resolution of $50\times50\times50$ per cell for the SDF (note that this effective resolution is used in a sparse volumetric representation in OpenVDB).

\begin{figure}
    \centering
    \includegraphics[width=\linewidth]{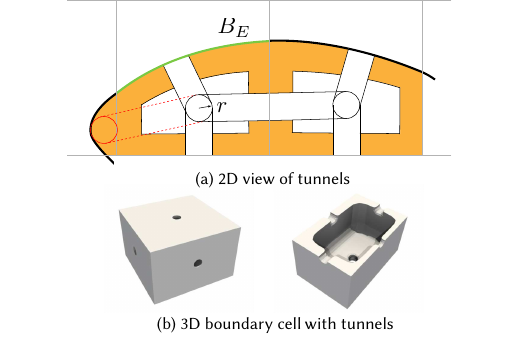}
    \caption{Tunnels connecting boundary cells. (a) The gray lines represent the grid partition, $B_E$ for one boundary cell is shown in green, the tunnels are cylinders with radius $r$, solid parts are shown in yellow. The red circle is too large to fit into the boundary cell, so we fill the cell completely. (b) A boundary cell with internal void and tunnels (left) and its cross-section (right).}
    \label{fig:boundary-tunnels}
\end{figure}

By construction, our boundary cells have a solid layer covering the model surface, which will prevent uncured resin or unmelted powder to be removed after fabricating the model. To make the model fabricable, we add tunnels to each boundary cell (Figure~\ref{fig:boundary-tunnels}).  We first go over boundary cells with internal voids and consider the parts $B_E$ of their surfaces that belong to the outer surface. 
If the sphere centered at the barycenter of the internal void and  with radius equal to the tunnel radius intersects $B_E$, constructing a tunnel to the exterior is impossible, and we completely fill this  cell.   We connect the centers of every two adjacent boundary cells with voids by a tunnel, and these segments form a graph consisting of disjoint connected components.
For each such connected  component, we create a cylindrical tunnel from one void center to the object's surface. To remove the materials inside interior cells, at least two tunnels are created from internal cells to adjacent boundary cells with voids.  To avoid the decrease in volume caused by creating tunnels, we pick the smallest possible radius as long as the 3D printing materials can be removed.

\subsection{2D Algorithm}
\label{sec:2d}

Our 2D pipeline closely follows the 3D pipeline, using  Triangle \cite{shewchuk1996triangle} instead of fTetWild, and using our own implementation of marching squares as OpenVDB does not support 2D domains. An advantage of Triangle is that it allows us to represent the boundaries exactly, without introducing a geometric approximation error. Tunnels are not added to the 2D boundary cells as they are not necessary for fabrication. %

\subsection{Construction of $\mtg$}
\label{sec:hom}

We use the method introduced in \cite{tozoni2020low} for creating  microstructure families for the interior cells both for 2D and 3D. However, we do not use their rhombic shape parameter, as in our case we will use only square/cubic cells. We briefly review this construction here. 

First, \emph{a homogenization map} $H(\p)$ is obtained for a low-parametric family of microstructures, with parameters $p_i$ shown in Figure~\ref{fig:parameters}, consisting of graph vertex positions and radii at these vertices determining how thick the microstructure is at this location. The method of \cite{Panetta:2017:WCS} 
is used to construct the implicit curve/surface bounding the microstructure geometry  from the graph and radii.

We run a sweep of the parameter values in a given range $[p^{min}_i,$ $ p^{max}_i]$ for each parameter $p_i$. Consider $B_p$ the  $n$-dimensional box of admissible parameter values. Each combination of values is used to generate a microstructure cell geometry using the implicit surface definition of \cite{Panetta:2017:WCS}.   The value of $H(\p)$ is obtained as the effective value of the elasticity tensor parameters (with additional symmetry conditions on our structures, there are only 3 independent parameters in the tensor, Young's modulus $E$, Poisson's ratio $\nu$ and the shear modulus $G)$, so the function $H(\p)$ maps $\p$ into $\mathbb{R}^3$. 

The map $H(\p)$ is not bijective as it typically maps a higher-dimensional space into a lower-dimensional space.  In  \cite{tozoni2020low}, a sequence of dimension reductions is used to identify a subspace $D$ in the space of shape parameters equal to the dimension of the elasticity material parameter space (3 in our case). 
The map $H(\p)$ restricted to this subspace may still be not bijective, but its image $H(B_p)$ can be partitioned into parts on which it is invertible; specifically, we construct simplices in the subspace of shape parameters and  split $B_p \cap D$ into connected subdomains $D_j$ by simplex orientation. 
On each of $D_j$, we can invert $H(\p)$.  We further restrict it to the two-dimensional space of isotropic material parameters parameterized by $(E,\nu)$ to obtain the  \emph{material-to-geometry} map $\mtg_i(E,\nu): H(D_i) \rightarrow  B^p$. 
Figure~\ref{fig:coverage} presents the coverage regions for our 2D and 3D families. Notice that for 3D we have two different material coverage sheets, with the blue one covering most of the space and being chosen as our main material range.

\begin{figure}
    \centering
    \includegraphics[width=0.96\linewidth]{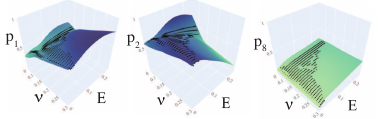}
    \caption{Examples of shape parameter dependence on materials in our 3D family. These maps correspond to our main coverage sheet.}
    \label{fig:shape-param}
\end{figure}

Figure~\ref{fig:shape-param} shows examples on the final map from material properties ($\nu$ and $E$) to geometric parameters $p_i$. 

\begin{figure*}[htb!]
  \includegraphics[width=0.95\textwidth]{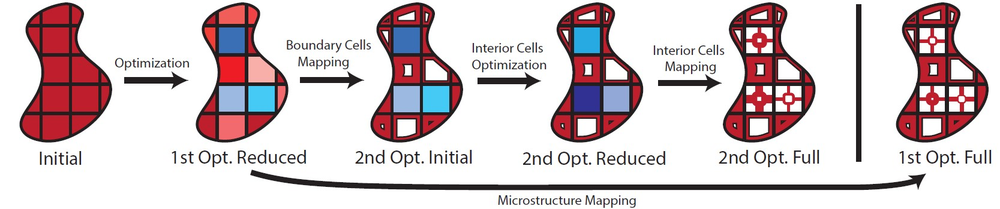}
  \caption{Summary of our method, showing the main steps of our algorithm. Quantitative results of each step are shown in Table \ref{tab:examples-numerical}. Dark red represents our base material. Different tones of red represent variable material on the boundary, while tones of blue are used for material properties of interior cells.}
  \label{fig:summary-steps}
\end{figure*}

\begin{figure*}[htb!]
\makebox[\linewidth][c]{%
\setlength\tabcolsep{4pt}
\renewcommand{\arraystretch}{0}
\resizebox{.981\linewidth}{!}{
\begin{tabular}{@{}ccccc@{}}
$E$ & $\nu$ & pattern & simulated & fabricated (photograph) \\
\includegraphics[angle=0,origin=c,height=1.15cm]{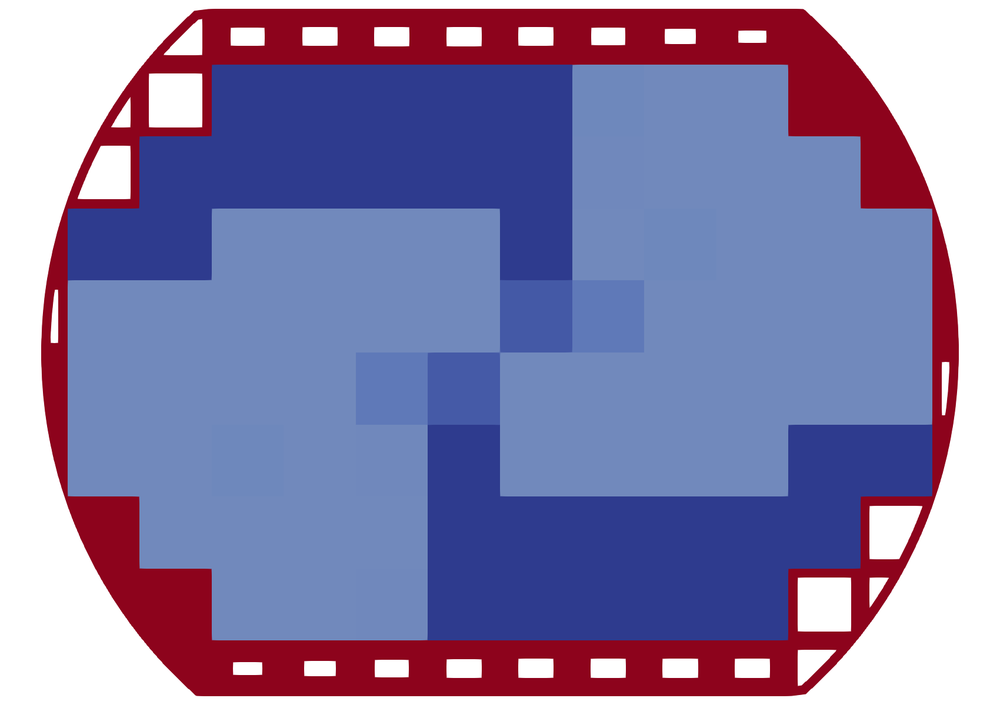} &
\includegraphics[angle=0,origin=c,height=1.15cm]{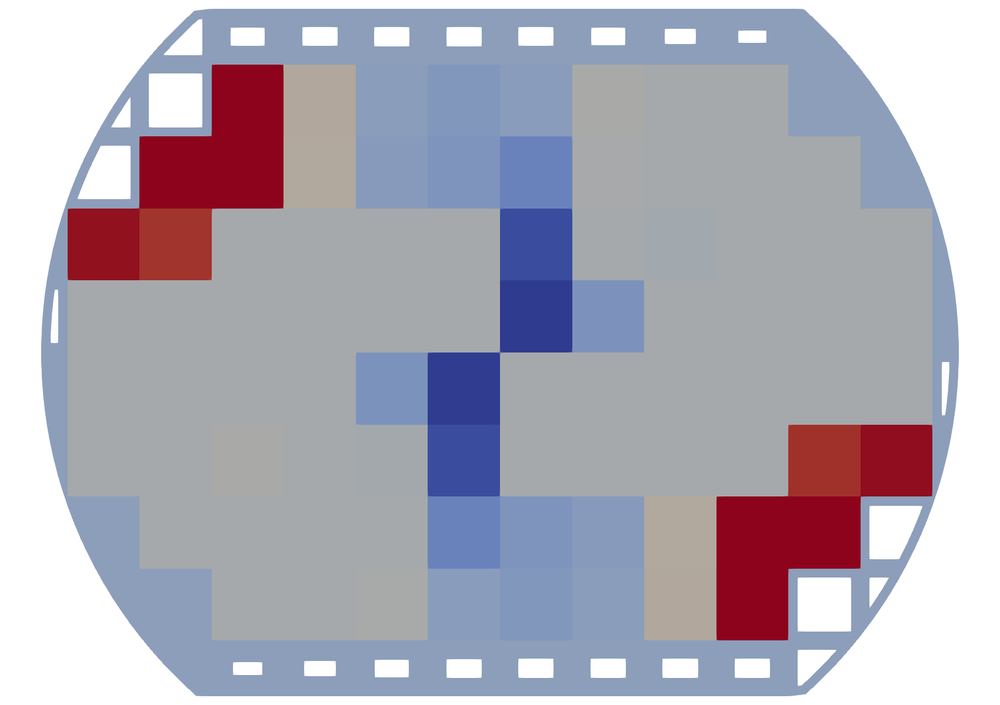} &
\includegraphics[angle=0,origin=c,height=1.15cm]{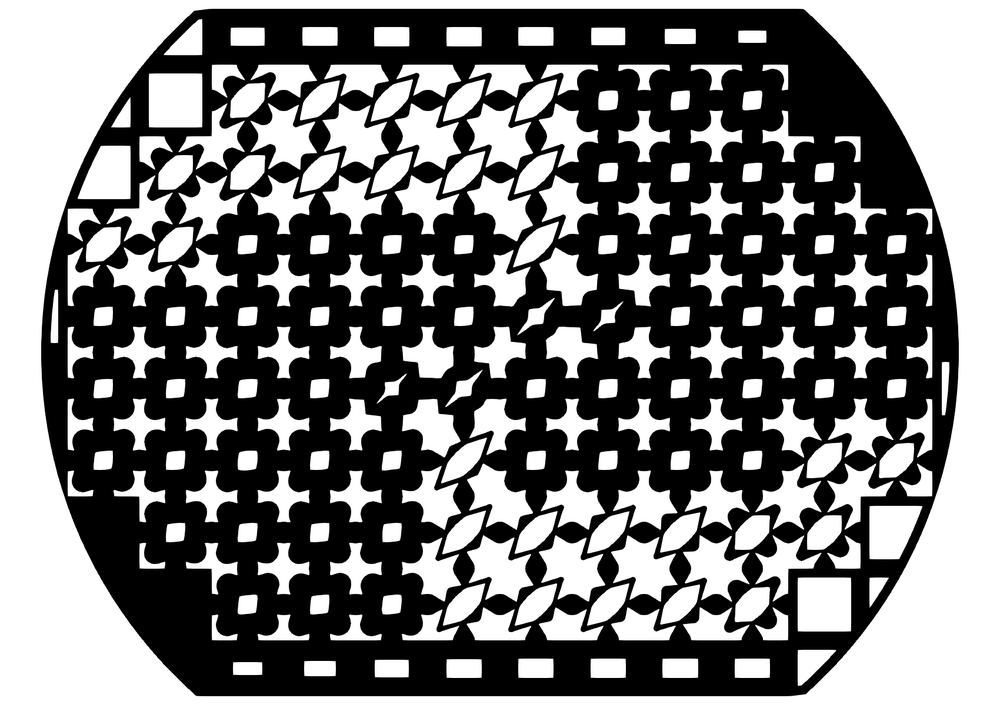} &
\includegraphics[angle=0,origin=c,height=1.15cm]{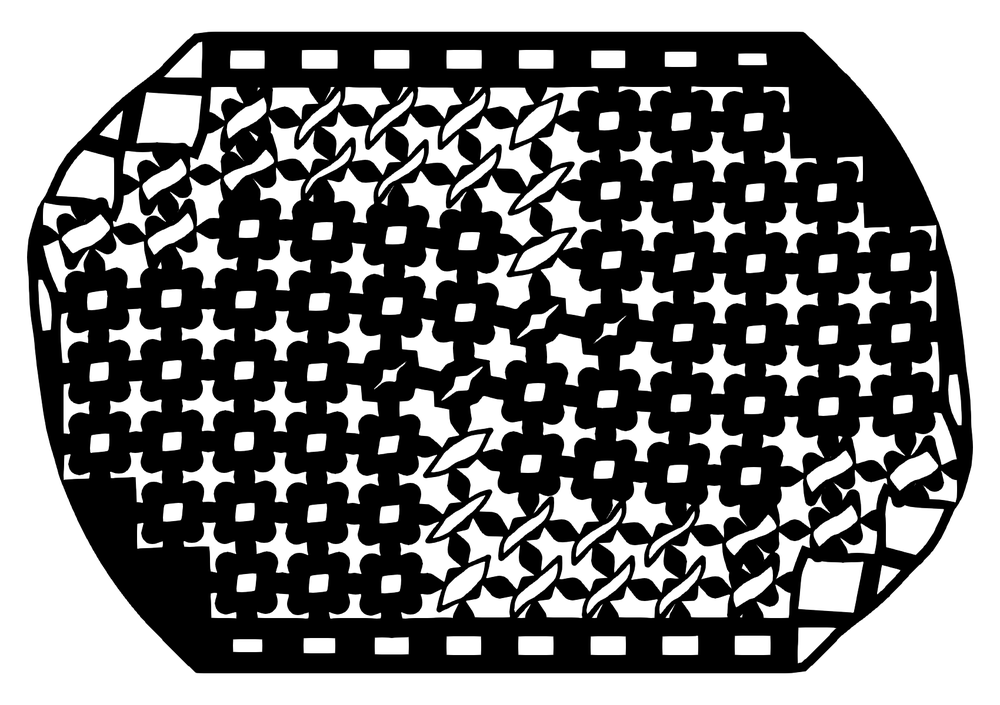} &
\includegraphics[angle=0,origin=c,height=1.2cm]{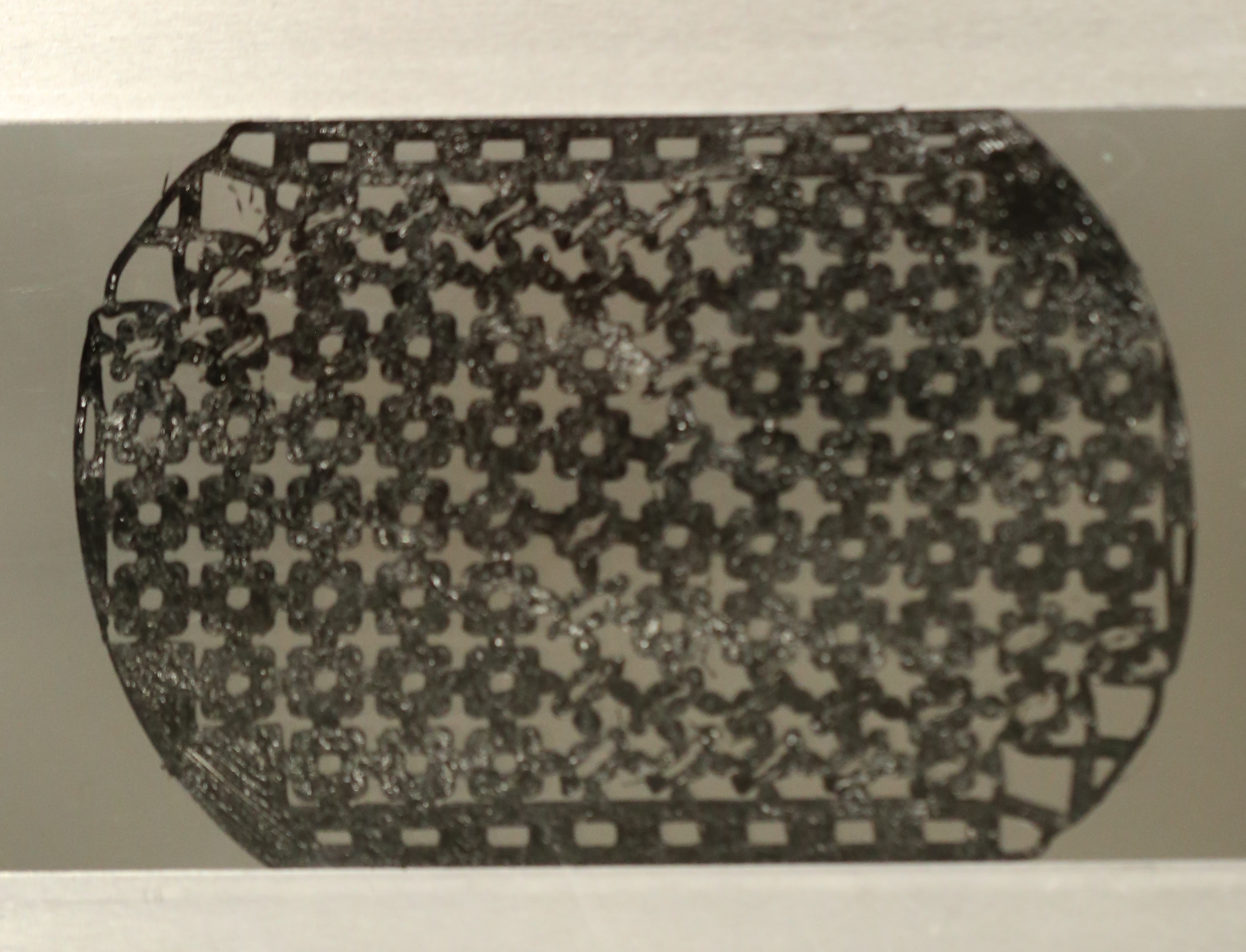} \\
\includegraphics[angle=0,origin=c,width=2.5cm]{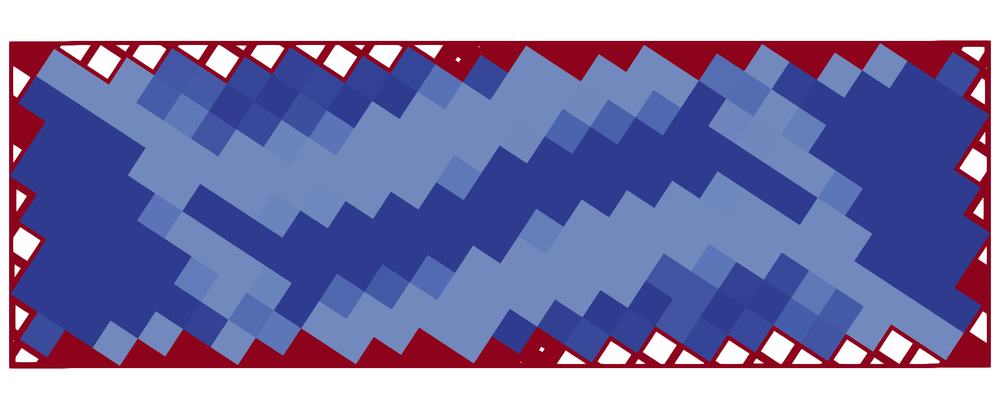} &
\includegraphics[angle=0,origin=c,width=2.5cm]{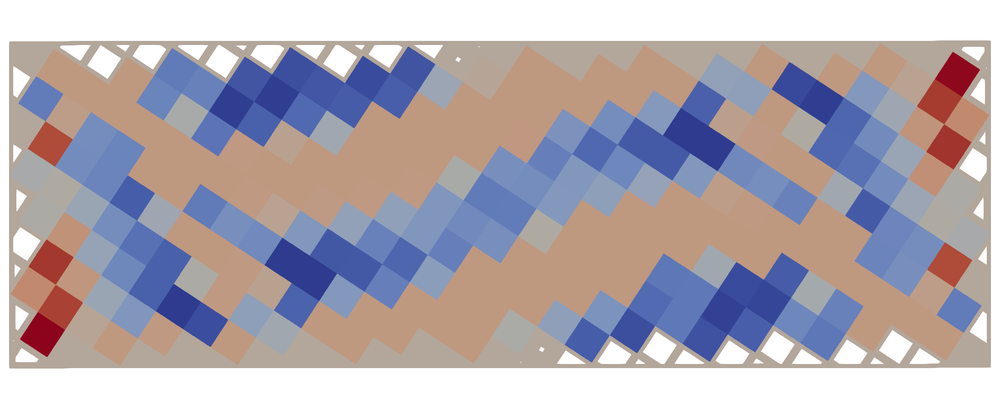} &
\includegraphics[angle=0,origin=c,width=2.5cm]{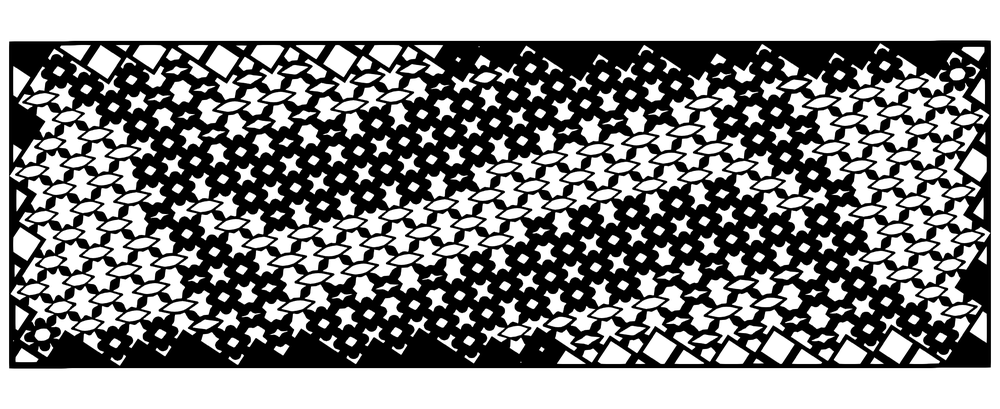} &
\includegraphics[angle=0,origin=c,width=2.5cm]{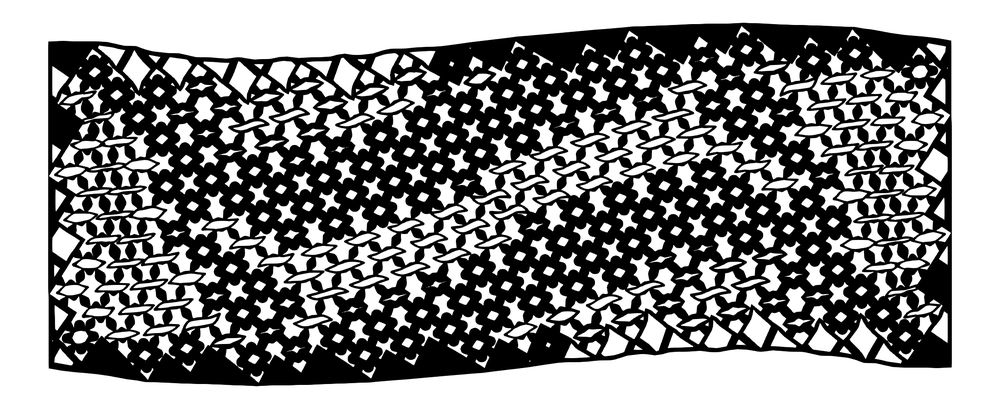} &
\includegraphics[angle=0,origin=c,width=2.4cm]{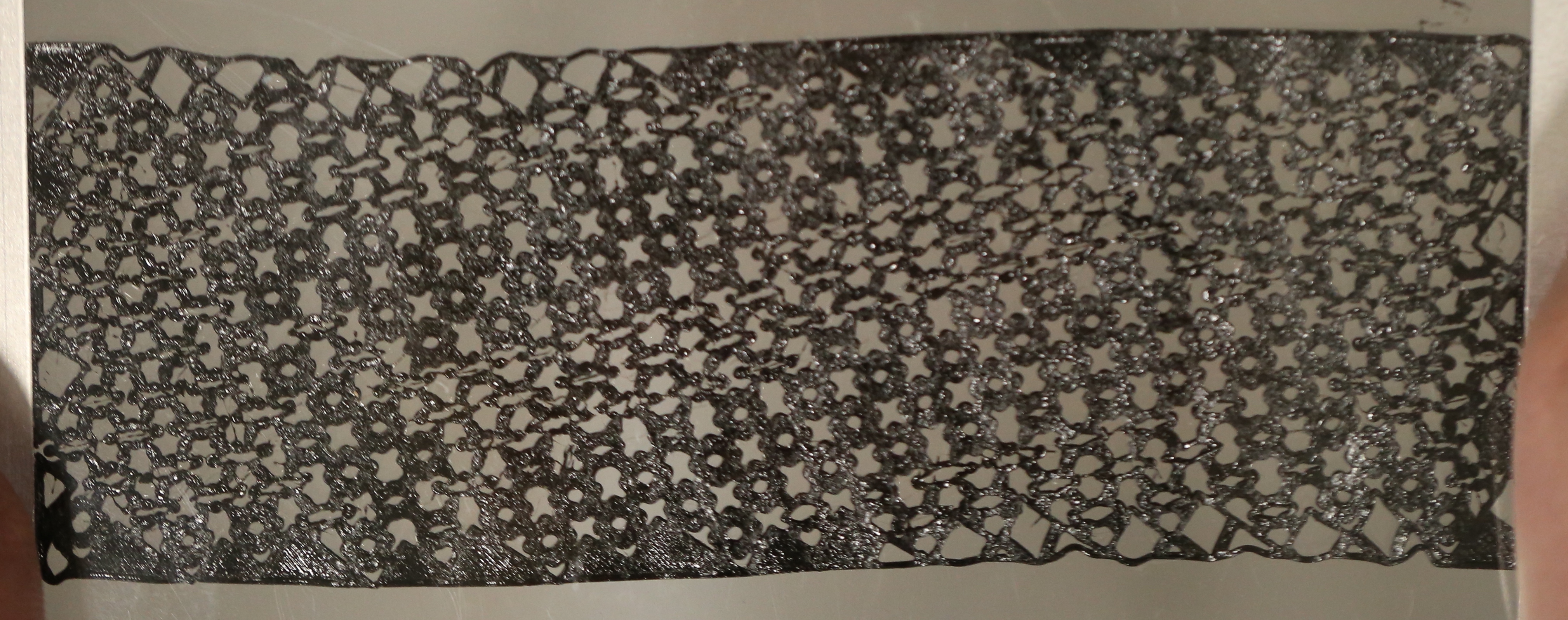} \\
\includegraphics[angle=0,origin=c,height=2cm]{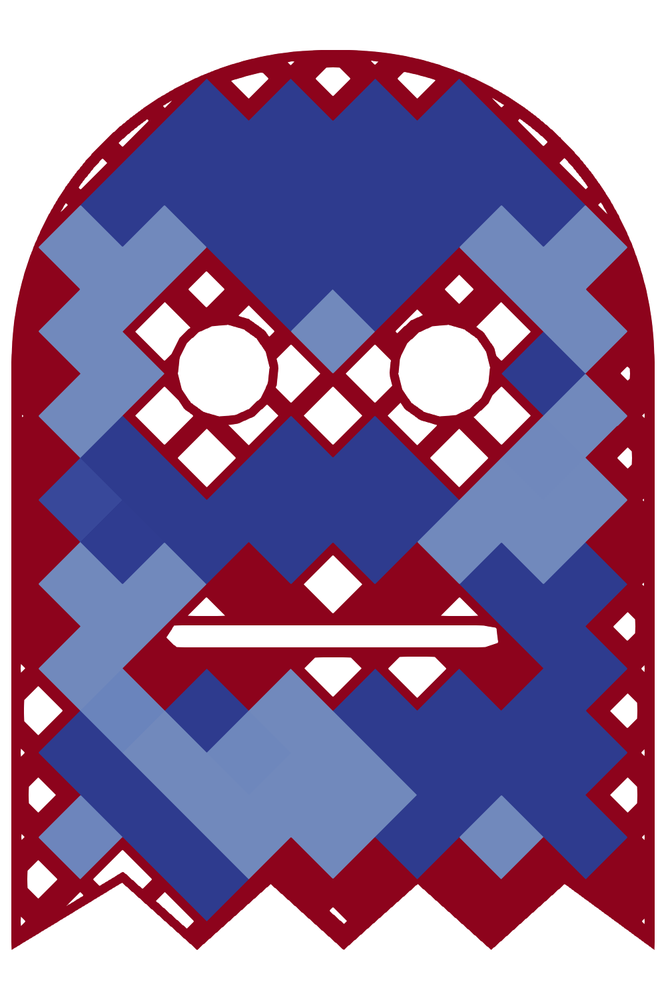} &
\includegraphics[angle=0,origin=c,height=2cm]{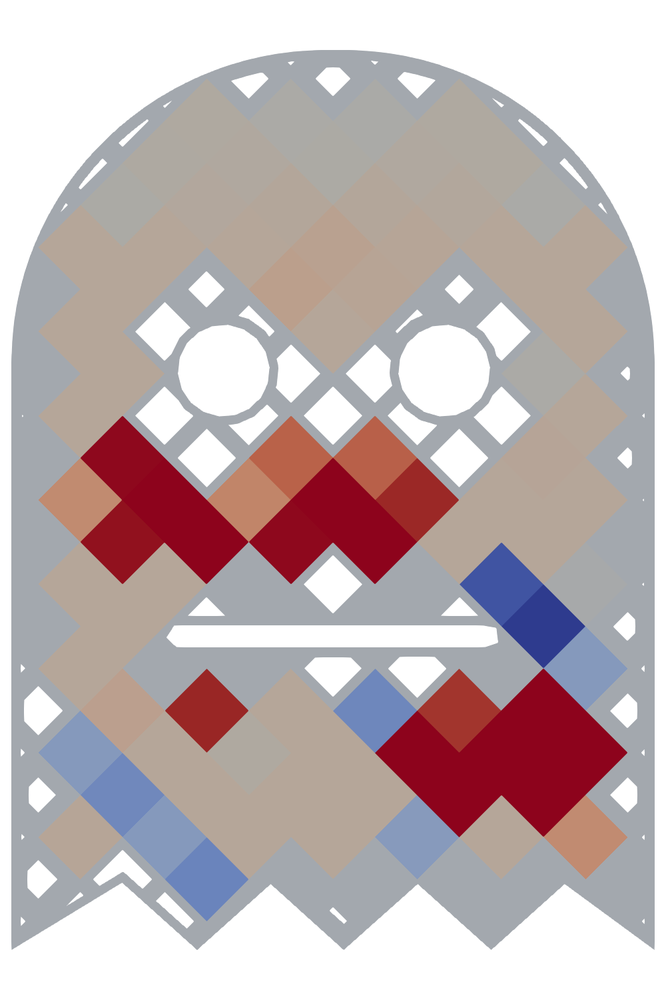} &
\includegraphics[angle=0,origin=c,height=2cm]{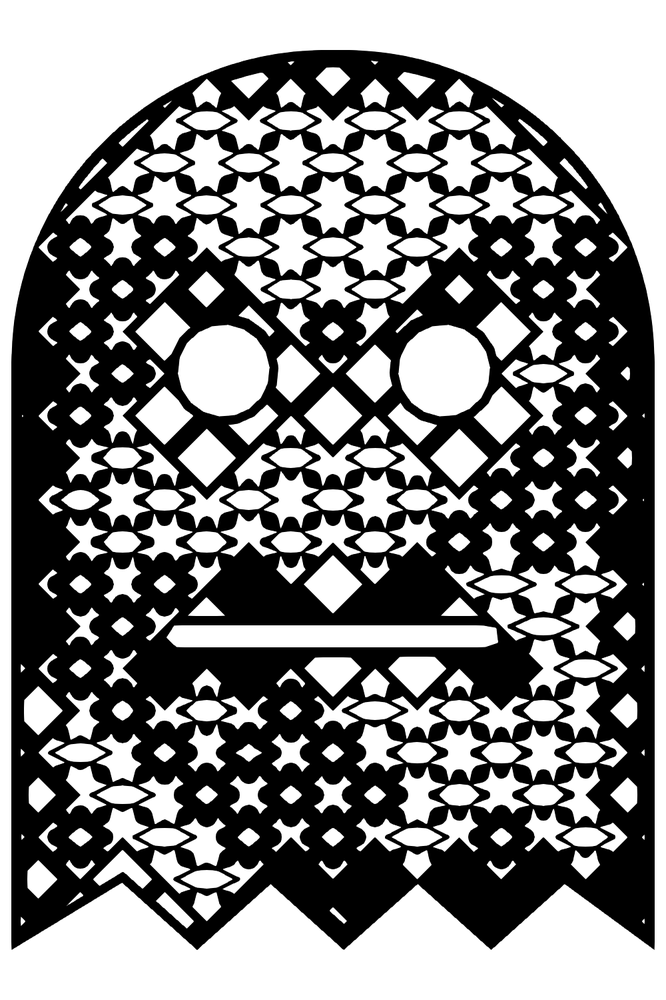} &
\includegraphics[angle=0,origin=c,height=2cm]{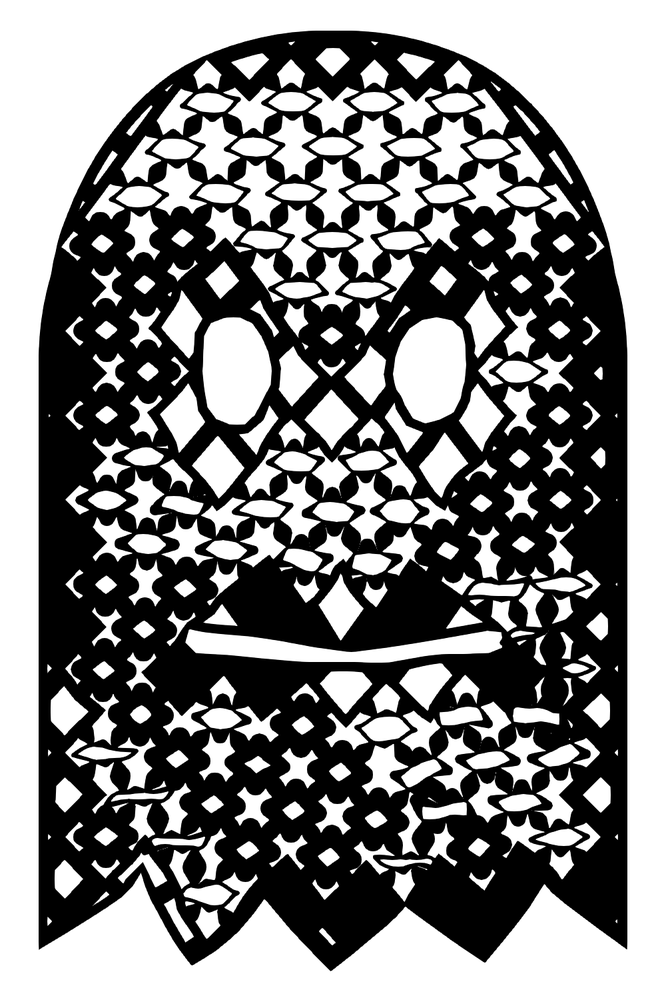} &
\includegraphics[angle=0,origin=c,height=2cm]{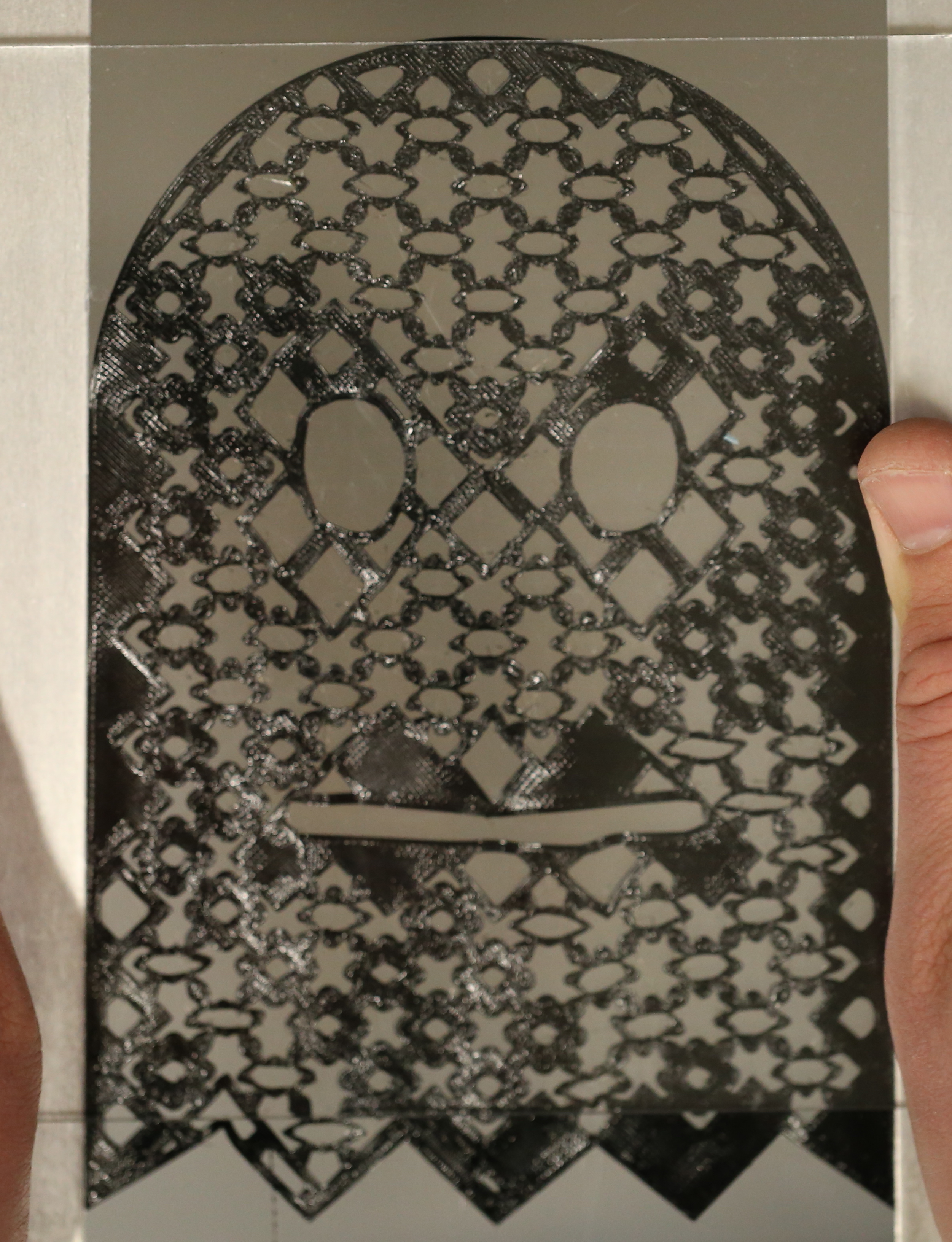} \\
\includegraphics[angle=0,origin=c,width=2.5cm]{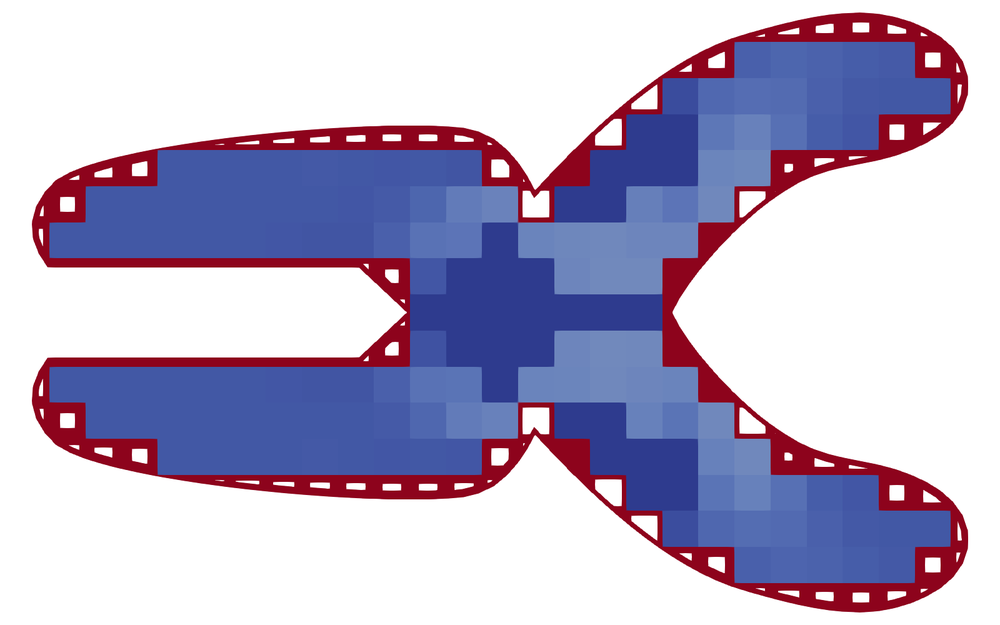} &
\includegraphics[angle=0,origin=c,width=2.5cm]{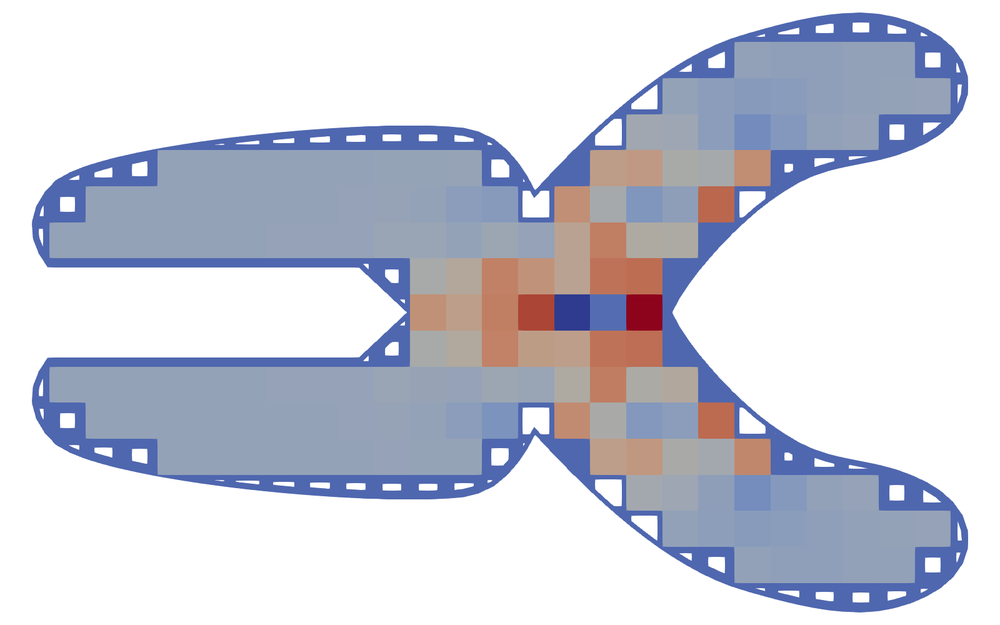} &
\includegraphics[angle=0,origin=c,width=2.5cm]{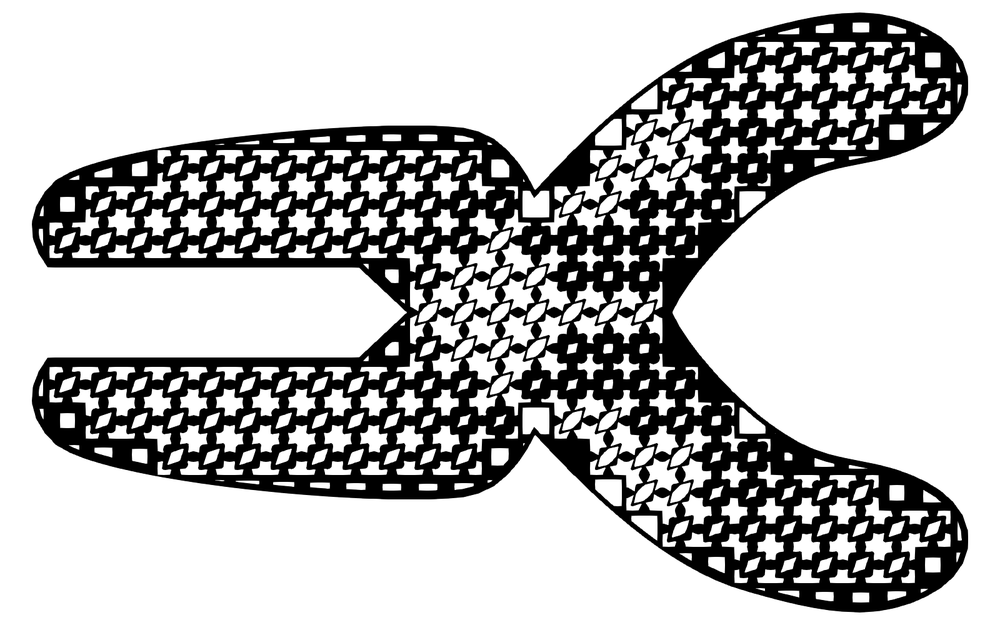} &
\includegraphics[angle=0,origin=c,width=2.5cm]{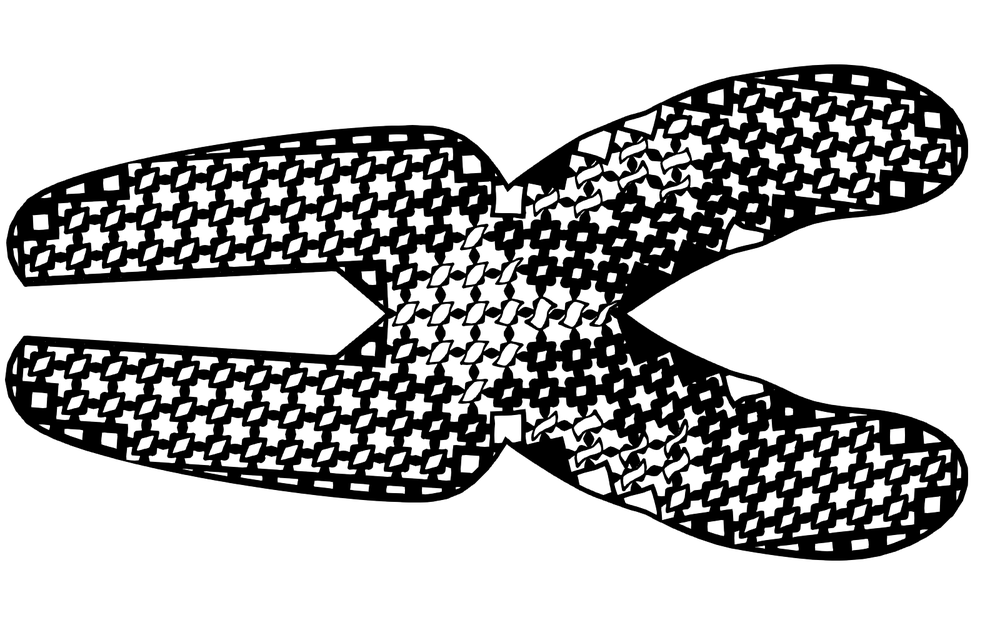} &
\includegraphics[angle=0,origin=c,width=2.5cm]{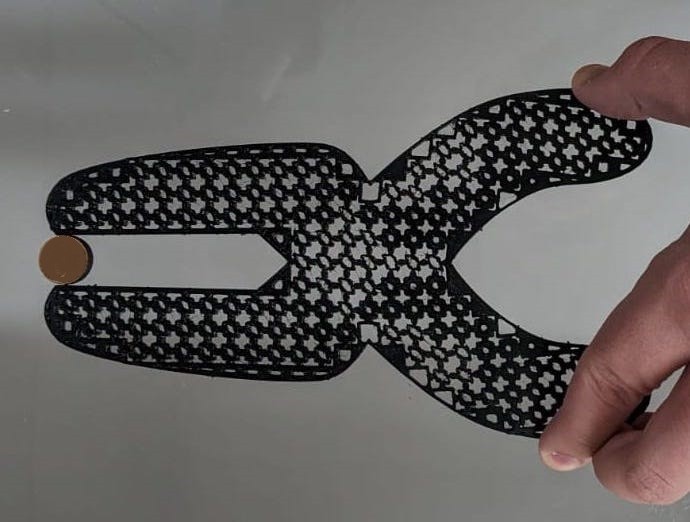} \\
\includegraphics[angle=0,origin=c,height=1.5cm]{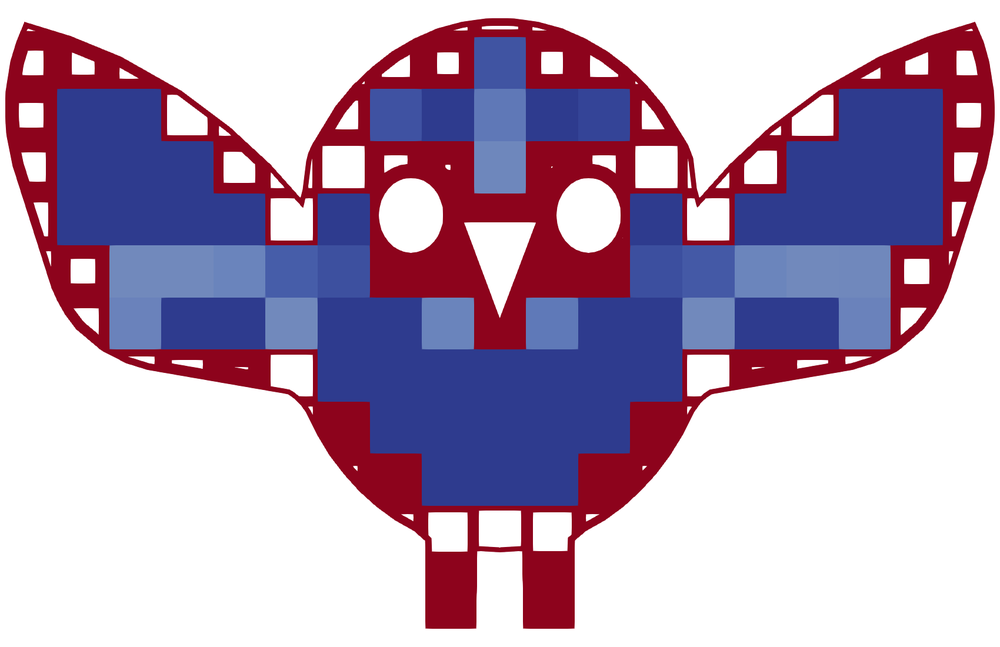} &
\includegraphics[angle=0,origin=c,height=1.5cm]{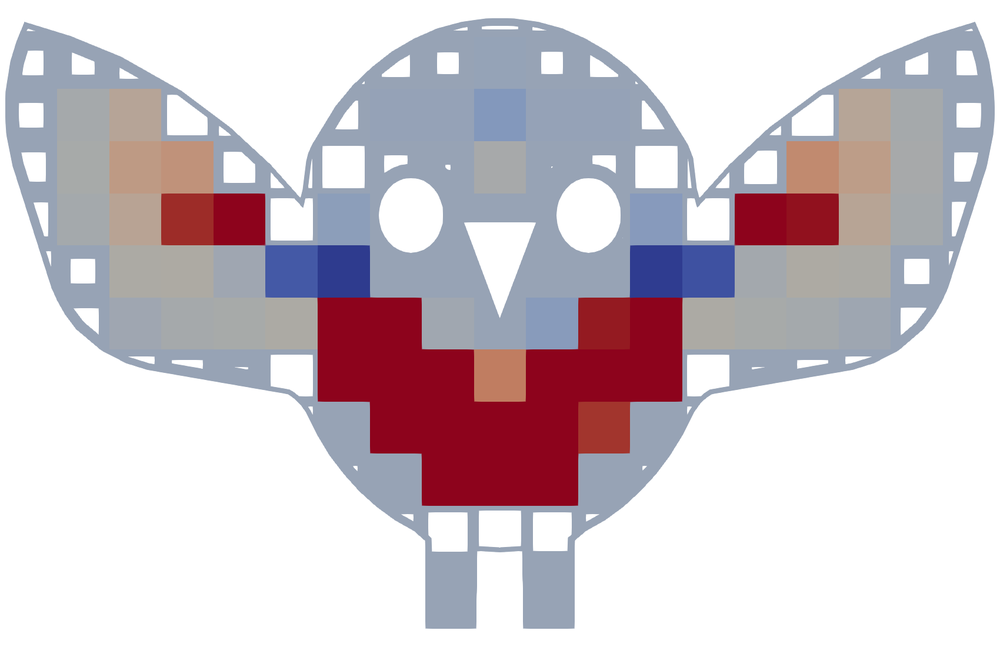} &
\includegraphics[angle=0,origin=c,height=1.5cm]{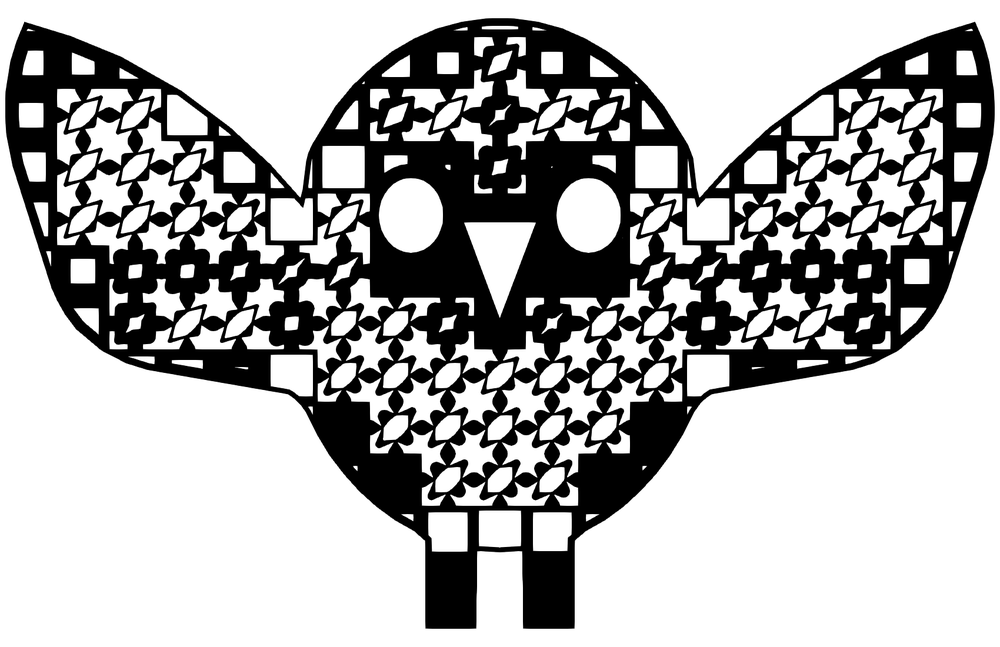} &
\includegraphics[angle=0,origin=c,height=1.5cm]{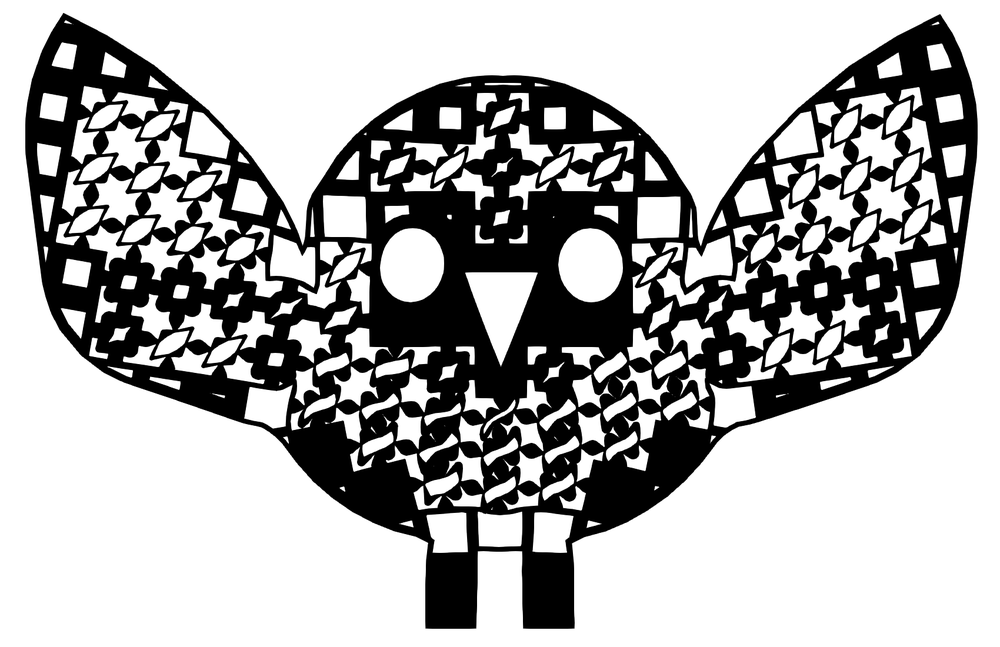} &
\includegraphics[angle=0,origin=c,height=1.5cm]{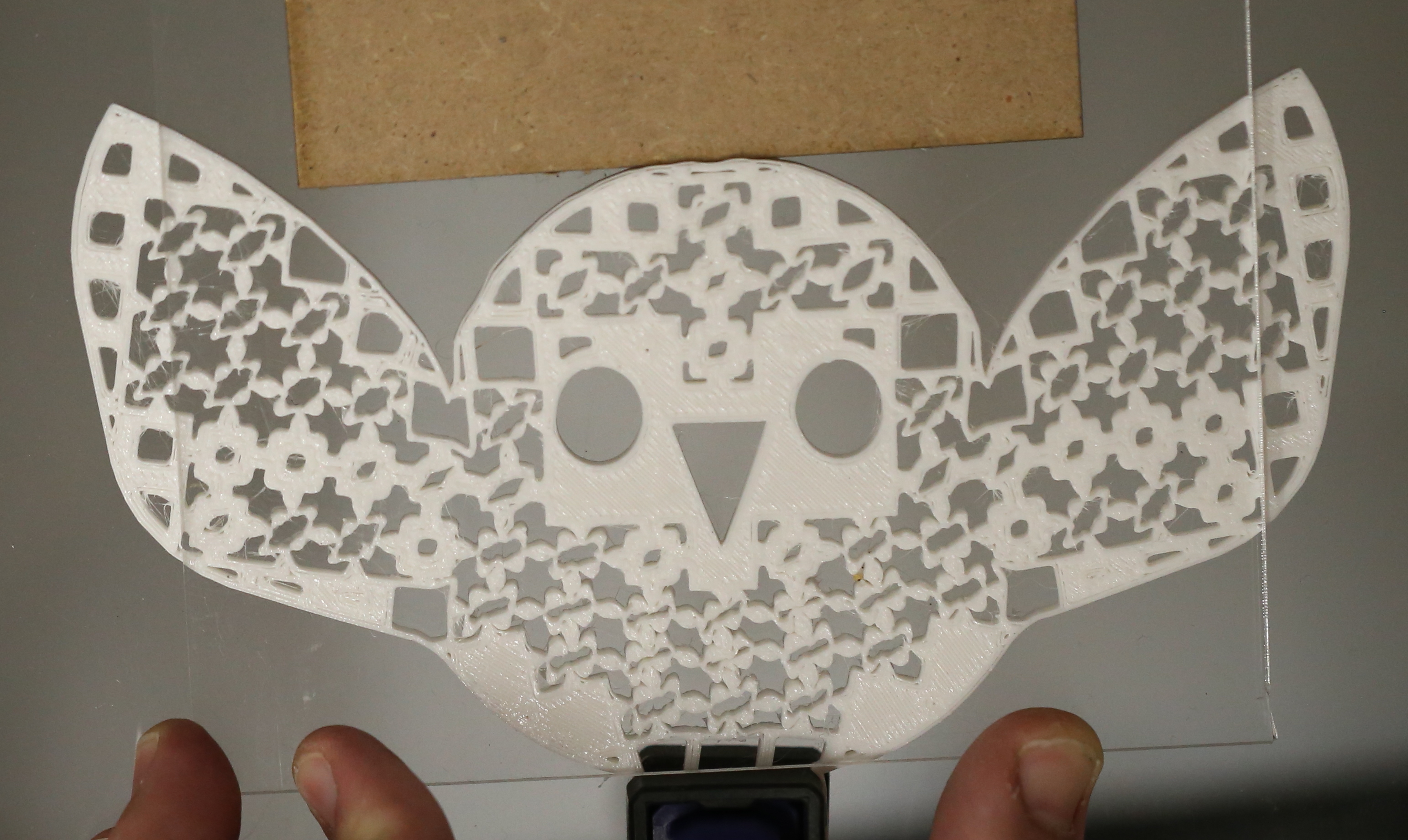} \\
\includegraphics[angle=180,origin=c,height=2cm]{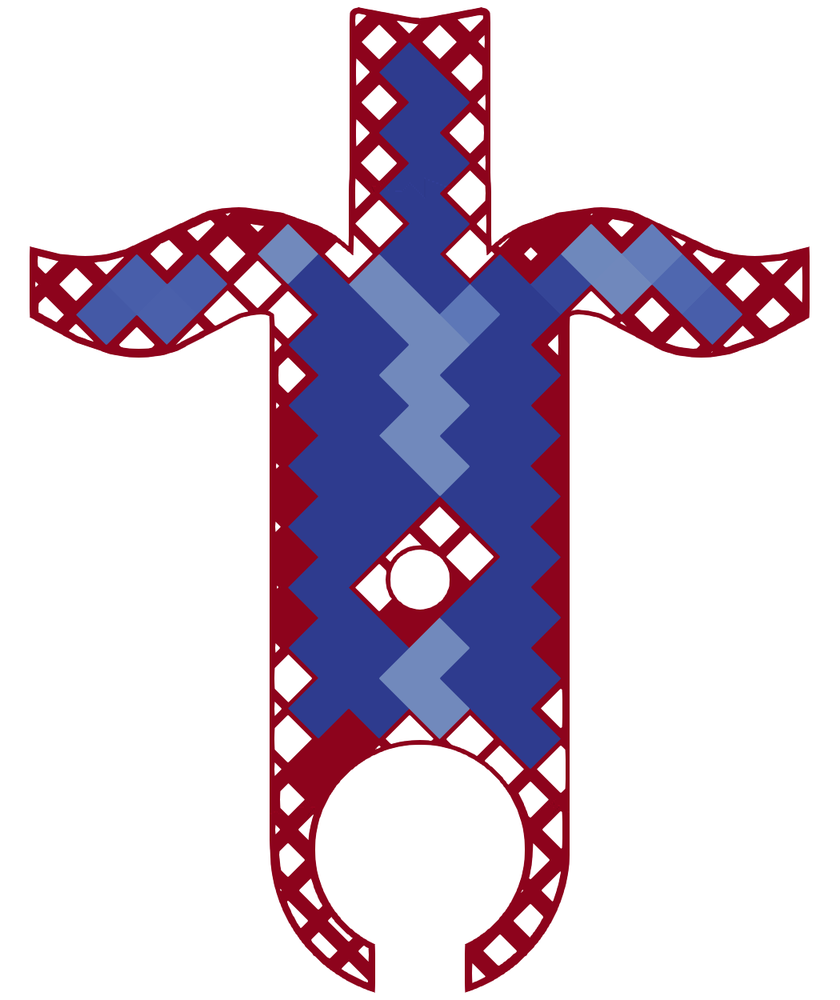} &
\includegraphics[angle=180,origin=c,height=2cm]{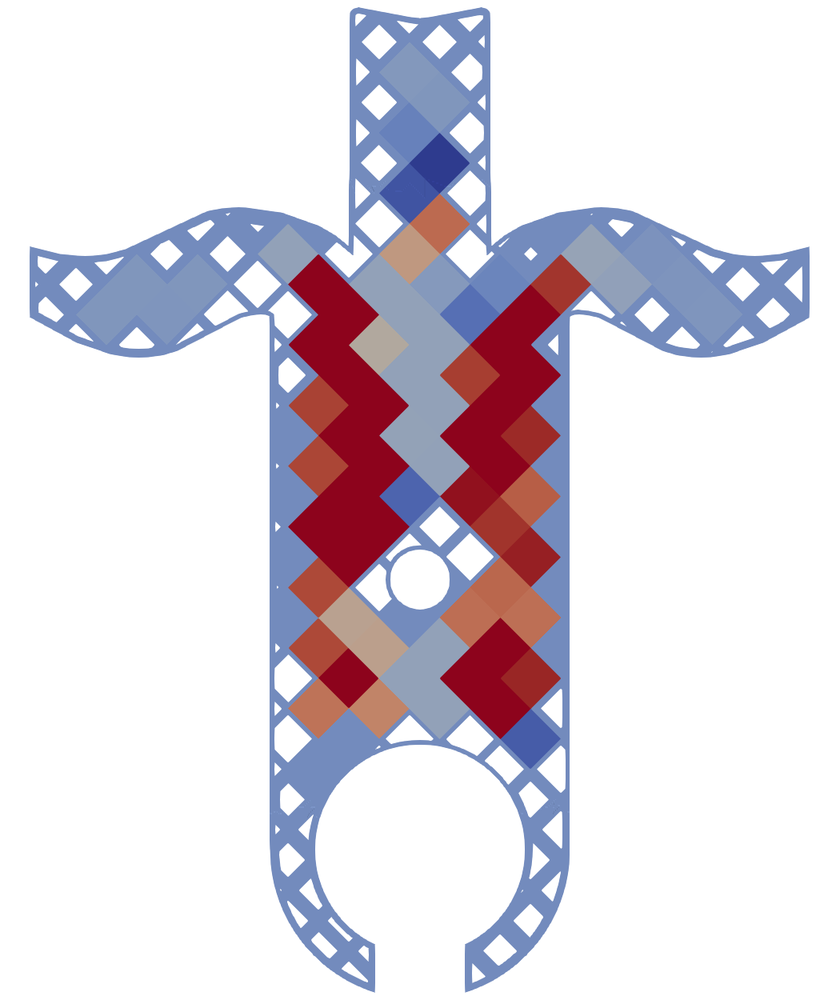} &
\includegraphics[angle=180,origin=c,height=2cm]{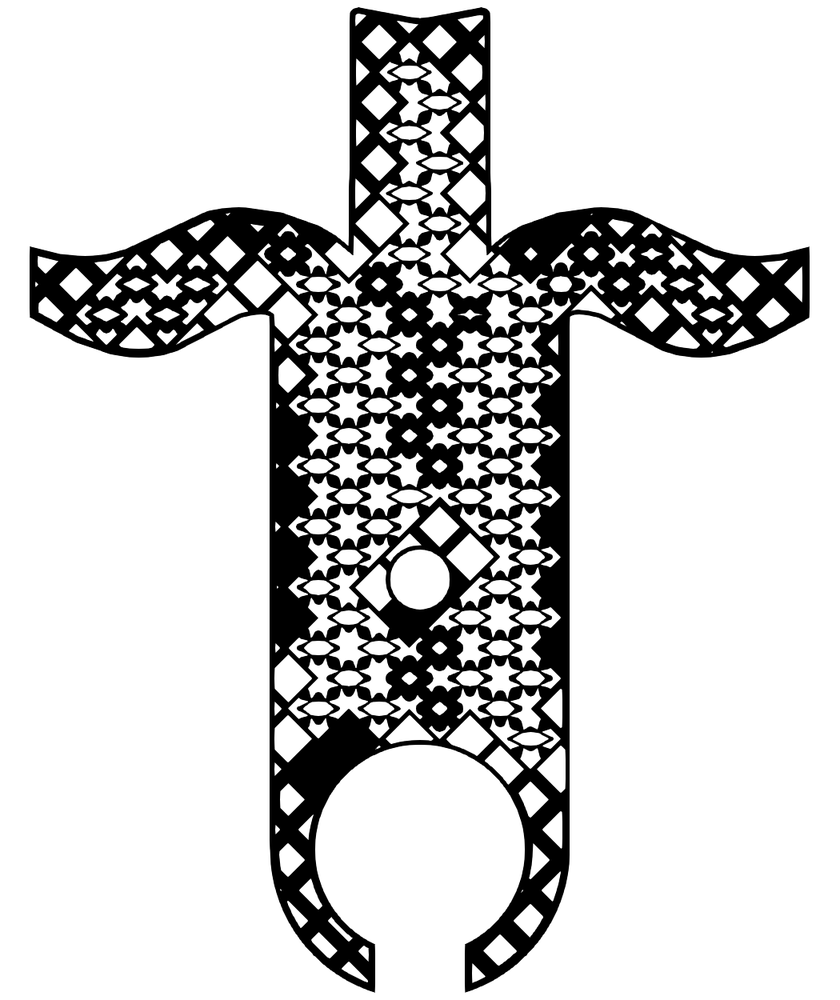} &
\includegraphics[angle=180,origin=c,height=2cm]{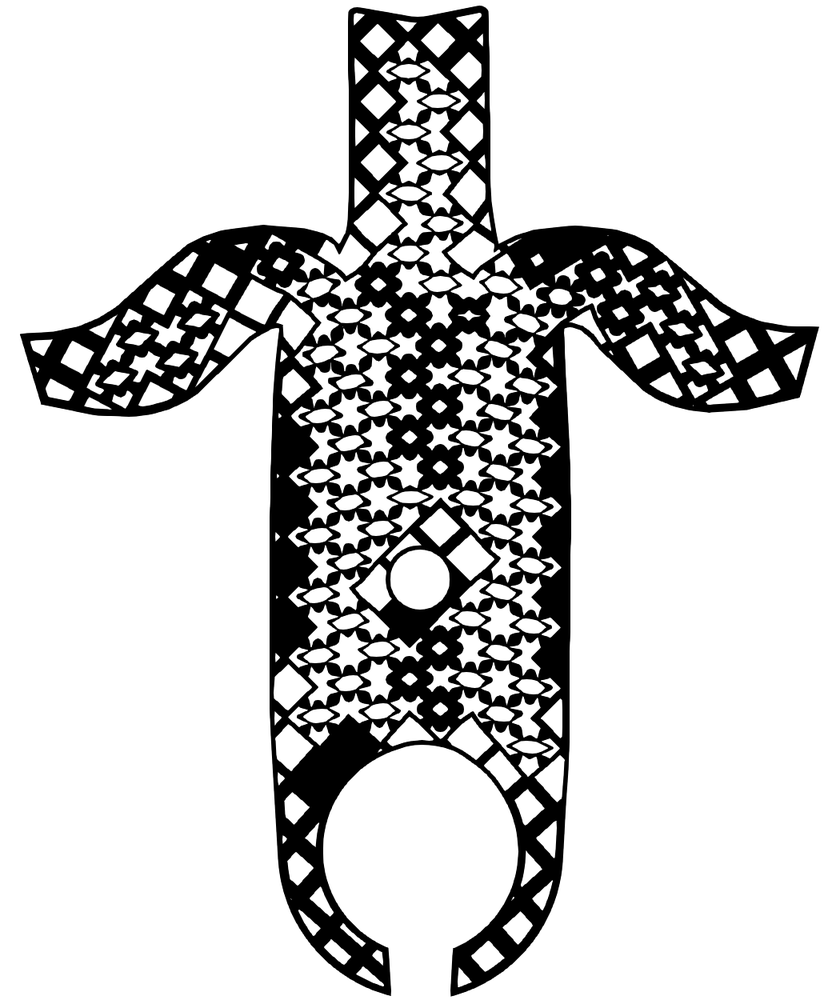} &
\includegraphics[angle=0,origin=c,height=2cm]{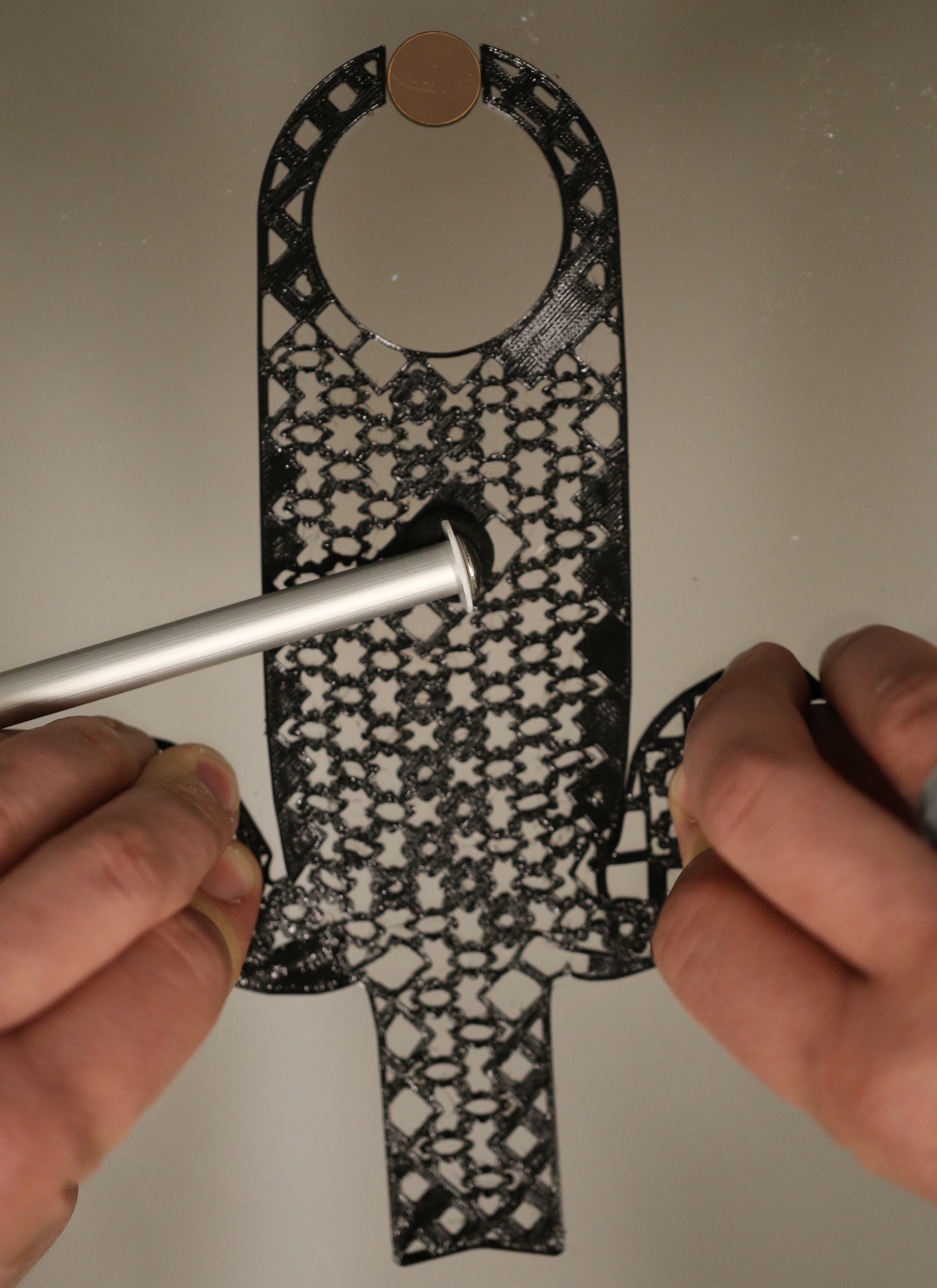}
\end{tabular}}
}
\caption{
A gallery of 2D examples. From left to right: optimized material distribution (Young’s modulus and Poisson’s ratio), final geometry at rest, deformed geometry (simulated) and photograph of deformed pattern (fabricated). From top to bottom: 
\textbf{Disk.} The objective is to obtain a shear-like deformation, bulging on the top left and bottom right, whenever the disk is compressed in the vertical direction. 
\textbf{Sine bar.} The bar is optimized to match a sine wave shape. %
\textbf{Happy Ghost.} The pacman ghost shape is optimized to smile when compressed on the sides. We add target displacements such that the middle region (of both the top and bottom) of the mouth  displace down.
\textbf{Pliers.} The pliers is optimized to close when the handles are pulled together.
\textbf{Bird.} The shape is optimized to flap its wings whenever its head is pushed down. In this case, we add a Dirichlet boundary condition on the bird's feet to keep it in place. %
\textbf{Sword Gripper.} 
The gripper has a rounded hole in the middle that is used to hold the shape in place and should not deform during the operation. The mechanism closes whenever the handle moves up.
}
\label{fig:2D_examples}
\end{figure*}

\section{Evaluation}
\label{sec:results}

Our algorithm is implemented in C++ and uses Eigen~\cite{eigenweb} for the linear algebra routines, PolyFEM~\cite{polyfem} for finite element simulation in 3D and MeshFEM~\cite{panetta2015elastic} in 2D, fTetwild~\cite{hu2020fast} and triangle \cite{shewchuk2005triangle} for meshing, and Pardiso~\cite{pardiso-7.2a,pardiso-7.2b,pardiso-7.2c} for solving linear systems. We run our experiments on a cluster node with 
an Intel Cascade Lake Platinum 8268 processor limited to 32 threads.

\subsection{2D examples}

\begin{figure}
\centering
    \includegraphics[width=0.7\linewidth]{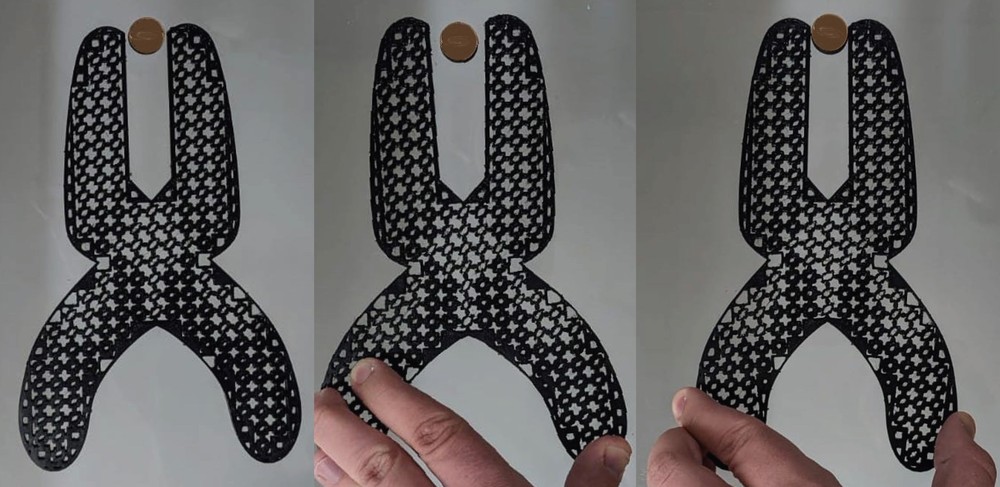}
\caption{
Fabricated pliers (black TPU). Rest shape on the left; moving the handles apart in the middle; compressing the handles to hold the object on the right.
}\label{fig:pliers-fabricated}
\end{figure}

\begin{figure}
\centering
    \includegraphics[width=0.7\linewidth]{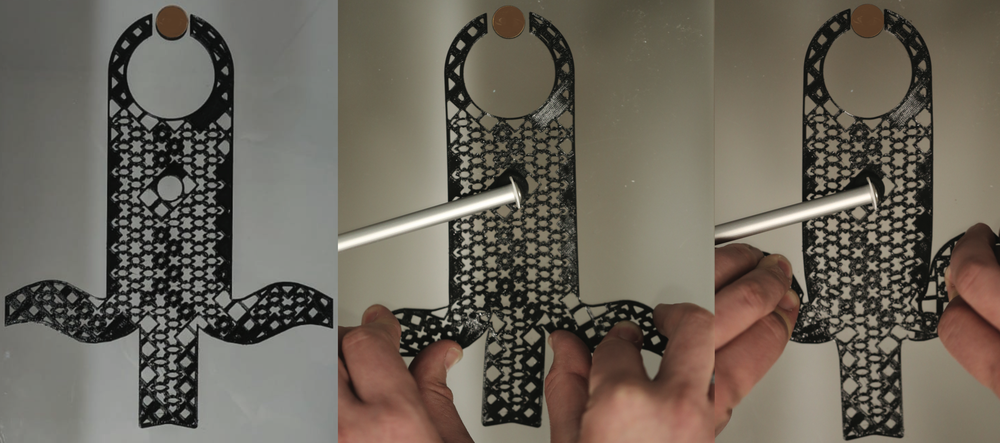}
\caption{
Fabricated sword gripper (black TPU). Rest shape on the left; moving handles back in the middle; moving handles to the front, holding the object, on the right.
}\label{fig:sword-fabricated}
\end{figure}

We experimented with six 2D models shown in Figure~\ref{fig:2D_examples}. Visually, our results closely match the specified target, both in direct simulation (done on a dense mesh of the final geometry) and in our physical validations. For physical validation, we fabricated the models using FDM printing with a TPU 95A filament on a Prusa i3 MK3S and an Ultimaker 3 (see Figures~\ref{fig:pliers-fabricated} and \ref{fig:sword-fabricated}
). 
Numerical results, breaking down the effect of the different optimization stages, are available in Table~\ref{tab:examples-numerical} and discussed in more detail below. 

 The target displacement for the six examples is described in the caption of Figure \ref{fig:2D_examples}. Additionally, we show two deformation sequences for the plier (Figure \ref{fig:pliers-fabricated}) and sword gripper (Figure \ref{fig:sword-fabricated}).

\begin{figure*}[htb!]
\makebox[\linewidth][c]{%
\setlength\tabcolsep{1pt}
\renewcommand{\arraystretch}{0}
\begin{tabular}{@{}ccccc@{}}
model & $E$ & $\nu$ & pattern & simulated \\
\midrule
\includegraphics[angle=0,origin=c,width=3cm]{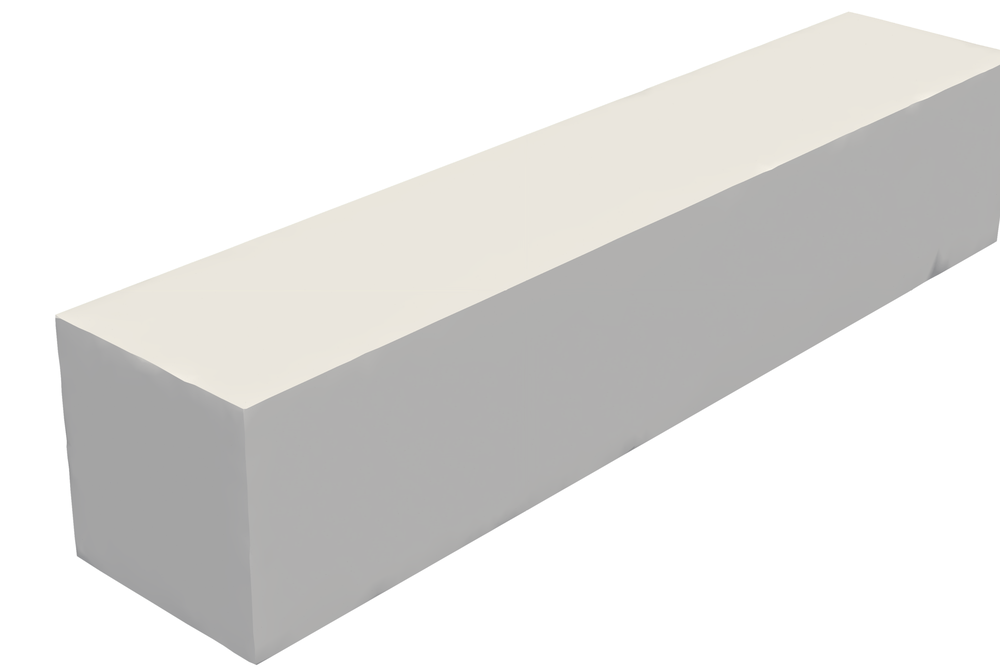} &
\includegraphics[angle=0,origin=c,width=3cm]{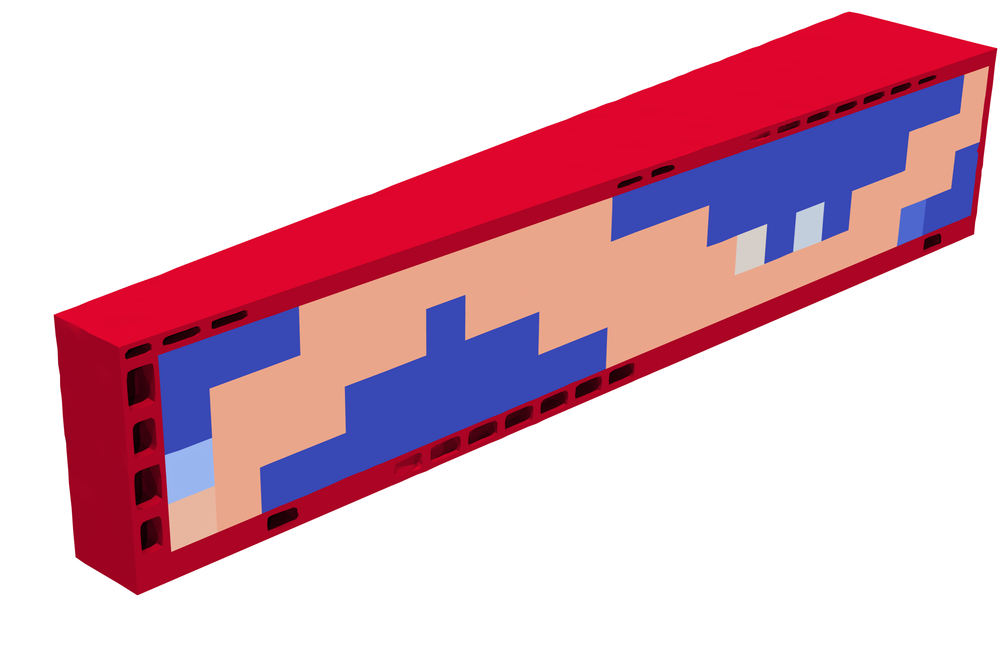} &
\includegraphics[angle=0,origin=c,width=3cm]{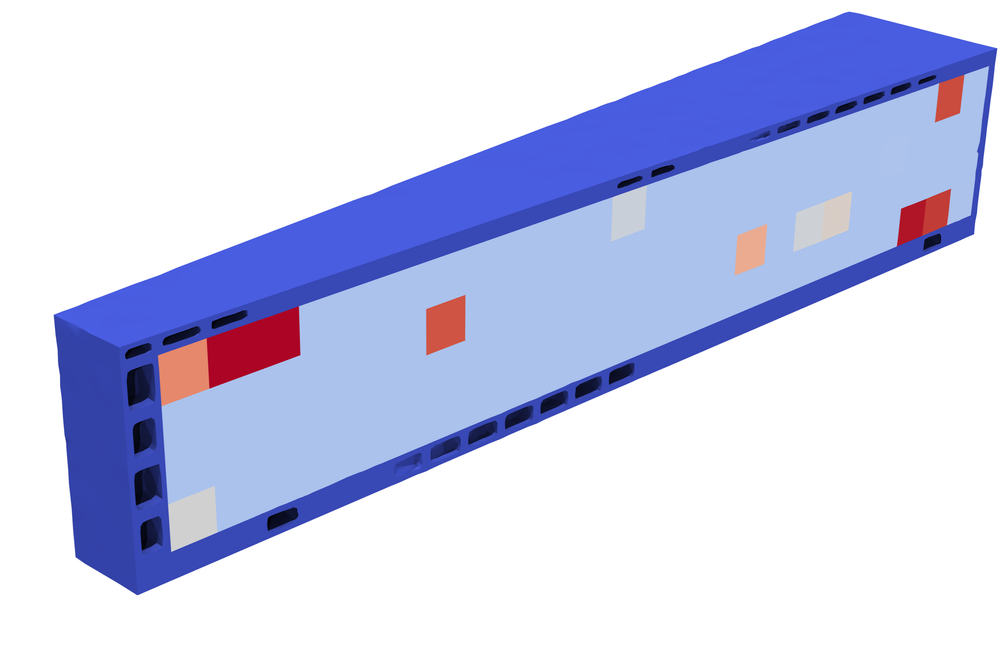} &
\includegraphics[angle=0,origin=c,width=3cm]{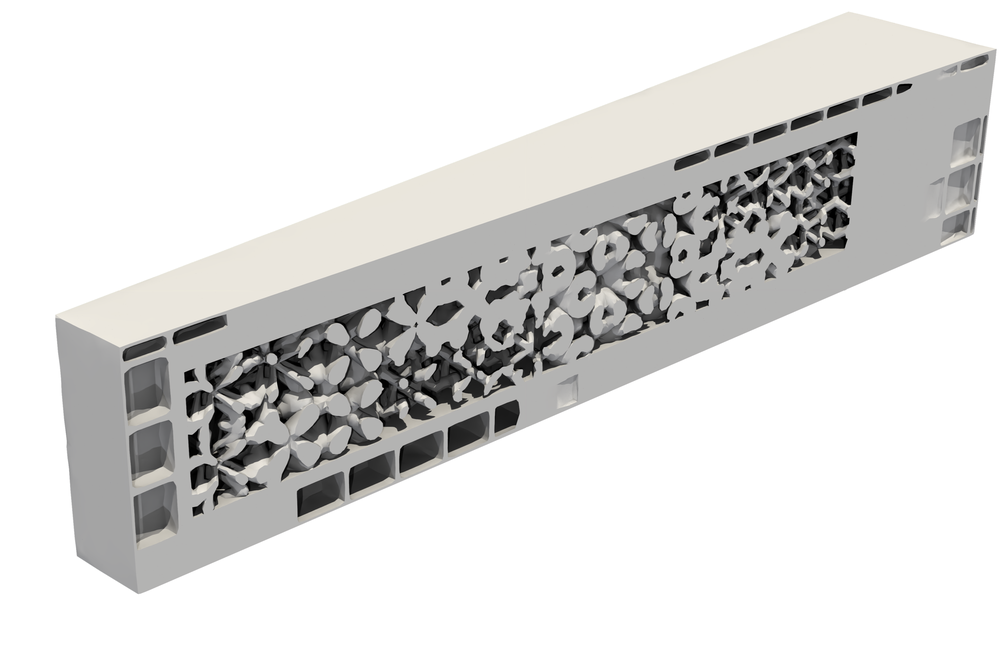} &
\includegraphics[angle=0,origin=c,width=3cm]{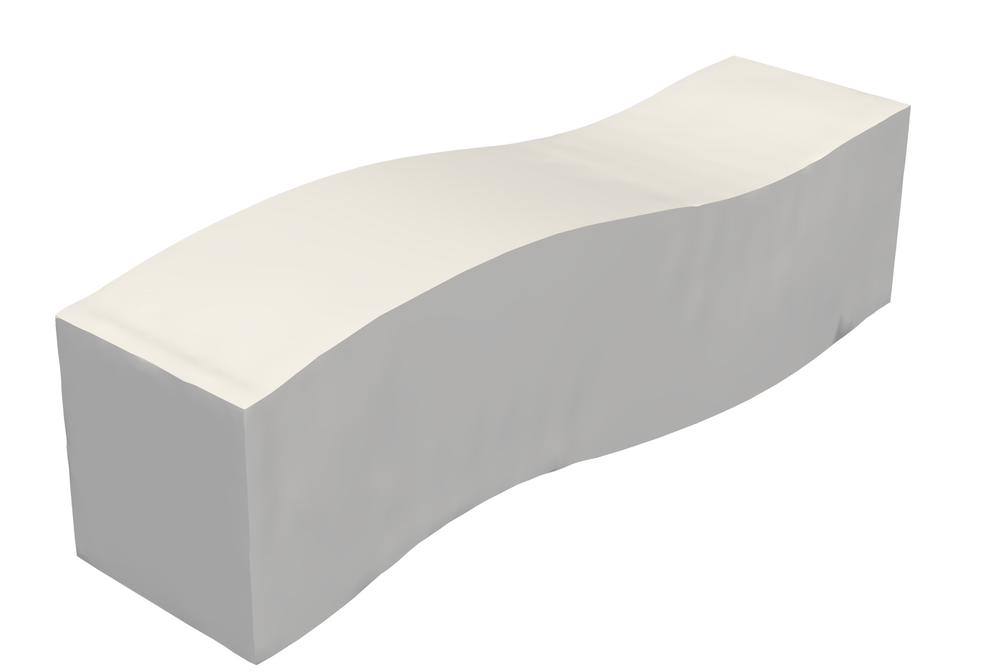}
 \\
 \includegraphics[angle=0,origin=c,width=3cm]{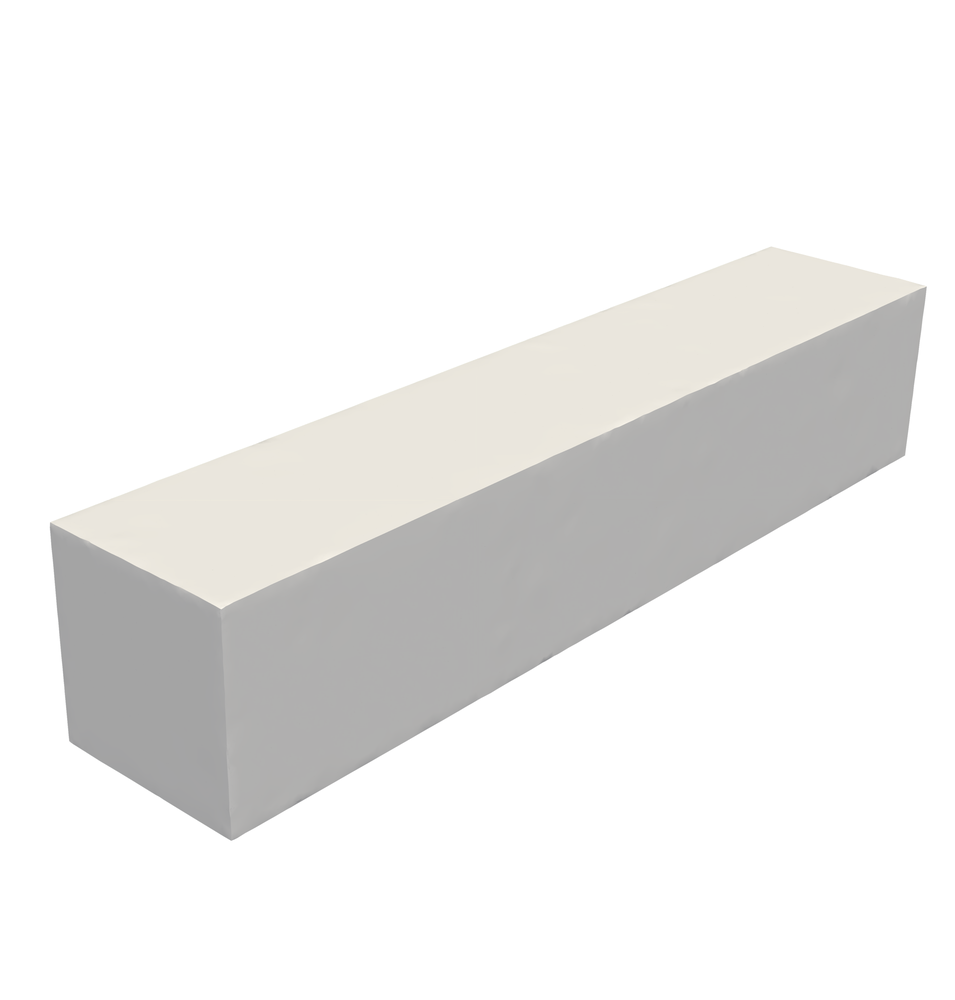} &
\includegraphics[angle=0,origin=c,width=3cm]{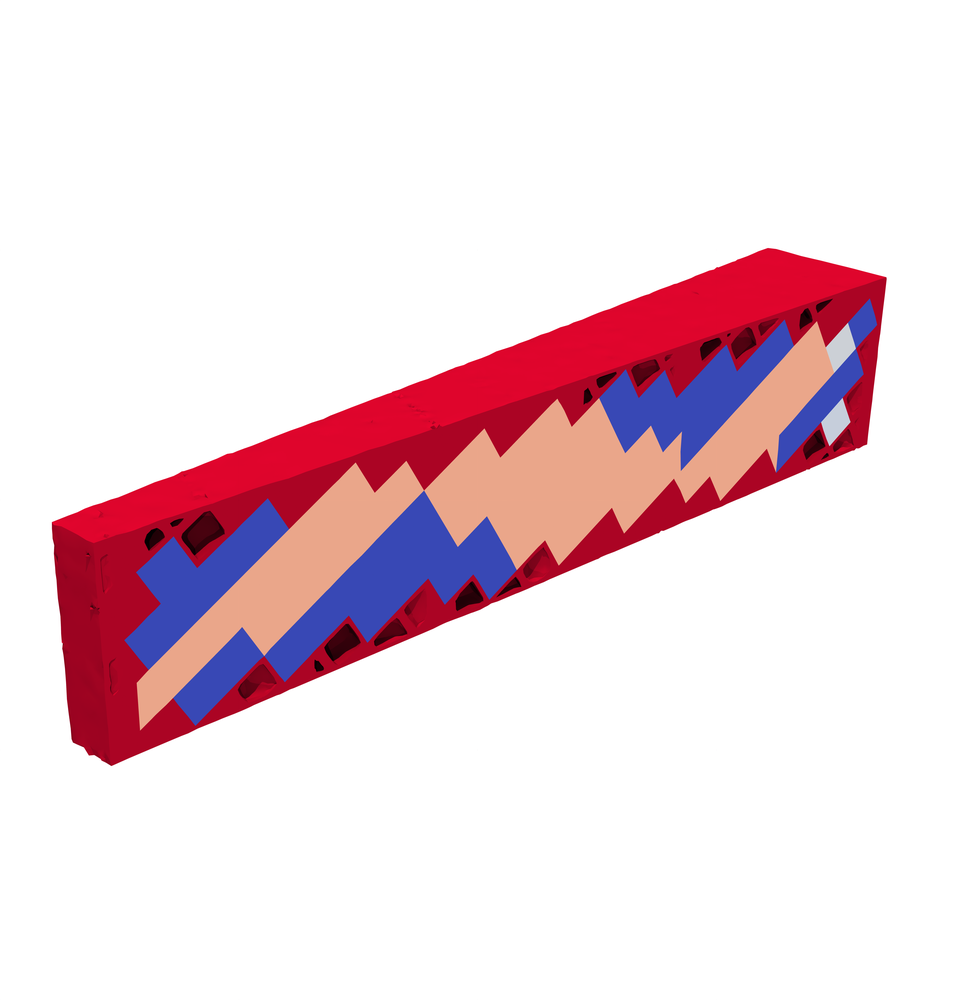} &
\includegraphics[angle=0,origin=c,width=3cm]{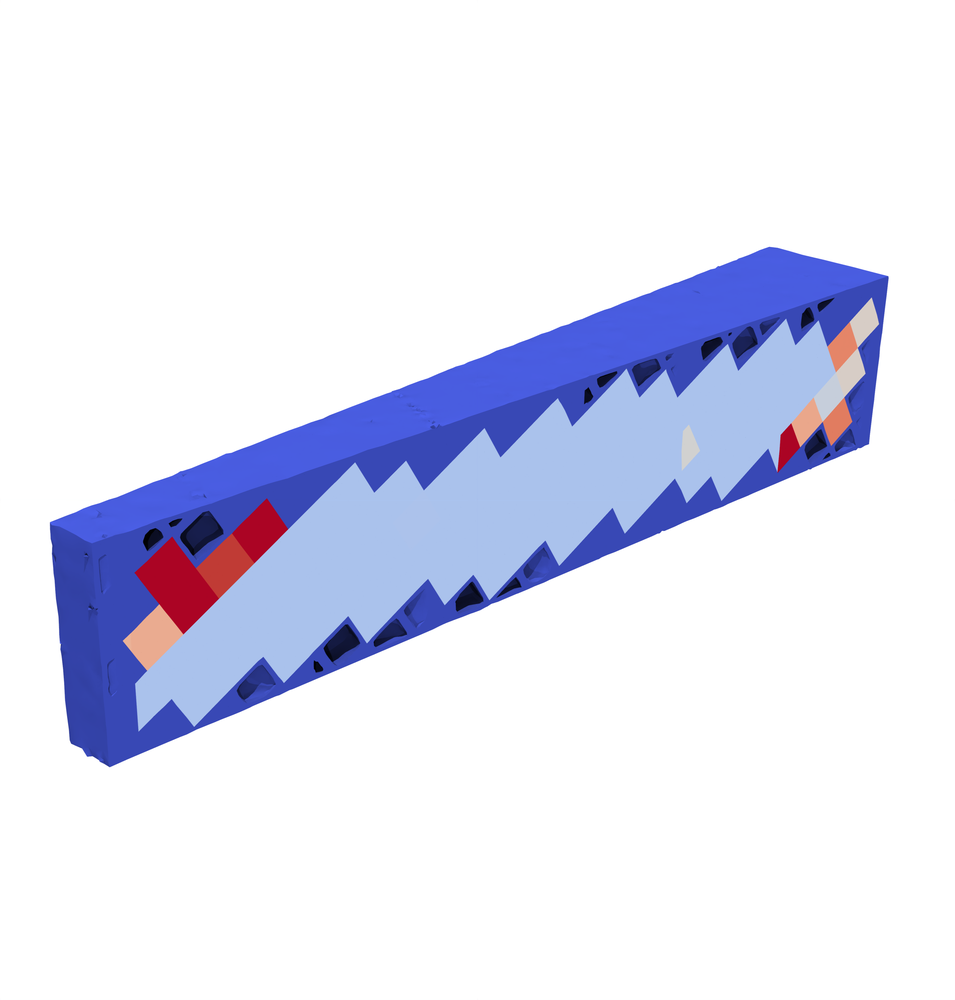} &
\includegraphics[angle=0,origin=c,width=3cm]{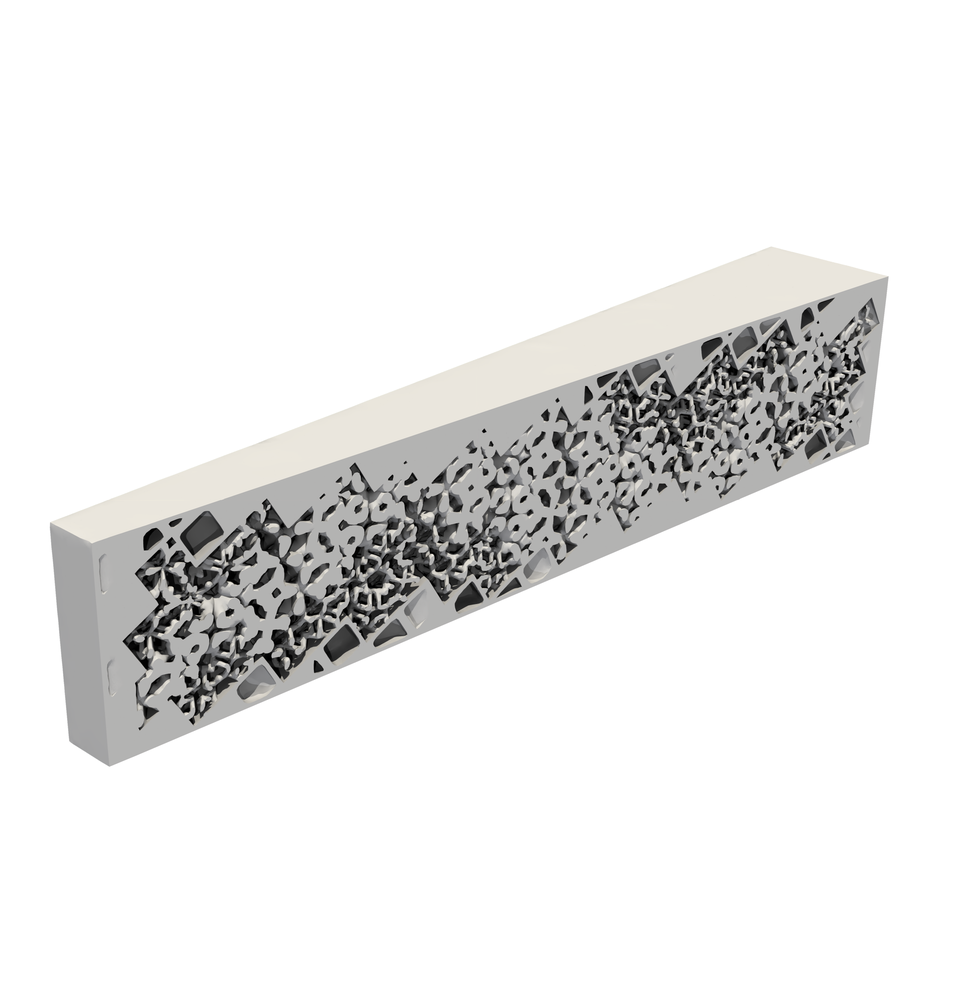} &
\includegraphics[angle=0,origin=c,width=3cm]{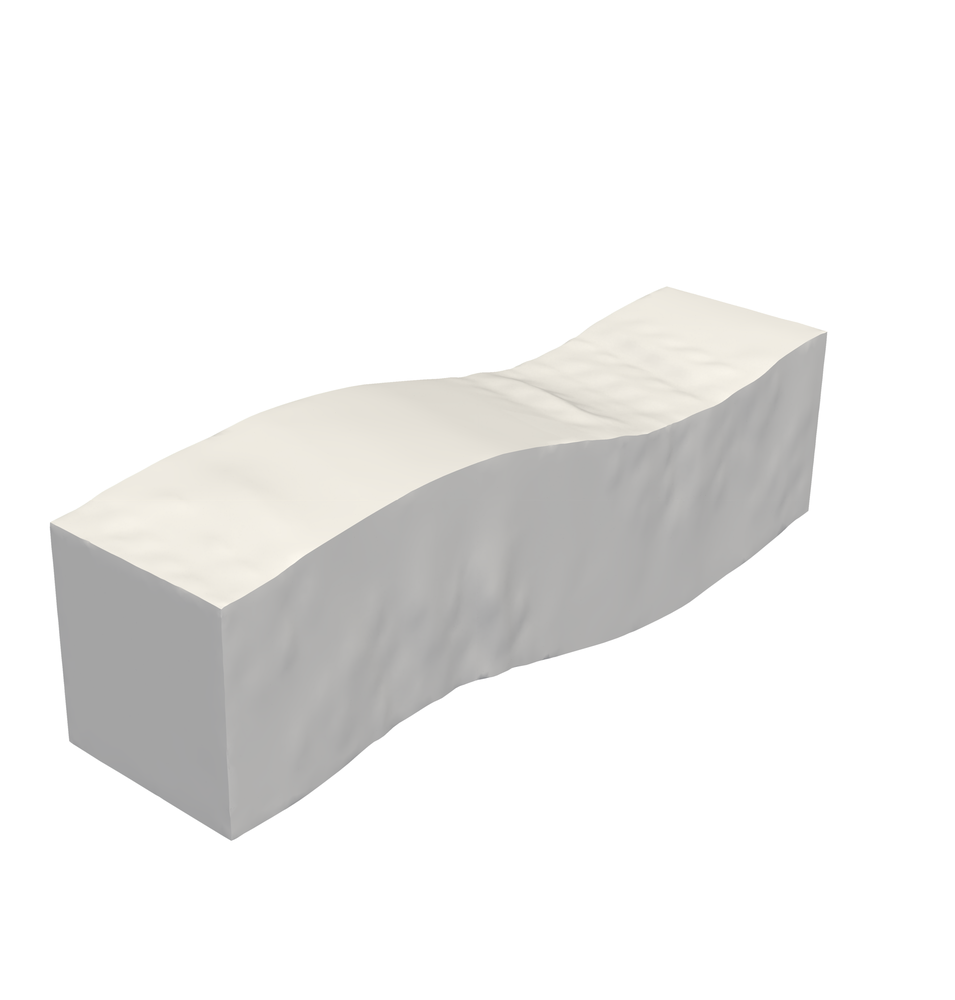}
 \\
  \includegraphics[angle=0,origin=c,width=3.5cm]{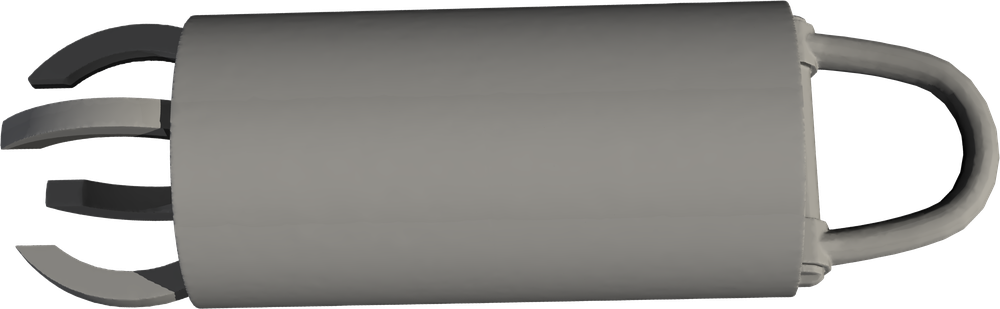} &
\includegraphics[angle=0,origin=c,width=3.5cm]{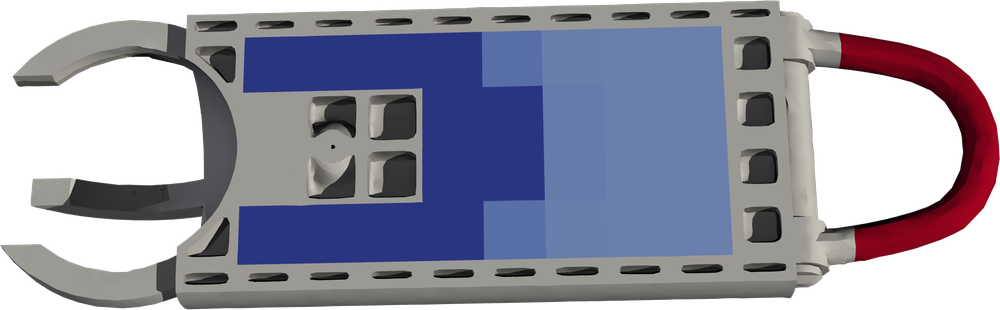} &
\includegraphics[angle=0,origin=c,width=3.5cm]{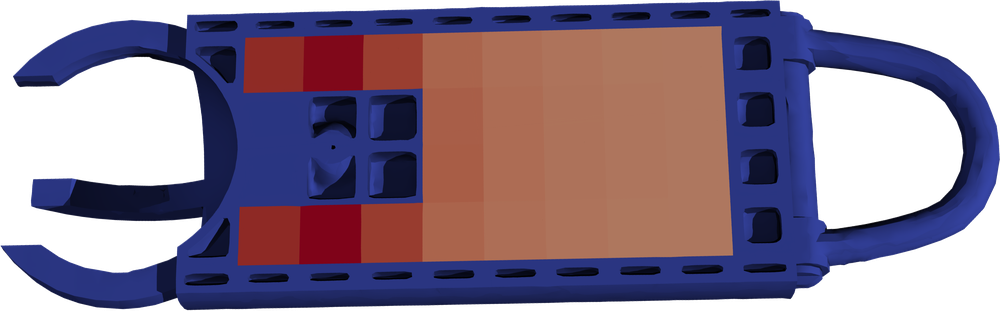} &
\includegraphics[angle=0,origin=c,width=3.5cm]{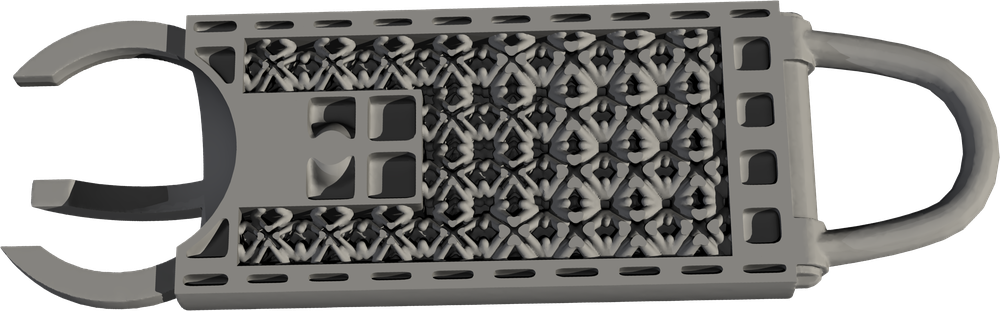} &
\includegraphics[angle=0,origin=c,width=3.5cm]{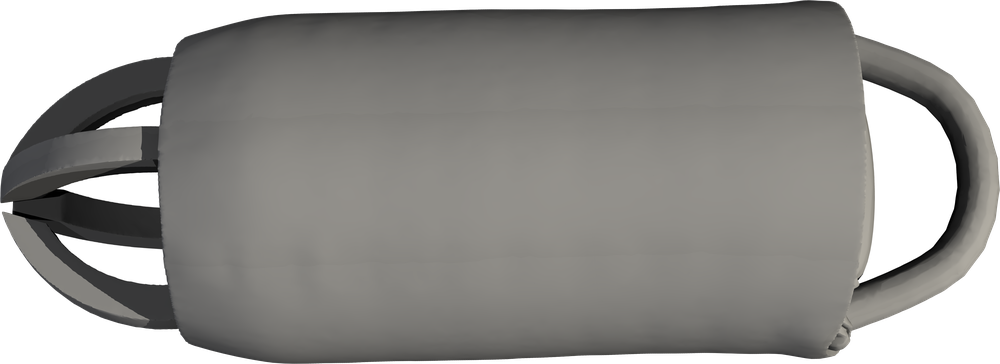}
 \\
   \includegraphics[angle=0,origin=c,width=3cm]{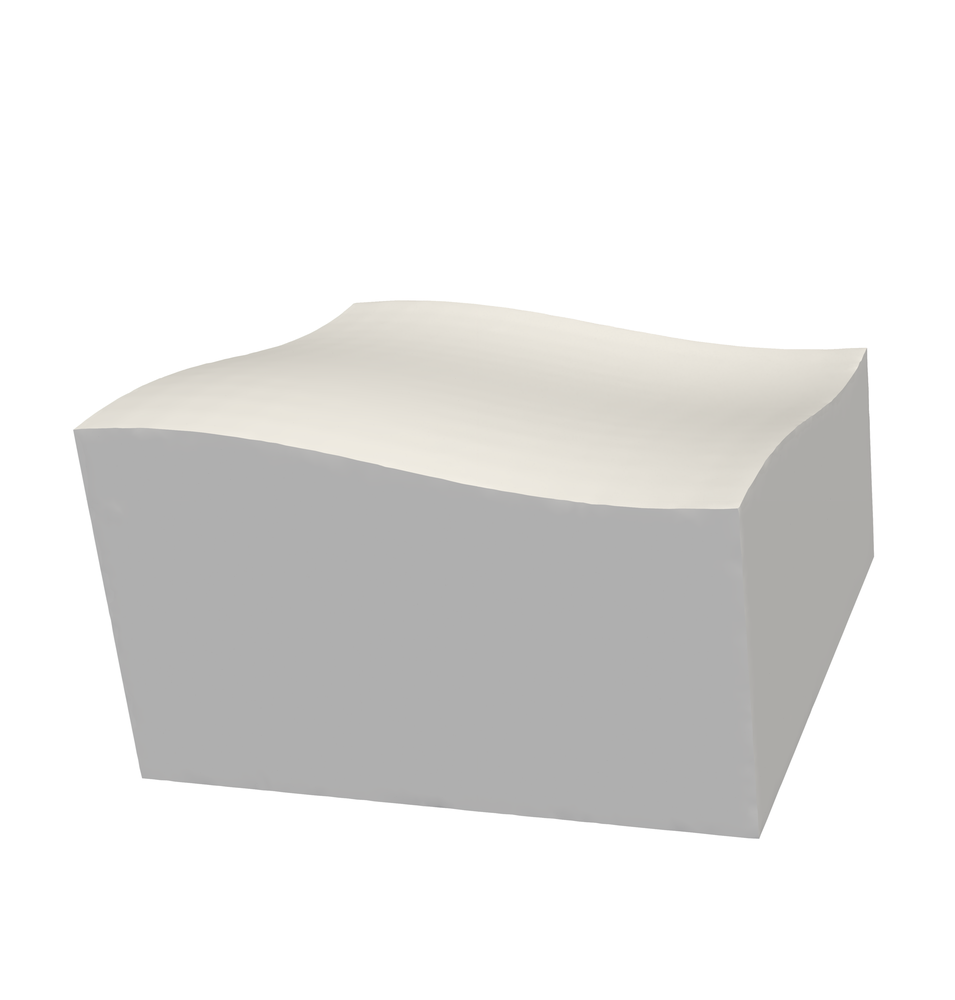} &
\includegraphics[angle=0,origin=c,width=3cm]{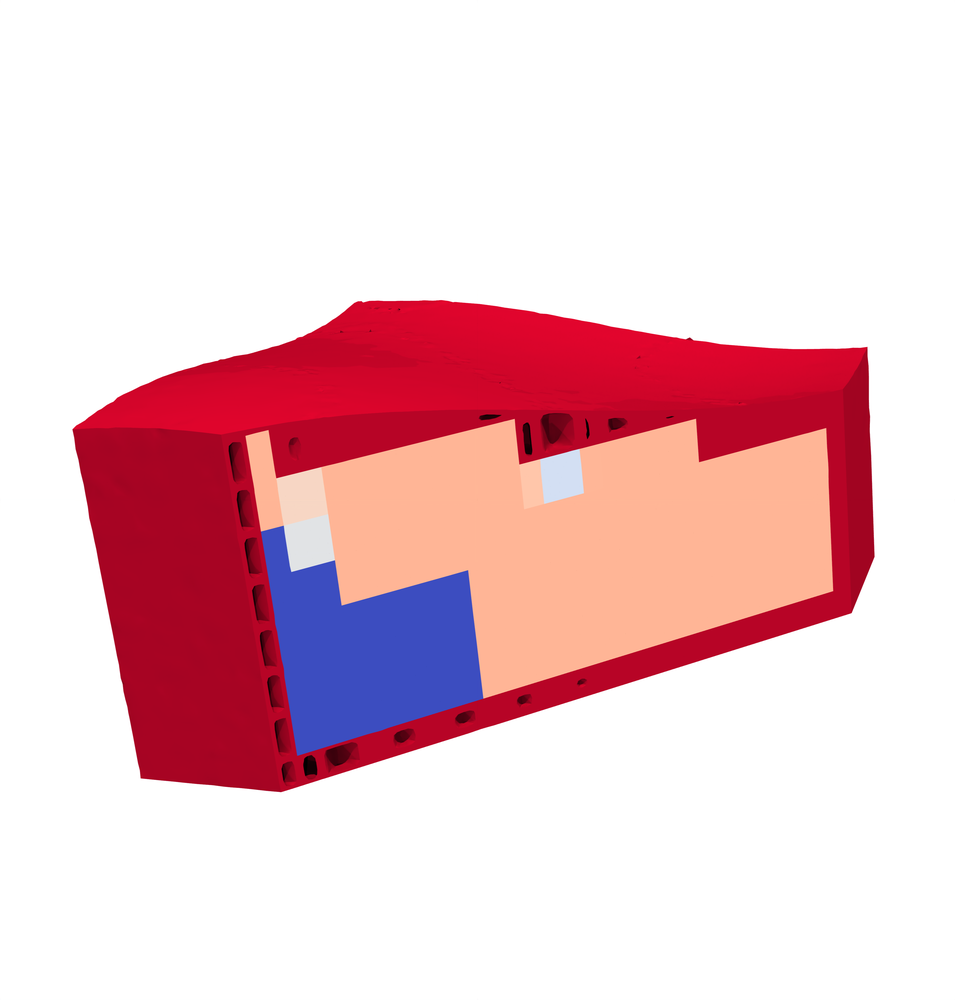} &
\includegraphics[angle=0,origin=c,width=3cm]{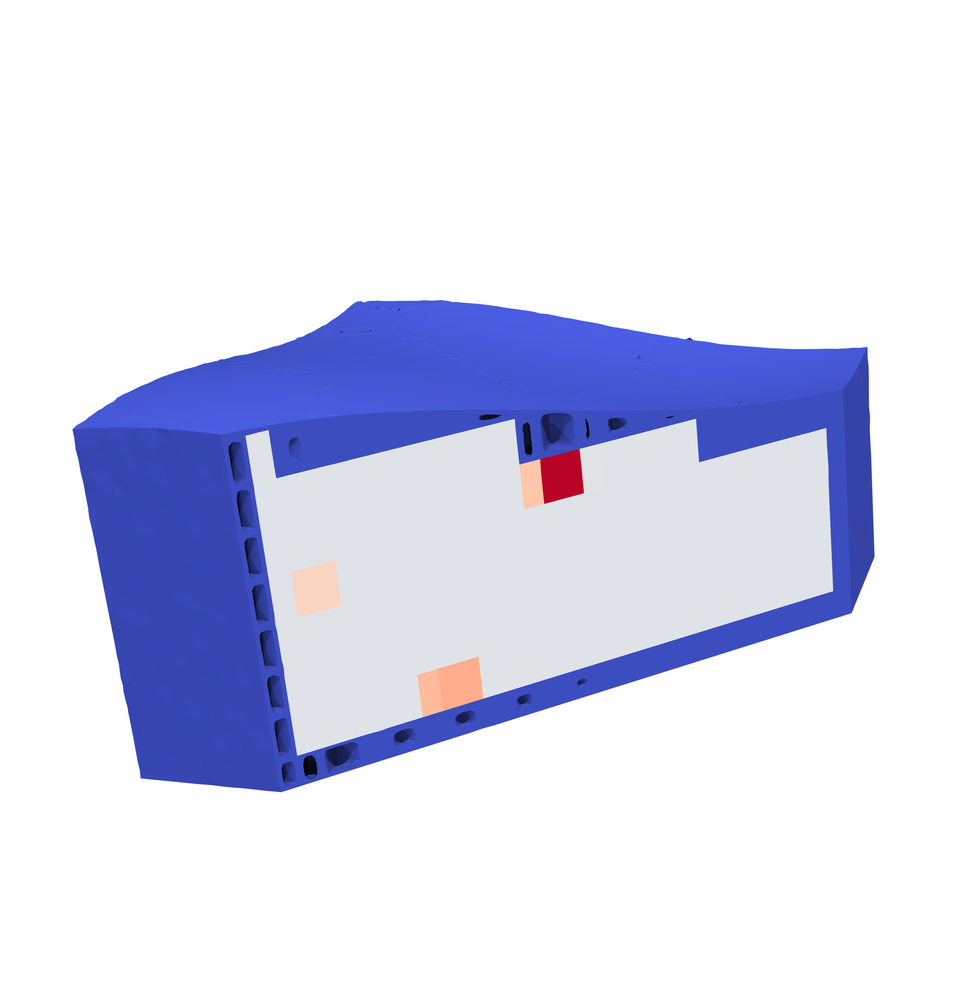} &
\includegraphics[angle=0,origin=c,width=3cm]{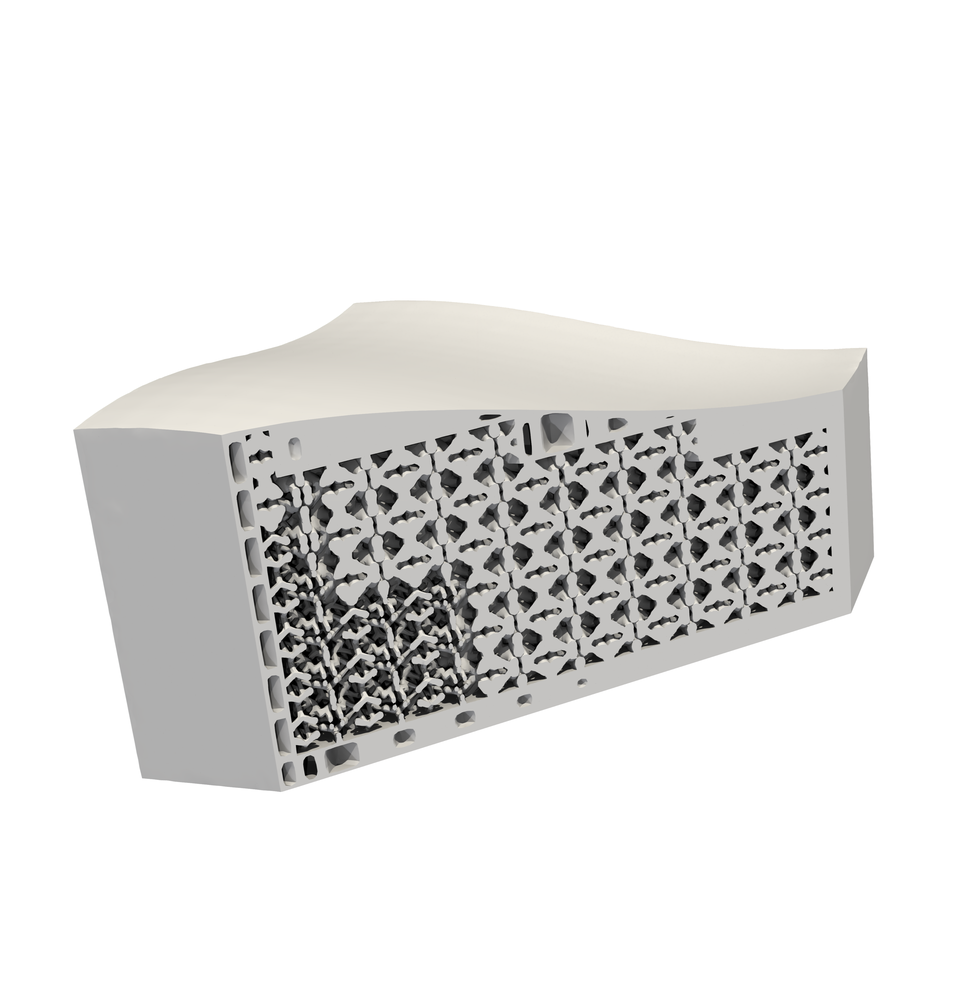} &
\includegraphics[angle=0,origin=c,width=3cm]{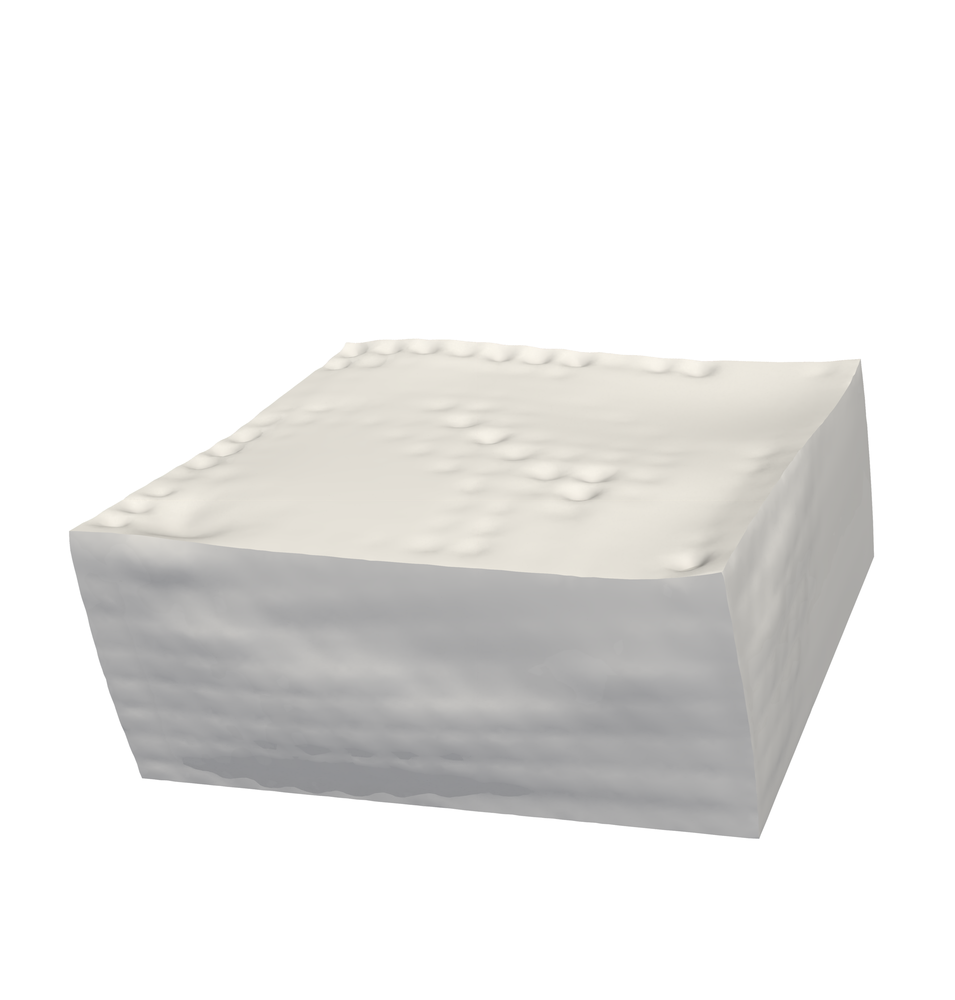}
 \\
    \includegraphics[angle=0,origin=c,width=3cm]{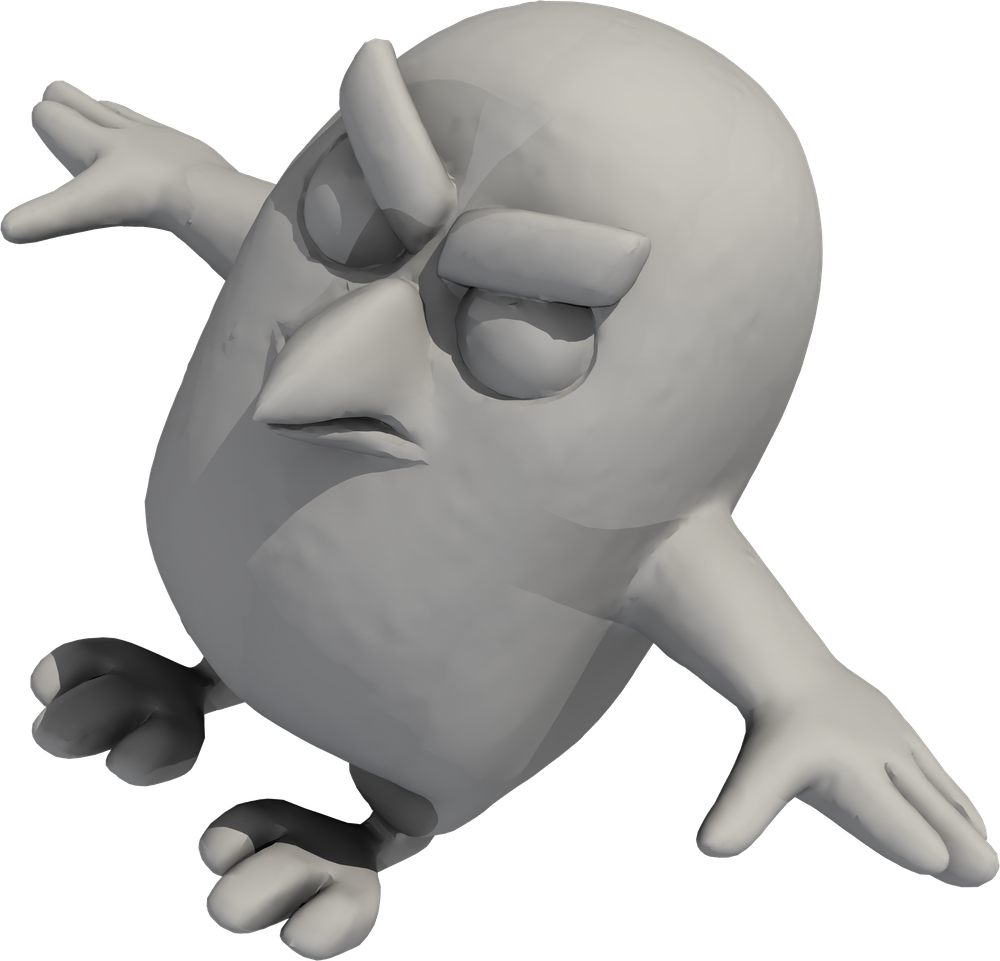} &
\includegraphics[angle=0,origin=c,width=3cm]{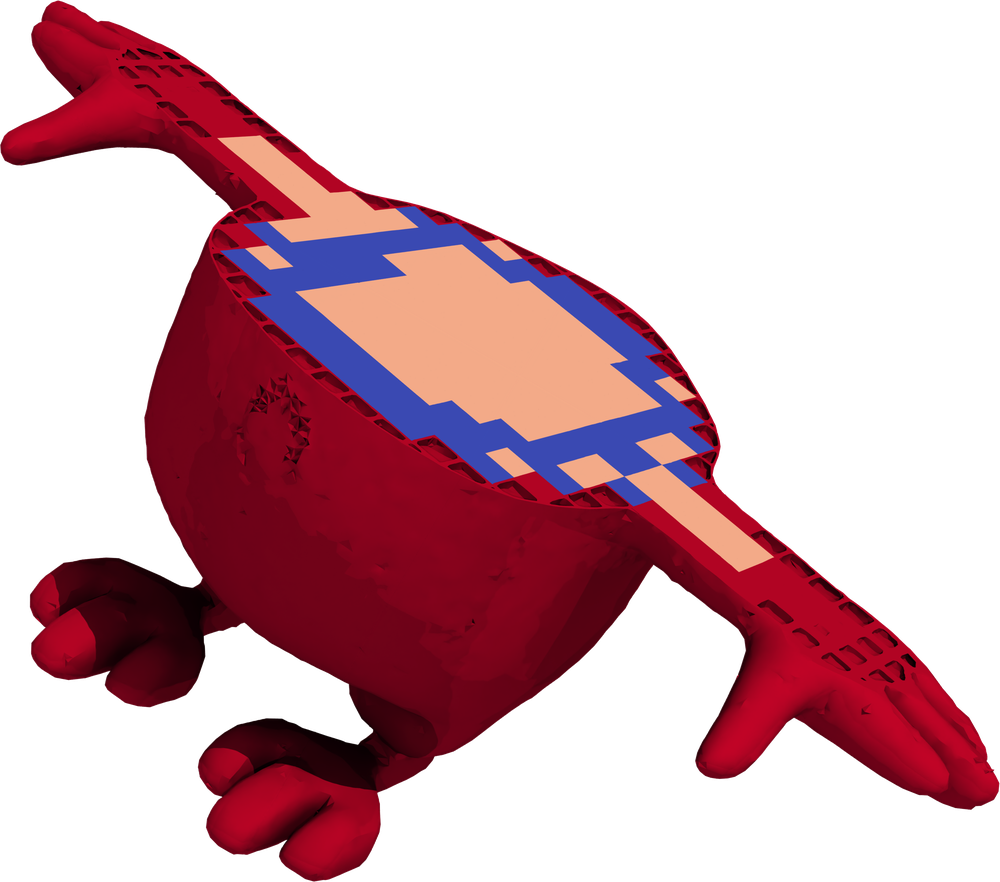} &
\includegraphics[angle=0,origin=c,width=3cm]{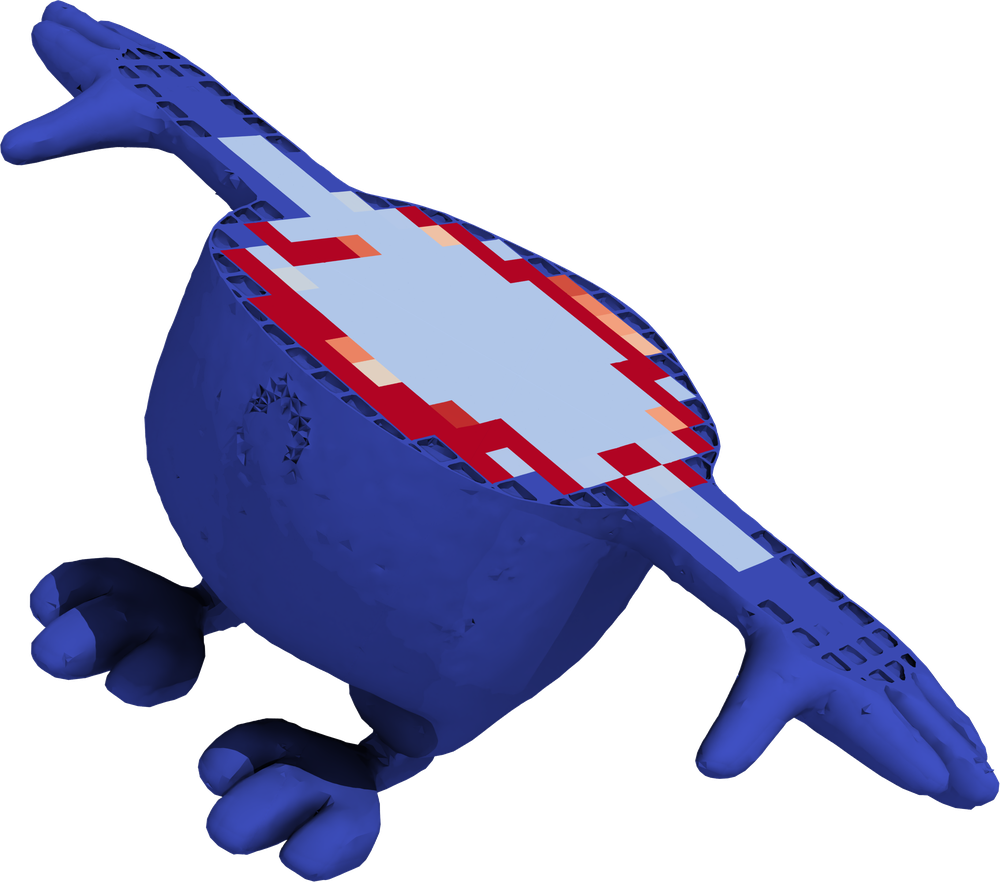} &
\includegraphics[angle=0,origin=c,width=3cm]{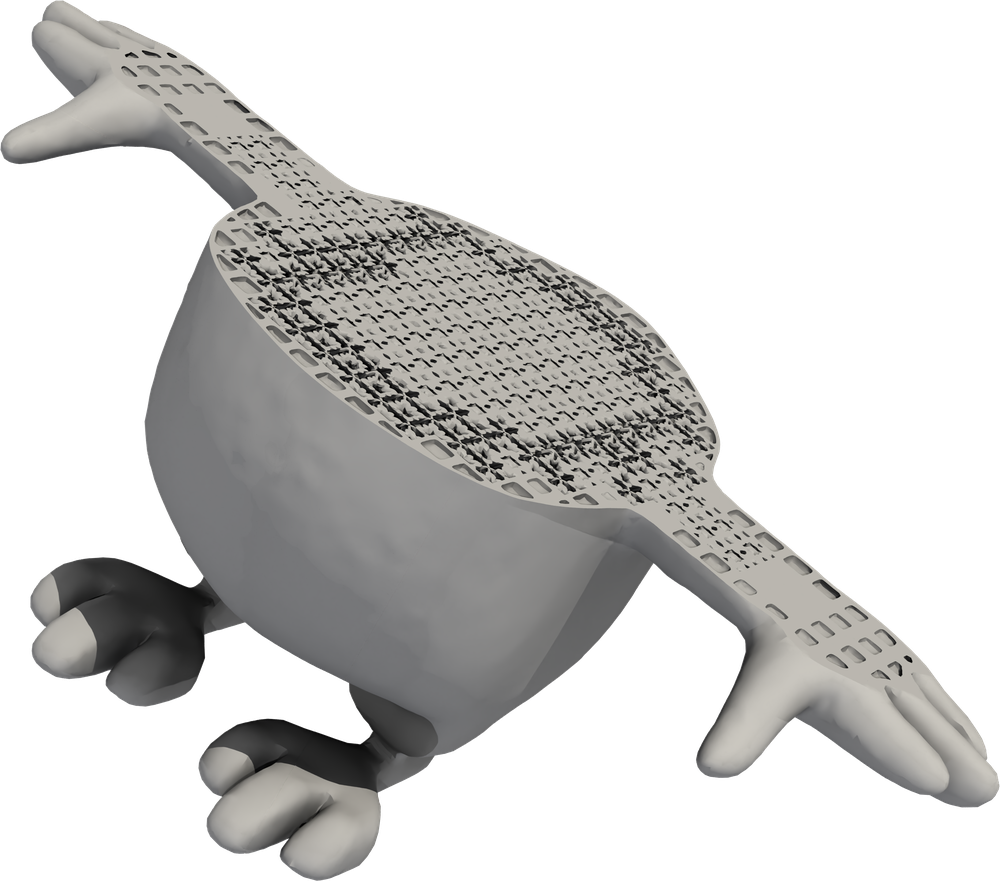} &
\includegraphics[angle=0,origin=c,width=3cm]{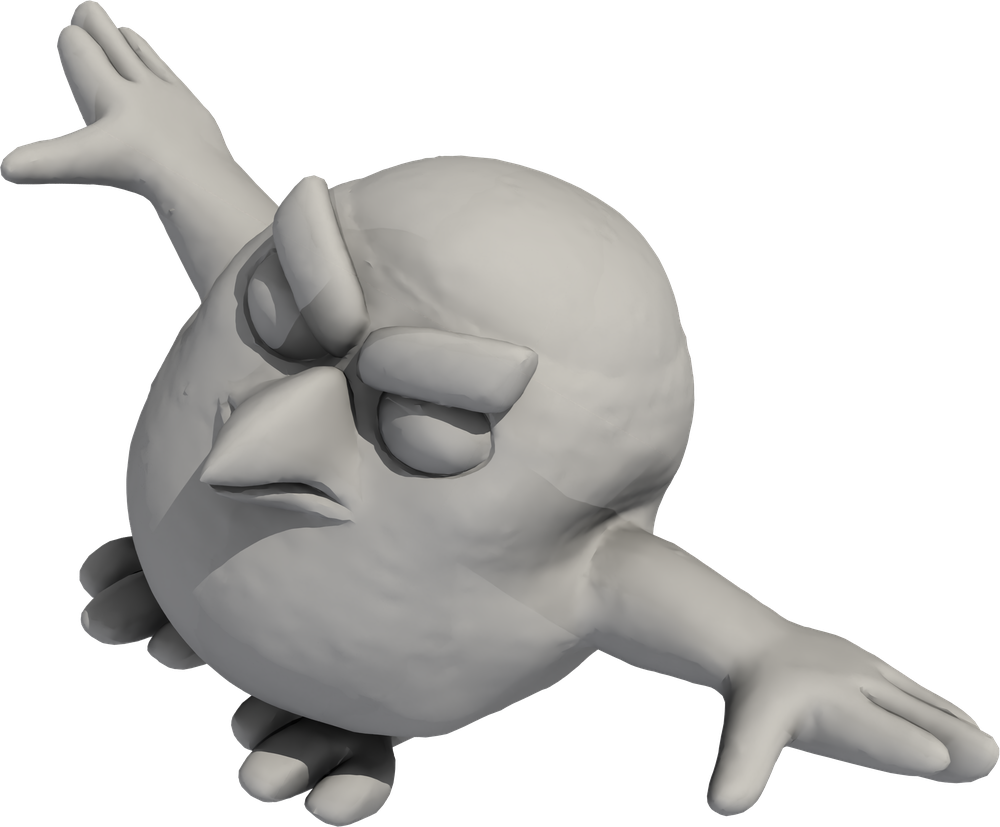}
 \\
\end{tabular}

}

\caption{
A gallery of 3D examples. From left to right: original model, optimized material distribution (Young’s modulus and Poisson’s ratio), final geometry at rest and deformed geometry (simulated).
}
\label{fig:3D_examples}
\end{figure*}

\begin{figure*}[htb!]

\makebox[\linewidth][c]{%
\setlength\tabcolsep{4pt}
\renewcommand{\arraystretch}{0}
\begin{tabular}{@{}cccccccccc@{}}
    \includegraphics[angle=0,origin=c,width=1.4cm]{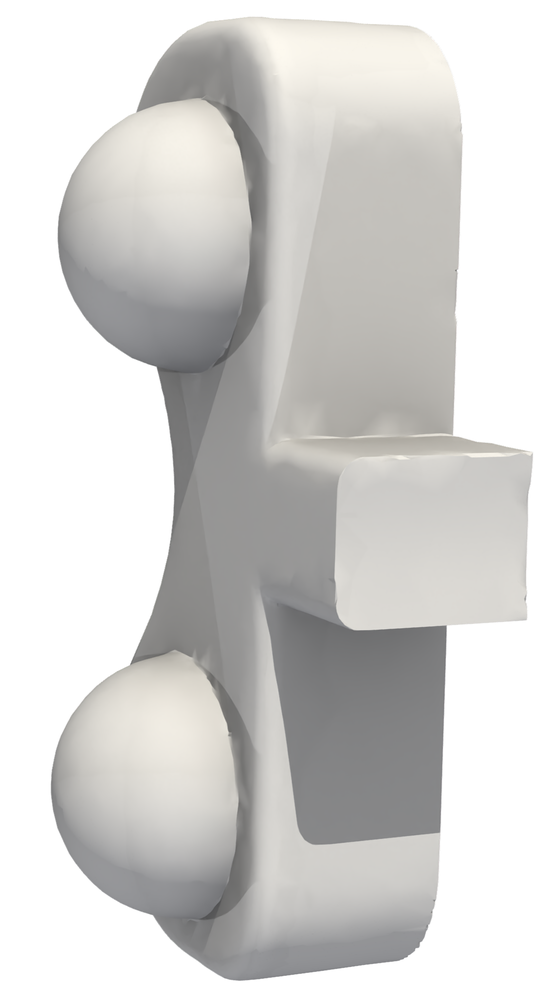} &
    \includegraphics[angle=0,origin=c,width=1.2cm]{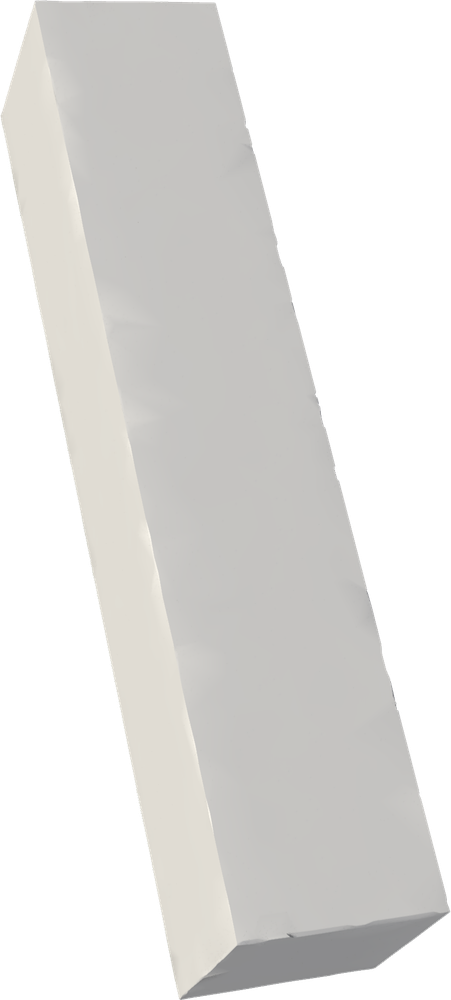} &
    \includegraphics[angle=0,origin=c,width=1.2cm]{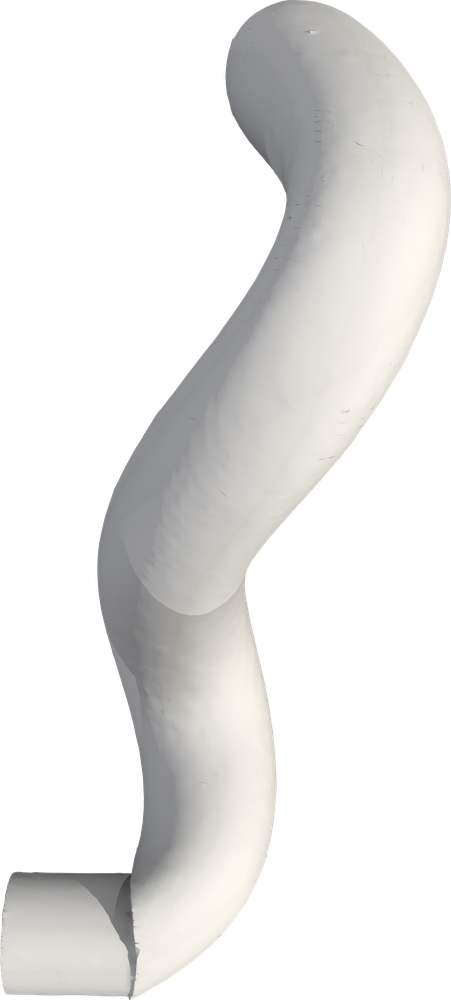} &
    \includegraphics[angle=0,origin=c,width=1.5cm]{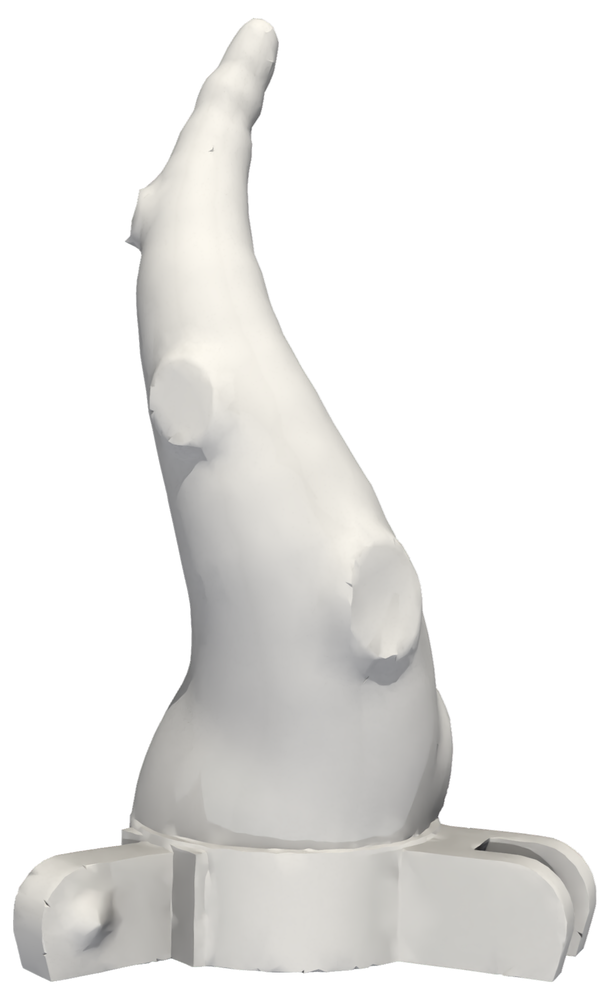} &
    \includegraphics[angle=0,origin=c,width=1.7cm]{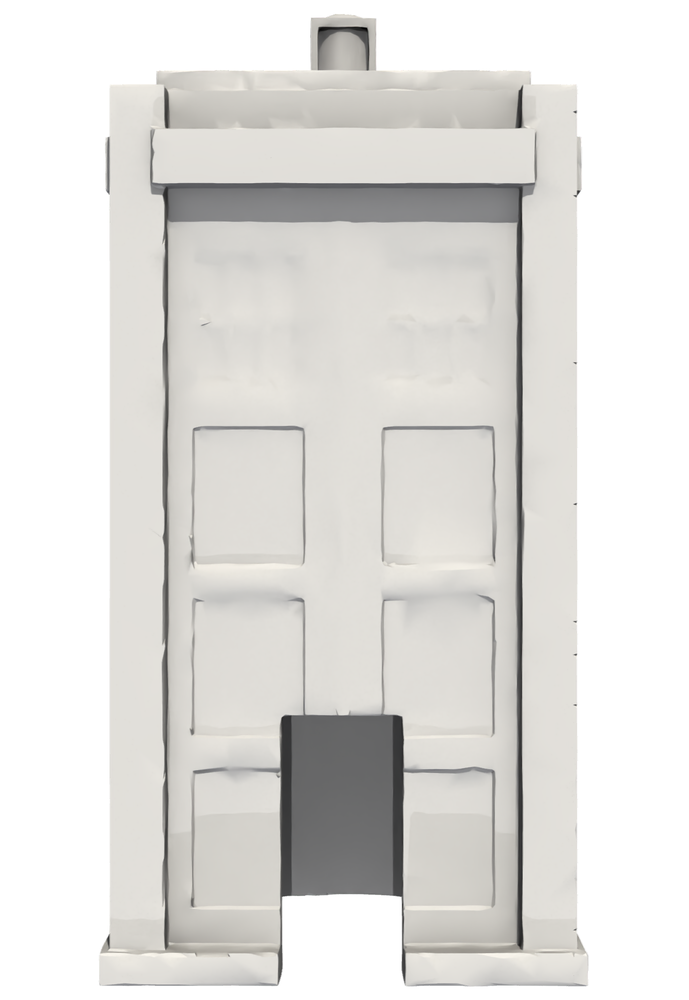} &
    \includegraphics[angle=0,origin=c,width=1.3cm]{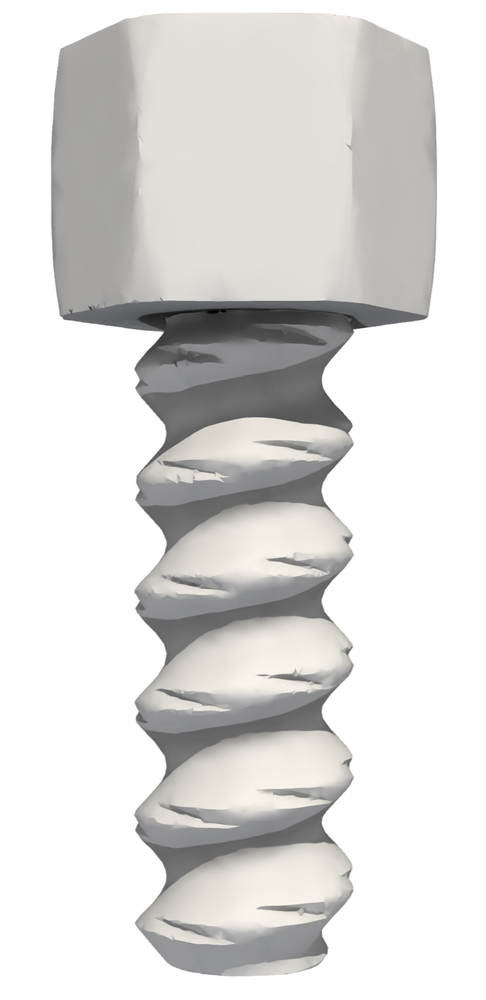} &
    \includegraphics[angle=0,origin=c,width=1.7cm]{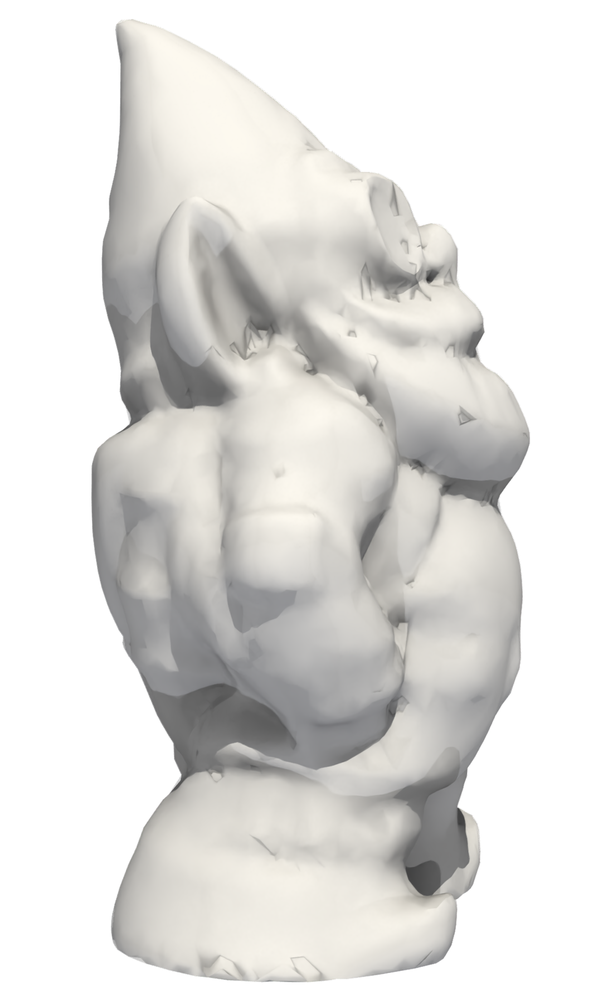} &
    \includegraphics[angle=0,origin=c,width=1.7cm]{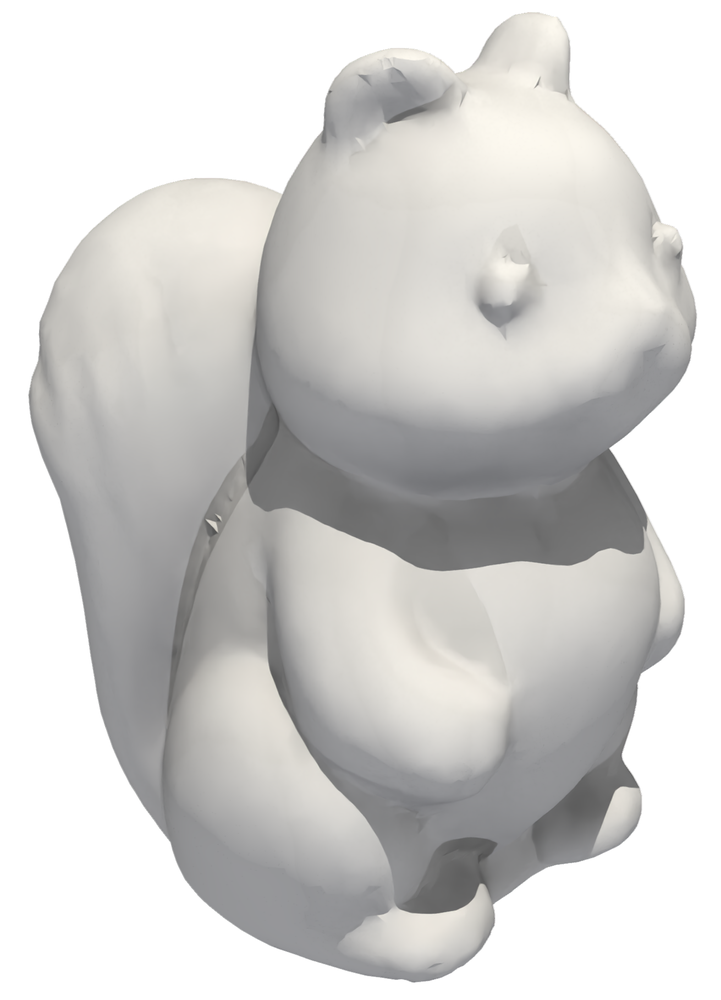} &
    \includegraphics[angle=0,origin=c,width=1.2cm]{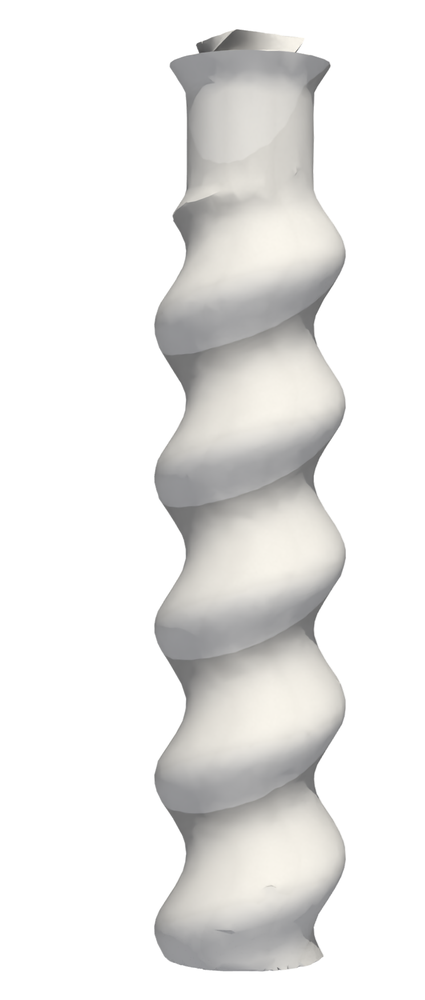} &
    \includegraphics[angle=0,origin=c,width=1.5cm]{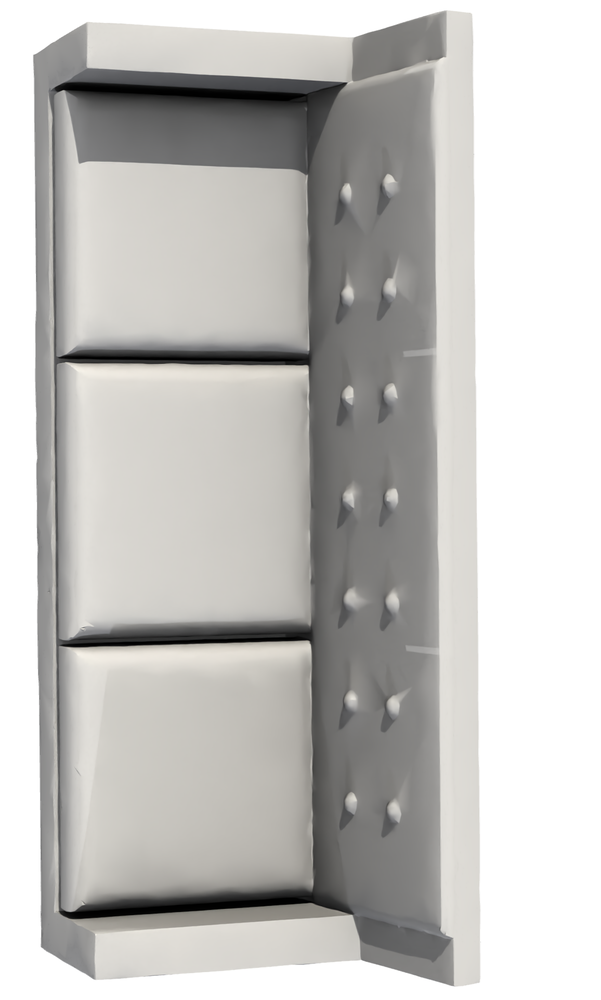} \\

    \includegraphics[angle=0,origin=c,width=1.4cm]{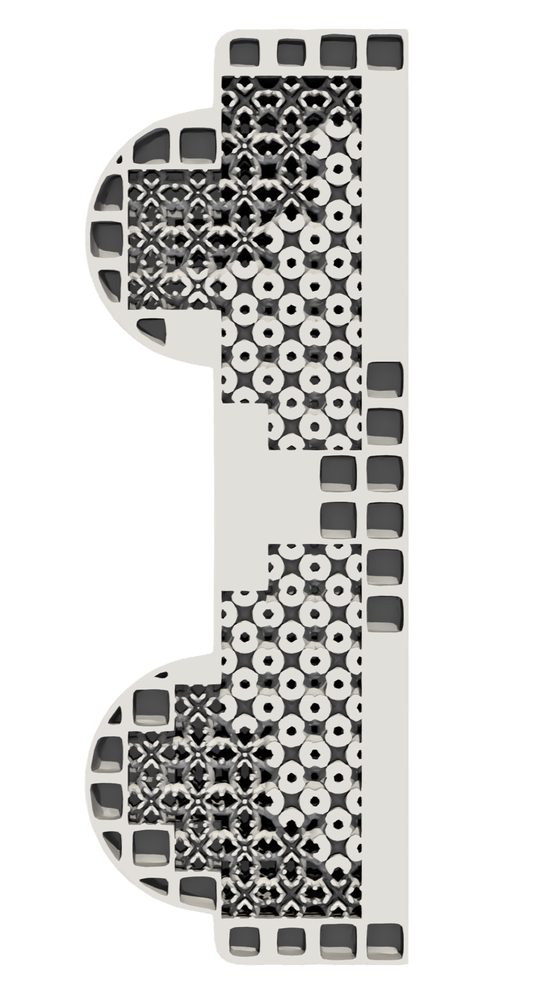} & \includegraphics[angle=0,origin=c,width=1.2cm]{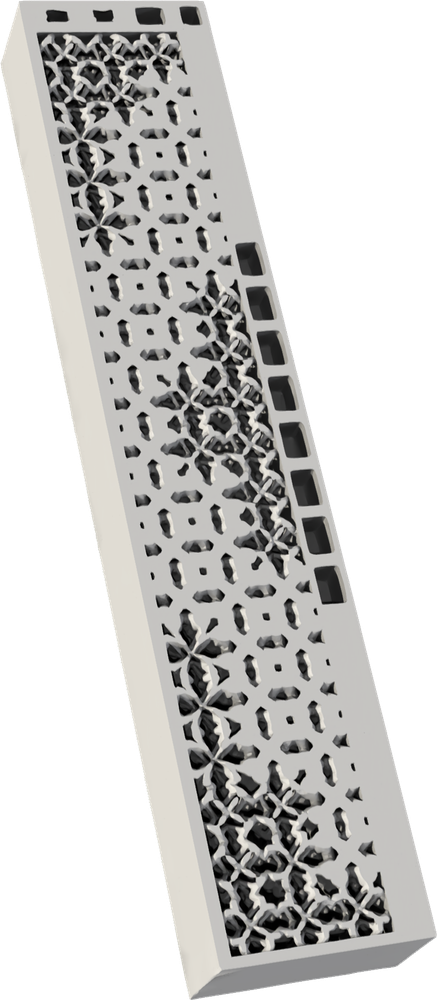} & \includegraphics[angle=0,origin=c,width=1.2cm]{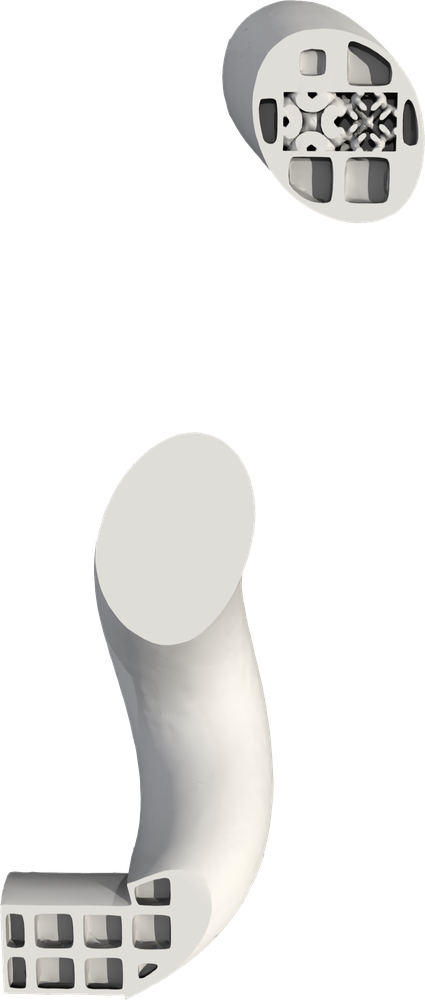} & \includegraphics[angle=0,origin=c,width=1.5cm]{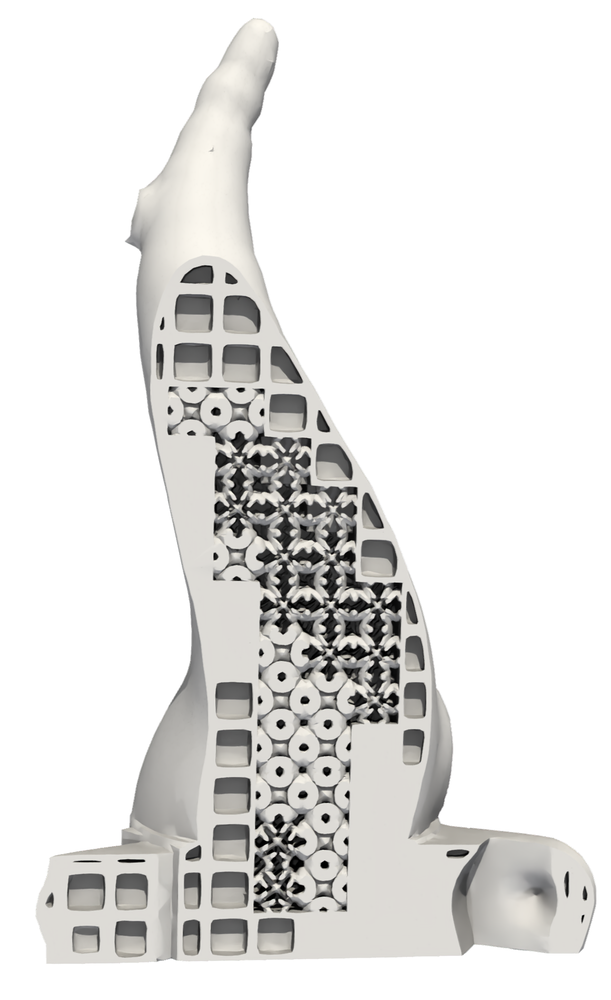} & \includegraphics[angle=0,origin=c,width=1.7cm]{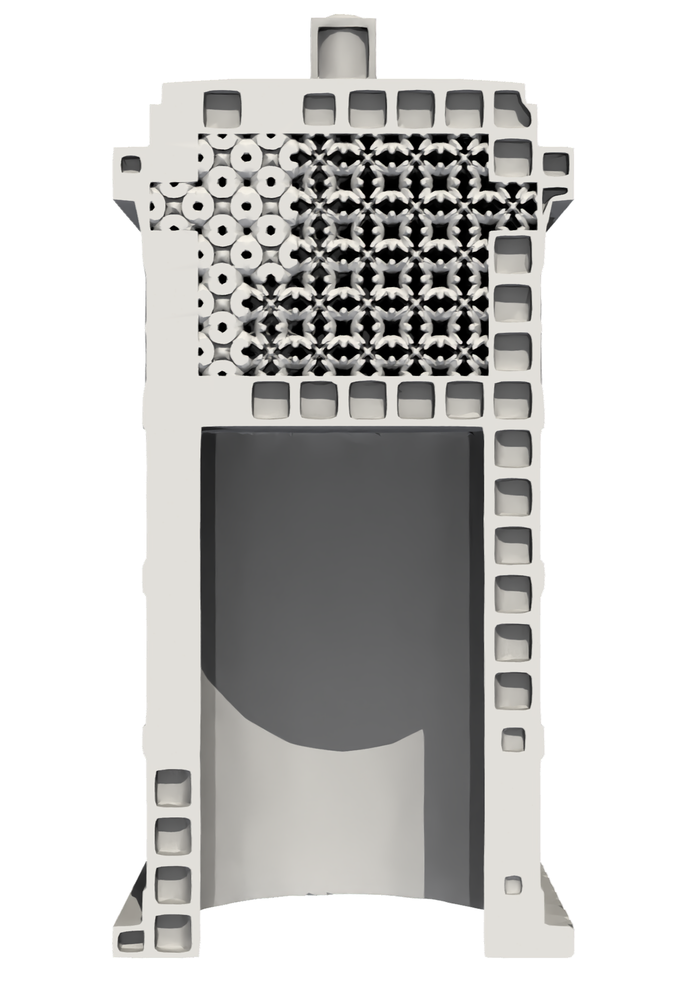} & \includegraphics[angle=0,origin=c,width=1.3cm]{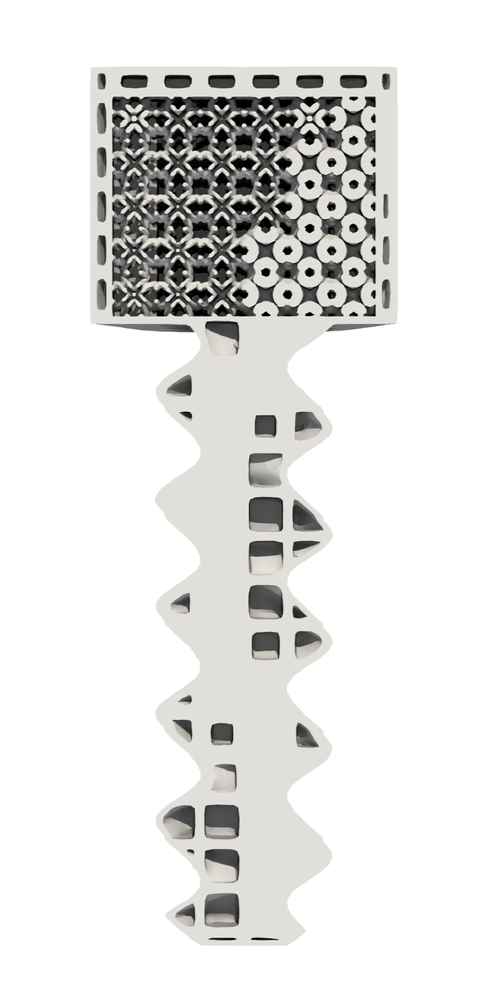} & \includegraphics[angle=0,origin=c,width=1.7cm]{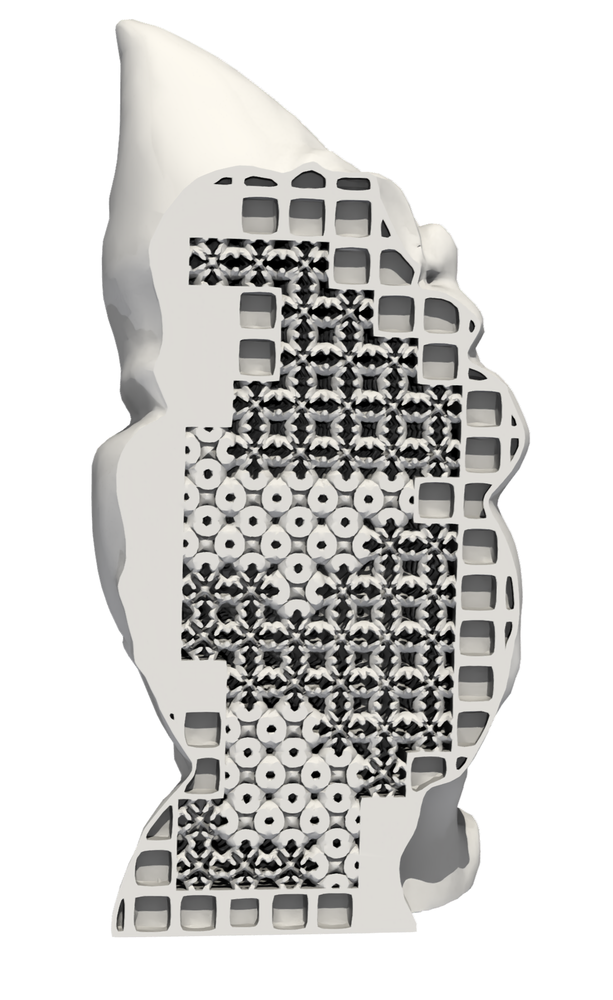} & \includegraphics[angle=0,origin=c,width=1.7cm]{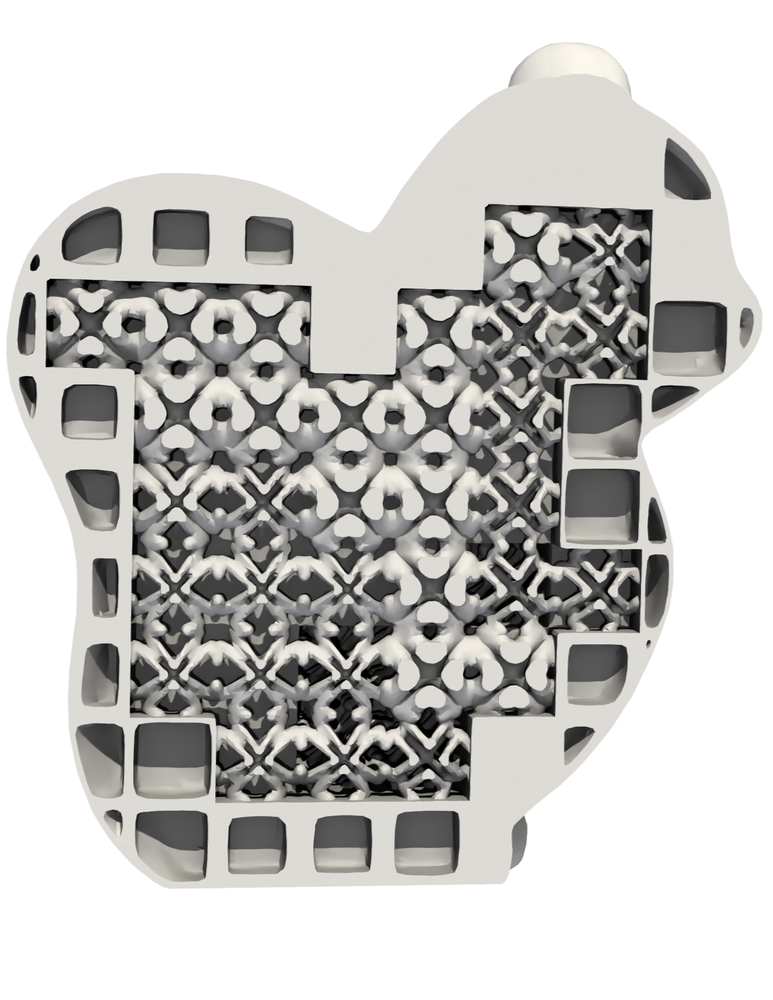} & \includegraphics[angle=0,origin=c,width=1.2cm]{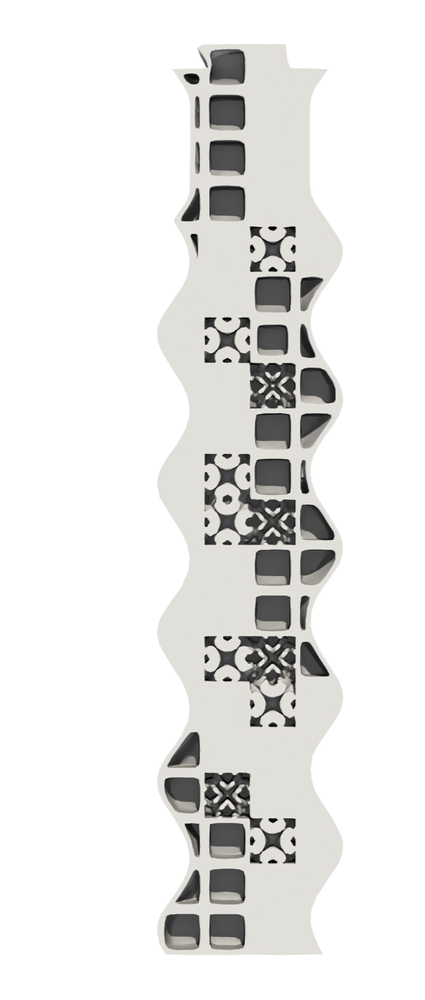} & \includegraphics[angle=0,origin=c,width=1.5cm]{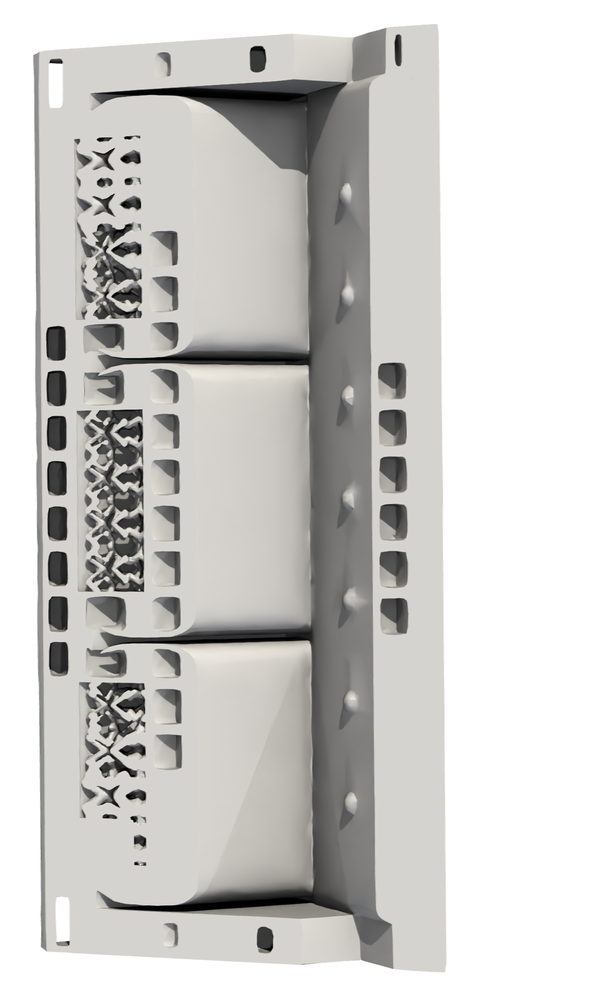} \\

    \includegraphics[angle=0,origin=c,width=1.4cm]{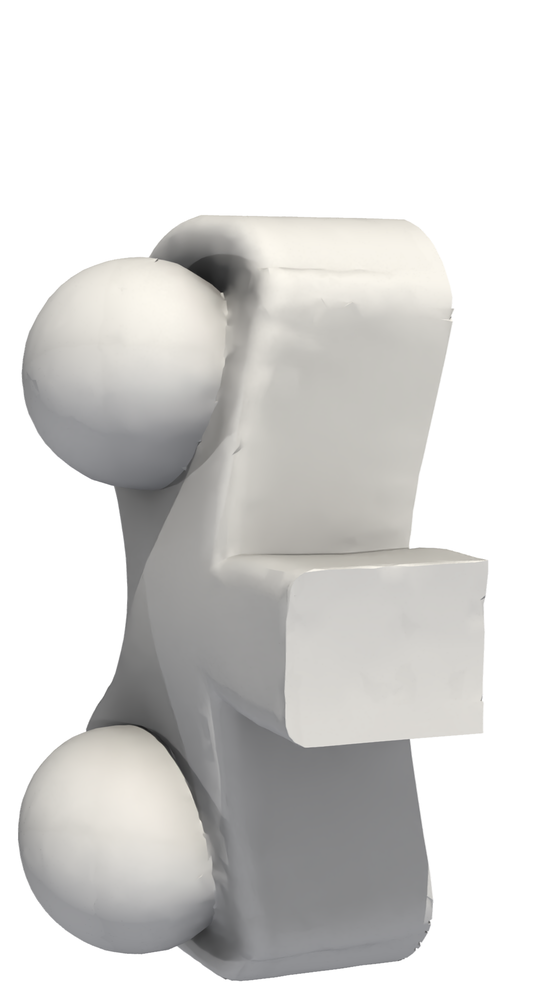} & \includegraphics[angle=0,origin=c,width=1.2cm]{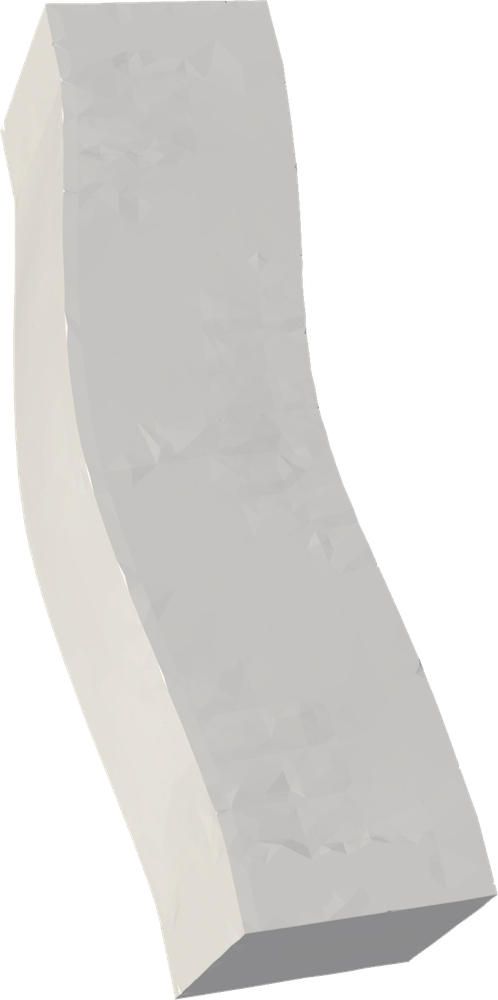} & \includegraphics[angle=0,origin=c,width=1.2cm]{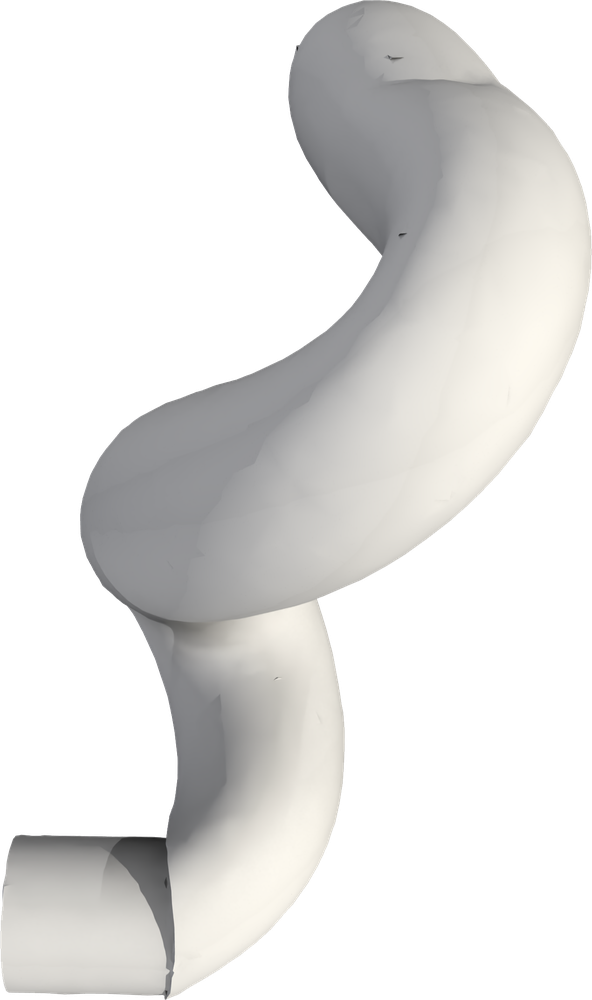} & \includegraphics[angle=0,origin=c,width=1.5cm]{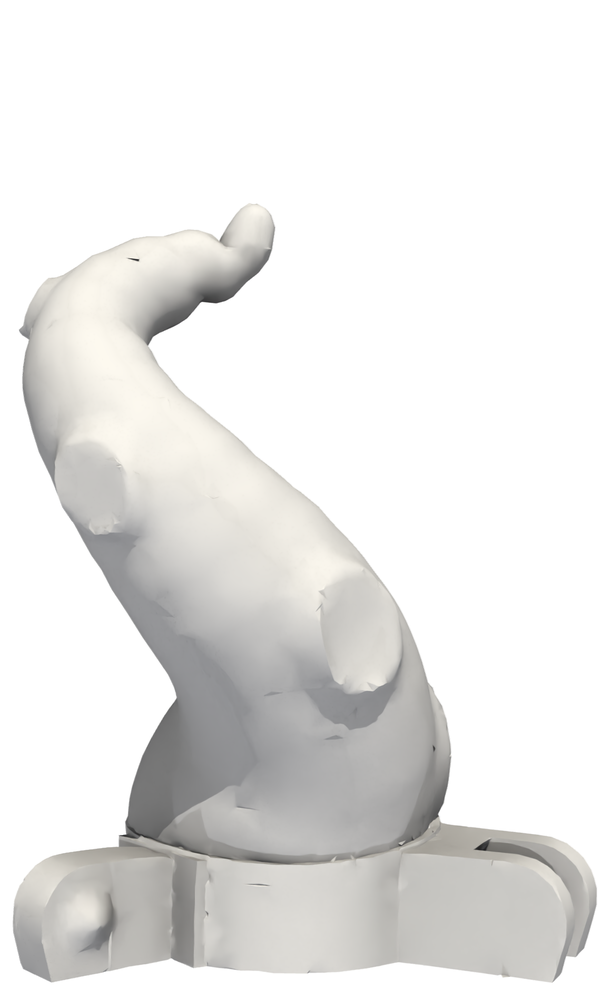} & \includegraphics[angle=0,origin=c,width=1.7cm]{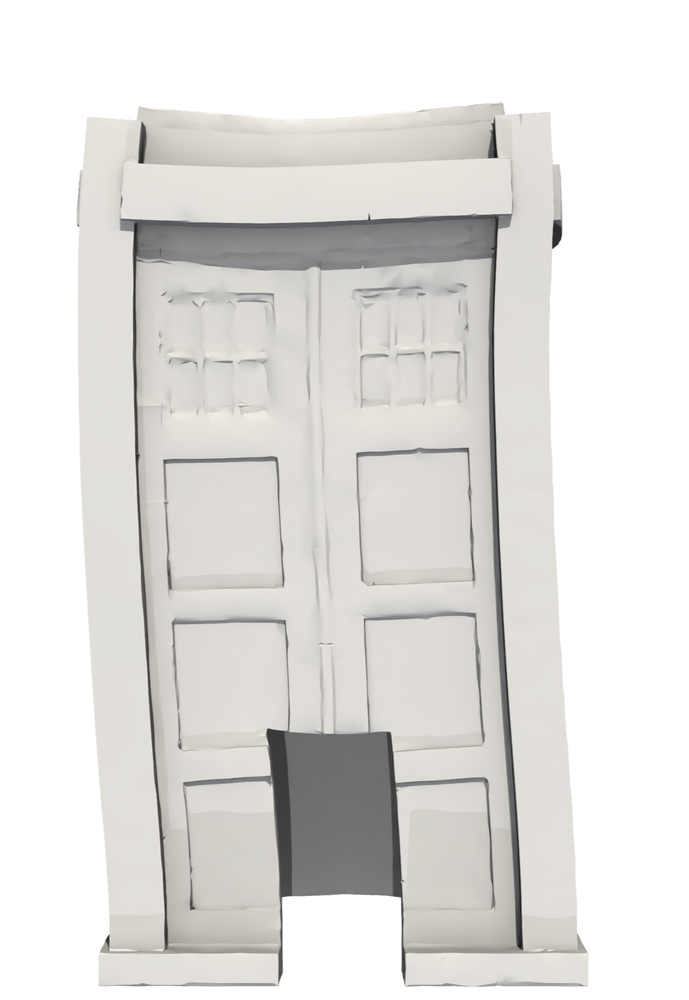} & \includegraphics[angle=0,origin=c,width=1.2cm]{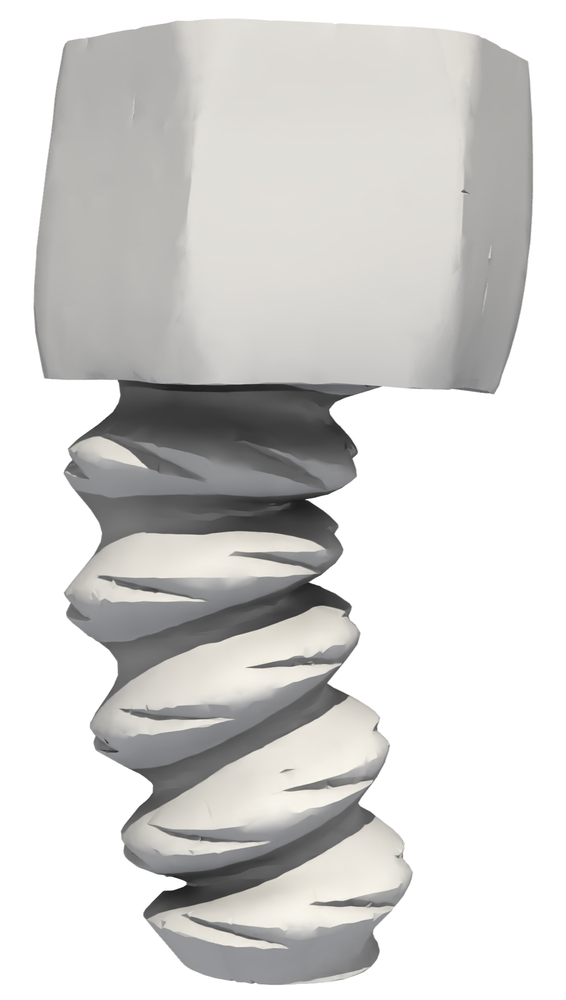} & \includegraphics[angle=0,origin=c,width=1.7cm]{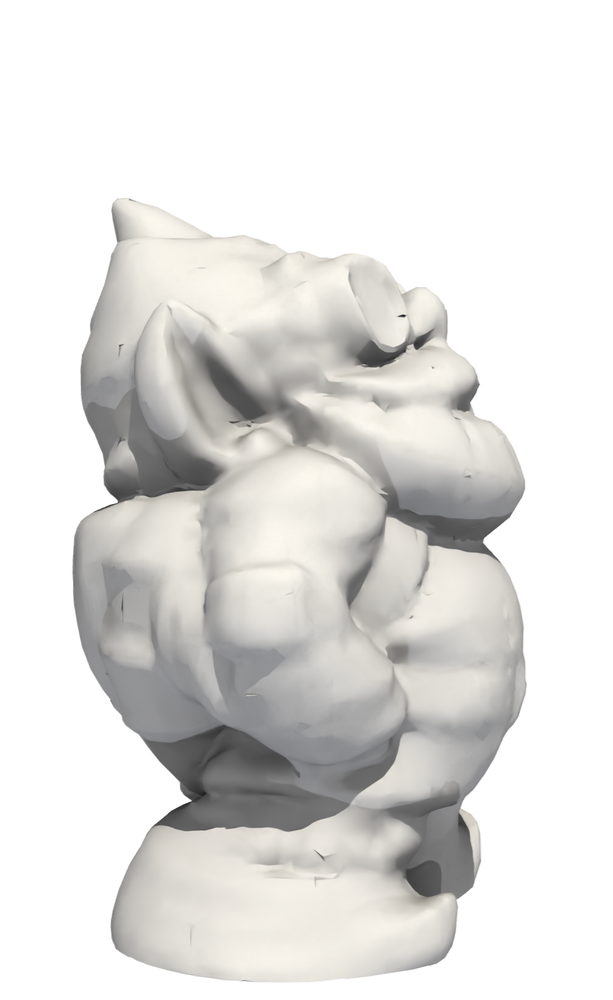} & \includegraphics[angle=0,origin=c,width=1.7cm]{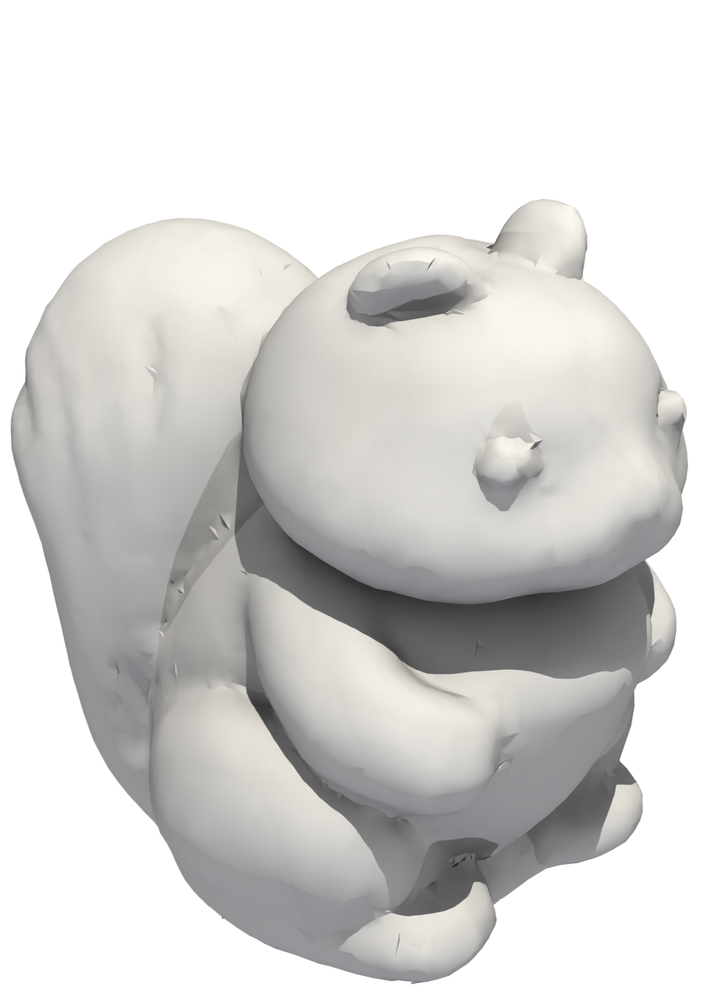} & \includegraphics[angle=0,origin=c,width=1.2cm]{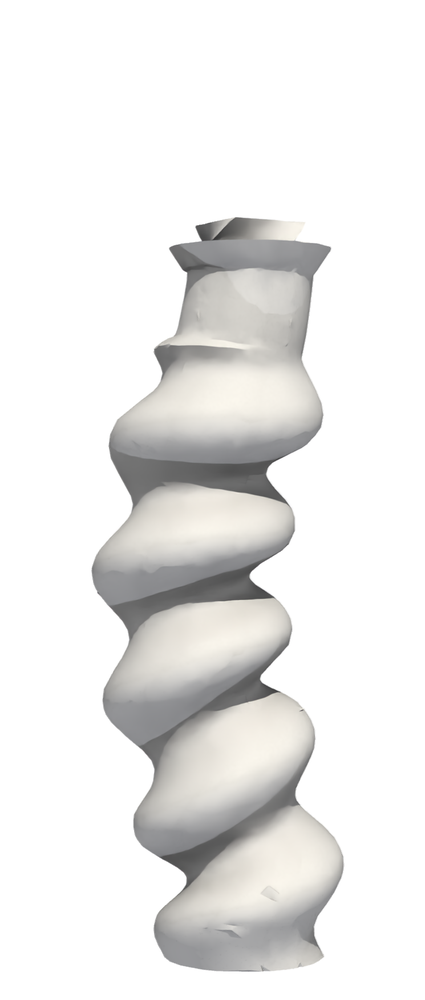} & \includegraphics[angle=0,origin=c,width=1.5cm]{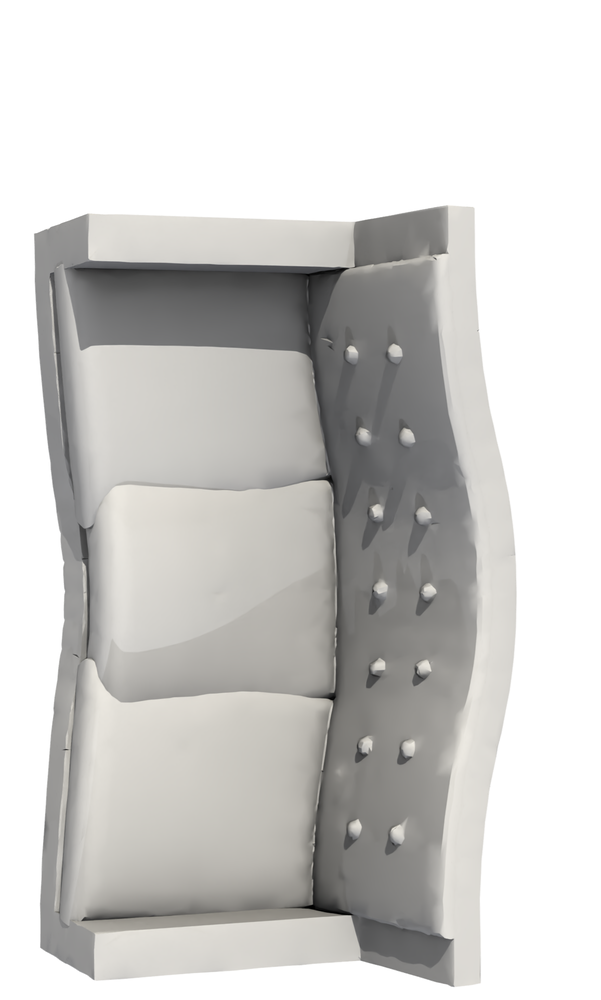}
\end{tabular}

}

\caption{
3D examples computed in batch mode. Top: rest shape. Middle: optimized pattern. Bottom: deformed shape.
}
\label{fig:3D_dataset}
\end{figure*}

\subsection{3D Examples}

Our algorithm reliably handles complex geometrical models. We show a gallery of our models and some fabricated results below. Numerical results for these models are shown in Table \ref{fig:3D_examples}, which show a similar trend to the 2D results discussed above. For physical validation, we fabricated the gripper and bar example with TPU 88A, using the 3D printing service Xometry. 

\begin{figure}
    \centering
\includegraphics[width=0.36\linewidth]{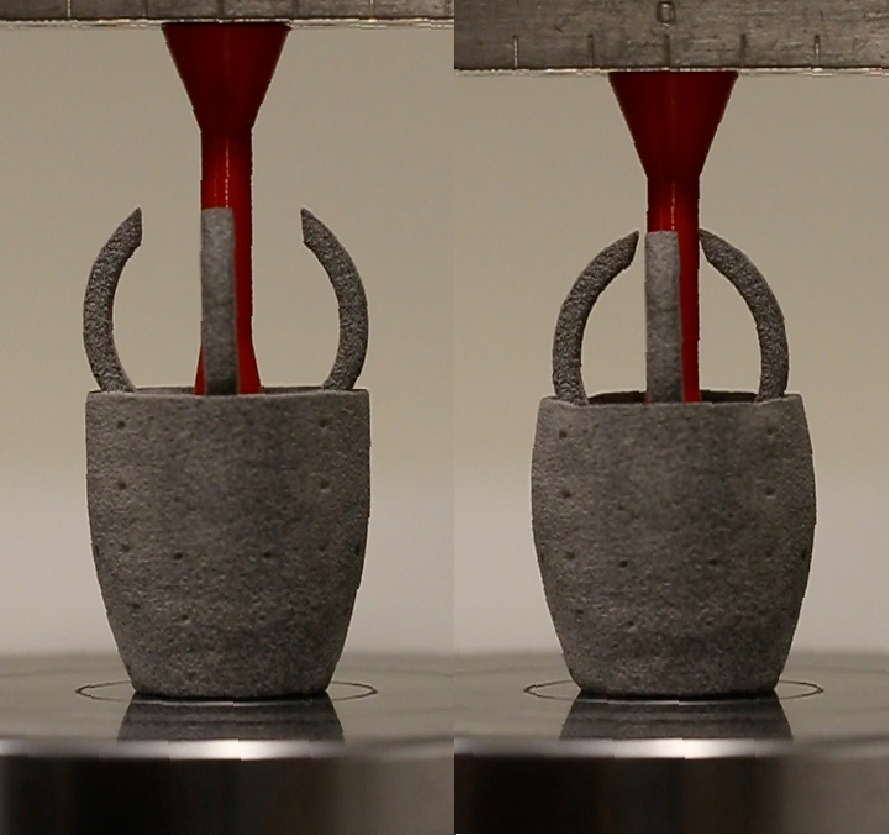}\hspace{0.05in}\includegraphics[width=0.42\linewidth]{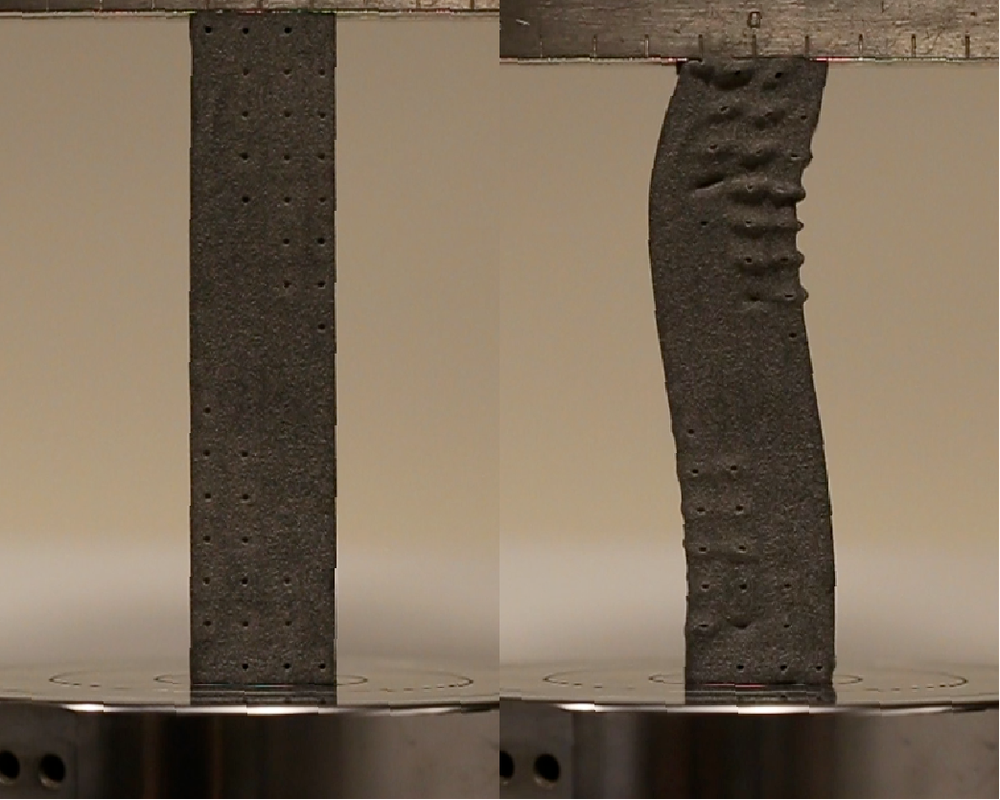}
\caption{Left image: Fabricated gripper. Rest shape on the left;  force applied to the handle on the right.
Right image: Fabricated bar. Rest shape on the left; compressed bar deforms in a controllable way, on the right.}\label{fig:3d-exp}
\end{figure}

\paragraph{Sine Bar.}
The first two rows in Figure~\ref{fig:3D_examples} show an example of a bar optimized to match a sine function when compressed. In the first row, the tiling orientation is aligned with the bar, while in the second row, we rotate the tiling to obtain similar deformation. This example shows that our method is not very sensitive to the orientation. Figure~\ref{fig:3d-exp} shows the fabricated result of the first row, which matches closely with our simulated result.

\paragraph{Gripper.}
The third row in Figure~\ref{fig:3D_examples} shows an example of a gripper, with four hands on the left and a handle on the right. We assign very stiff material to the handle (to avoid any deformation) and apply a compression force to it. The structure is optimized so that the gripper closes when the force is applied to the handle.  We show a fabricated gripper in Figure~\ref{fig:3d-exp}, with force applied. 

\paragraph{Shoe sole.}
The fourth row in Figure~\ref{fig:3D_examples} shows an example of a box with a curved top surface. Its structure is optimized so that the deformed top surface is as flat as possible for a uniform applied force. 

\paragraph{Bird.}
The last row in Figure~\ref{fig:3D_examples} shows a bird model whose structure is optimized so that its wings go upwards as its head is pushed down. 

\paragraph{Batch Examples.}
To further illustrate the robustness and generality of our algorithm, we show an additional gallery of 10 examples in Figure \ref{fig:3D_dataset}, computed in batch mode using a fixed objective and fixed boundary conditions. The model surface in the top 5\% of the bounding box is fixed, and the surface in the bottom 5\% is compressed by 10\% using Dirichlet boundary conditions. The optimization maximizes the horizontal displacement under deformation integrated over the middle 10\% surface.
We observe that the algorithm generates intricate geometries (for reference, the size of the tetrahedral meshes used for the final simulation varies between 0.5 and 15 million tetrahedra), adapting to the complex boundaries while filling the interior with the microstructure geometry.

\subsection{Alternative methods}
\label{sec:baselines}

We compare our method against two simple baseline approaches to handle boundary cells and the method of \cite{tozoni2020low} (applicable only to 2D). 
In addition, we perform an ablation study to evaluate the effect of the second optimization.

\paragraph{Trimmed boundary microstructures.}
A simple approach to extend two-scale microstructures to shapes with a complex boundary is trimming the parts outside the boundary shape \cite{schumacher2015microstructures}. The downsides are obvious: arbitrarily cutting a microstructure may rxesult in dramatic uncontrollable changes in the effective material properties, as well as entirely disconnected structures, in addition to not having a surface layer, one of the major goals of our method. We show a comparison between this method and ours in Figure~\ref{fig:comparison-2D}, where it is possible to see that the method does not generally behave well -- the results strongly depend on how the grid lines intersect the object.

\paragraph{Solid boundary cells.}
An alternative, which produces a surface layer, is filling all boundary cells with the base material, i.e. forcing the volume fraction in boundary cells to be 1 in the optimization. However, this adds uncontrollable stiffness to the object, especially in thinner parts. The numerical evaluation in Table \ref{tab:examples-numerical} shows that our approach consistently outperforms this method in both reduced (i.e., using solid cells with variable material properties, as in the output of Step 2) and full (using full microstructure geometry) simulations. Additional comparisons can also be seen in Figures \ref{fig:comparison-2D} and \ref{fig:comparison-3D}.

\begin{figure}[t]
    \centering
    \includegraphics[width=0.9\columnwidth]{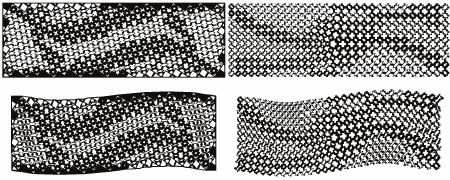}
\caption{
Bar example. Dirichlet conditions on both sides compress the object while targeting a sin wave shape at the top and bottom.  On the left, the result obtained with our cut-cell method; on the right, the original result obtained by~\cite{tozoni2020low}.}
\label{fig:bar-comparison}
\end{figure}

\begin{figure}
    \centering
    \includegraphics[width=0.7\linewidth]{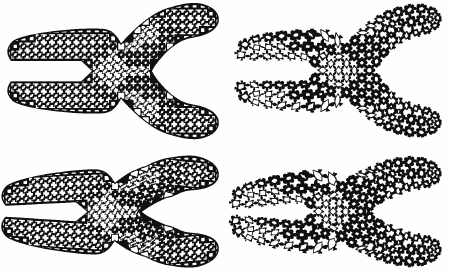}
\caption{
Pliers example (black TPU). We use Dirichlet conditions on the right to compress the handles with the target of closing the jaws. On the left, is the result obtained with our cutcell method; on the right, is the result of \cite{tozoni2020low}.
}
\label{fig:pliers-comparison}
\end{figure}

\paragraph{Rhombic microstructures, 2D only.}
~\cite{tozoni2020low} describes a method for partitioning a mesh to cells close to rhombic and conforming to the boundary, and constructs a suitable microstructure family.  This eliminates the need for a separate cut-cell microstructure family, but is hard to generalize to 3D as it requires high-quality conforming quad meshing. 
We compare four 2D optimization problems (disk, bar, ghost, and pliers, Figure~\ref{fig:2D_examples}) with this approach, making the best effort to replicate the problem setups, which is not entirely possible due to differences in formulation.  Specifically, our objective computes an integral of the target error over the object's boundary, which is the original surface for our case, while for  \cite{tozoni2020low} it corresponds to the microstructures' interface areas for every quad in the boundary since there is no solid boundary. 
We observe in  Figures \ref{fig:bar-comparison} and \ref{fig:pliers-comparison} that the results are similar between the two methods. %
We note that \cite{tozoni2020low} only applies to 2D, and its extension to 3D is highly non-trivial due to the requirement of a high-quality, isotropic, hexahedral mesher (see discussion in Section \ref{sec:related}). In contrast, our method generalizes well to 3D.

\begin{table}[t]
    \centering
    \resizebox{.96\linewidth}{!}{
    \begin{tabular}{lcccccc} %
         & Disk & Bar & Ghost  & Pliers & Bird & Sword \\ \hline
        Baseline  & 0.01213 & 0.01270 & 0.00792 & 0.00189 & 0.01207 & 0.02132 \\ \hline
        1st Opt. Reduced & 0.00710 & 0.00214 & 0.00469 & 0.00009 & 0.00847 & 0.01202 \\ 
        2nd Opt. Reduced & 0.00826 & 0.00260 & 0.00342 & 0.00006 & 0.00719 & 0.01240 \\ 
         \rowcolor{Gray}
        2nd Opt. Full & 0.01060 & 0.00565 & 0.00590 & 0.00089 & 0.01001 & 0.01834 \\ 
        1st Opt. Full  & 0.01055 & 0.01498 & 0.00579 & 0.00094 & 0.01030 & 0.01852 \\  \hline
        cells & 128 & 356 & 186 & 287 & 168 & 209 \\ 
        runtime & 538 & 687 & 1013 & 309 & 406 & 487 \\ \hline
    \end{tabular}}
    \bigskip
    \resizebox{.85\linewidth}{!}{
    \begin{tabular}{lcccccc} %
         & Bird & Gripper & Sine bar & \begin{tabular}{@{}c@{}}Sine bar \\ rotated\end{tabular} & Sole \\ \hline
        Baseline & 2.058 & 0.799 & 1.211 & 1.327 & 1.331 \\ \hline
        1st Opt. Reduced & 1.540 & 0.232 & 0.550 & 0.452 & 0.010 \\ 
        2nd Opt. Reduced & 1.080 & 0.224 & 0.635 & 0.499 & 0.030 \\ 
        \rowcolor{Gray}
        2nd Opt. Full & 1.620 & 0.030 & 0.586 & 0.437 & 0.259 \\
        1st Opt. Full & 1.729 & 0.085 & 0.550 & 0.432 & 0.289 \\ \hline
        cells & 4442 & 400 & 936 & 1060 & 1475 \\
        runtime & 20033 & 1738 &  2461 & 3062 & 3249 \\ %
    \end{tabular}}
    \caption{Numerical comparisons of our approach vs. a baseline using solid boundary cells filled with base material in 2D and 3D. We report the displacement objective term in Equation~\ref{eq:material-objective}, which represents the average point-wise error from the prescribed target for different stages of the pipeline, see Figure~\ref{fig:summary-steps} for illustrations.
     The last two rows show the number of cells and runtime (\unit{\sec}).
     }
    \label{tab:examples-numerical}
\end{table}

\paragraph{Numerical results and an ablation study for second optimization.}
Numerical comparisons for the examples in the paper are shown in Table \ref{tab:examples-numerical}. 
The first optimization stage is very effective for all models (\emph{1st Optimization Reduced} row); however, inserting cut-cell microstructures significantly increases the objective values in some cases (\emph{1st Optimization Full} row);
The second optimization stage  (\emph{2nd Optimization Reduced/Full} rows)
significantly reduces this effect in many cases (e..g, 2D Bar, 3D Bird, Gripper), but may slightly increase errors in some cases, (e.g. 2D Disk and Ghost).
This small objective increase is dominated by the objective increase due to insertion of the microstructure.  Because first and second optimizations produce different target material parameter values for interior cells, corresponding microstructures in two cases may have different errors in approximation of these material properties and random variation in these errors may result in the observed increase in the objective. 

Both the full pipeline with the second optimization and the reduced pipeline produce a large improvement compared to the baseline in all cases.

\section{Conclusions}
\label{sec:concl}

We have presented a \emph{fully automated} pipeline starting from an arbitrary 2D or 3D shape and prescribed boundary conditions, and target deformation,  to a complete manufacturable geometric shape with microstructure infill, retaining the input surface, and approximating the target behavior. 

\paragraph{Limitations and future work.}
Our approach has several limitations.  First, for thin objects and/or relatively coarse cells, most cells would be cut cells, and the quality of the results may be lower.  Some of our examples suffer from buckling of the surface layer, as no terms penalizing buckling are included in our optimization.
Second, we use the linear elasticity model in our examples.
Extending our work to nonlinear elasticity, including contact, would reduce the simulation-to-reality gap and could extend  the applicability of our method to more complicated scenarios. However, this would require a more
complex mapping between material properties and microstructure geometry.

In addition, in this work, we explored one family of structures for boundary cells. There are many possible ways to develop it further;  for example, rather than keeping it entirely fixed during the second optimization stage, one can combine material optimization for interior cells with low parametric shape optimization for the boundary cells.  The results can also be improved by using a broader spectrum of microstructure topologies.

% \bibliographystyle{ACM-Reference-Format}
% \bibliography{99-biblio}

\newpage

\newpage

\end{document}